\documentclass[a4paper,twocolumn,showpacs,superscriptaddress,floatfix]{revtex4-1}

\newcommand\aap{Astron. Astrophys.}

\usepackage{latexsym}
\usepackage{amssymb}
\usepackage{amsfonts}
\usepackage{amsmath}
\usepackage{bm}
\usepackage{graphicx}   
\usepackage[squaren]{SIunits} 
\usepackage{color}
\usepackage{subfigure}
\usepackage{times}
\usepackage{units}
\usepackage{hyperref}

\graphicspath{ {./} }

\newcommand{\kHz}{\usk\kilo\hertz} 
\newcommand{\kmeter}{\usk\kilo\meter} 
\newcommand{\msec}{\usk\milli\second} 
\newcommand{\parsec}{\mathrm{pc}}

\newcommand{\Fref}[1]{Fig.~\ref{#1}} 
\newcommand{\Sref}[1]{Sec.~\ref{#1}} 
\newcommand{\Tref}[1]{Table~\ref{#1}}

\newcommand{\mnras}{{Mon. Not. R. Astron. Soc.}}

\begin{document}

\title[GRMHD simulations of BNS mergers]{General relativistic magnetohydrodynamic simulations of binary neutron 
star mergers \\forming a long-lived neutron star}

\author{Riccardo Ciolfi}
\address{Physics Department, University of Trento, 
via Sommarive 14, I-38123 Trento, Italy}
\address{INFN-TIFPA, Trento Institute for Fundamental Physics
and Applications, via Sommarive 14, I-38123 Trento, Italy}
\address{INAF, Osservatorio Astronomico di Padova, Vicolo 
dell'Osservatorio 5, I-35122 Padova, Italy}

\author{Wolfgang Kastaun}
\address{Physics Department, University of Trento, 
via Sommarive 14, I-38123 Trento, Italy}
\address{INFN-TIFPA, Trento Institute for Fundamental Physics
  and Applications, via Sommarive 14, I-38123 Trento, Italy}

\author{Bruno Giacomazzo}
\address{Physics Department, University of Trento, 
via Sommarive 14, I-38123 Trento, Italy}
\address{INFN-TIFPA, Trento Institute for Fundamental Physics
  and Applications, via Sommarive 14, I-38123 Trento, Italy}

\author{Andrea Endrizzi}
\address{Physics Department, University of Trento, 
via Sommarive 14, I-38123 Trento, Italy}
\address{INFN-TIFPA, Trento Institute for Fundamental Physics
  and Applications, via Sommarive 14, I-38123 Trento, Italy}

\author{Daniel M. Siegel}\altaffiliation{NASA Einstein Fellow}
\address{Physics Department and Columbia Astrophysics
  Laboratory, Columbia University, New York, NY 10027, USA}

\author{Rosalba Perna}
\address{Department of Physics and Astronomy, Stony Brook University, Stony Brook, NY 11794, USA}

\date{\today}

\begin{abstract}
Merging binary neutron stars (BNSs) represent the ultimate targets for multimessenger 
astronomy, being among the most promising sources of gravitational waves (GWs), and,
at the same time, likely accompanied by a variety of electromagnetic 
counterparts across the entire spectrum, possibly including short gamma-ray 
bursts (SGRBs) and kilonova/macronova transients.
Numerical relativity simulations play a central role in the study of these events.
In particular, given the importance of magnetic fields, various aspects of this investigation 
require general relativistic magnetohydrodynamics (GRMHD).
So far, most GRMHD simulations focused the attention on BNS mergers leading to the 
formation of a hypermassive NS, which, in turn, collapses within few tens of ms into a
black hole surrounded by an accretion disk. However, recent observations suggest
that a significant fraction of these systems could form a 
long-lived NS remnant, which will either collapse on much longer timescales or 
remain indefinitely stable.
Despite the profound implications for the evolution and the emission properties 
of the system, a detailed investigation of this alternative evolution channel is 
still missing.
Here, we follow this direction and present a first detailed GRMHD study of 
BNS mergers forming a long-lived NS. We consider magnetized binaries
with different mass ratios and equations of state and analyze the 
structure of the NS remnants, the rotation profiles, the accretion disks, 
the evolution and amplification of magnetic fields, and the ejection of matter.
Moreover, we discuss the connection with the central 
engine of SGRBs and provide order-of-magnitude estimates for the 
kilonova/macronova signal. Finally, we study the GW emission, with particular 
attention to the post-merger phase. 
\end{abstract}


\maketitle


\section{Introduction}
\label{sec:intro}

With the discovery of binary black hole (BH) mergers by the Laser
Interferometer Gravitational Wave Observatory (LIGO),
the era of gravitational wave (GW) astronomy and multimessenger
astronomy including GWs has begun
\cite{LIGO:BBHGW:2016,LIGO_GW151226,Abbott2016d}. 
As the advanced LIGO and Virgo detectors approach design 
sensitivity in the next few years \cite{Aasi:2015,Accadia:2011:114002}, 
exciting new discoveries could be made, including binary neutron 
star (BNS) and NS--BH mergers \cite{Abadie2010,Abbott2016c}.
Due to the absence of baryonic matter in these systems, stellar-mass 
binary BH mergers are not expected to produce bright electromagnetic 
(EM) counterparts to their GW signal (but see, e.g.,~\cite{Perna2016}). 
Instead, mergers involving NSs
are expected to link the EM and GW skies. Furthermore, these 
mergers are also of wide interest as they offer a unique opportunity 
to constrain the equation of state (EOS) of matter at supranuclear 
densities (e.g., \cite{Shibata:2005:84021,Bauswein:2013:131101}) and 
provide a prime candidate astrophysical site for the production of heavy 
elements in the universe, via r-process nucleosynthesis in the matter 
ejected during and possibly after merger (e.g., \cite{Just2015a,Wu2016,Roberts2017a}).

Mergers involving NSs are expected to generate EM emission across the
entire EM spectrum and over a variety of timescales
\cite{Metzger2012}. Detection of EM
counterparts will enable the identification of the host
galaxy and its position within/relative to the host, which will
provide valuable information on binary formation channels, age of the
stellar population, and supernova birth kicks
\cite{Berger2014}. Additionally, by
measuring redshifts, EM counterparts can determine the distance to
the source and help alleviate degeneracies in the GW parameter
estimation between distance and inclination of the binary. 
Moreover, combined GW and EM observations can prove the 
connection between short gamma-ray bursts (SGRBs) and BNS or 
NS-BH mergers (see below), revealing crucial information on when 
and how a SGRB can be produced.
Finally, even without a GW detection, EM counterparts can reveal 
exclusive information on the very rich physics of the merger and 
post-merger evolution, especially if the merger remnant is a massive NS 
\cite{Metzger2014a,Siegel:2016a,Siegel:2016b}.

SGRBs are among the earliest proposed
counterparts to BNS and NS-BH mergers \cite{Paczynski1986,Eichler:1989:126,Narayan1992,Barthelmy2005a,Fox2005,Gehrels2005,Shibata2006b,Tanvir2013,Berger2013,Paschalidis:2015:14,Ruiz2016}. 
The standard paradigm explains the formation of a SGRB via a relativistic outflow
(jet) generated by a torus of matter accreting onto a remnant BH. Although
there is tentative evidence for this scenario on the basis of
previous general-relativisitc magnetohydrodynamic (GRMHD) simulations
\cite{Paschalidis:2015:14,Ruiz2016}, much still remains to be understood.
Moreover, if the merger leads to the formation of a long-lived NS instead of a
BH, which, as we argue below, can occur in an order unity fraction of
all BNS merger events (but not in NS-BH mergers), baryon pollution in the
surrounding of the merger site \cite{Hotokezaka2013a,Oechslin2007,Bauswein:2013:78,Kastaun:2015:064027,Dessart:2009:1681,Siegel:2014:6,Metzger2014c} 
can choke a relativistic outflow
\cite{Murguia-Berthier2014,Nagakura2014,Murguia-Berthier2017a} or even
prevent its formation in the first place. 
Ref. \cite{Ciolfi:2015:36} has proposed the ``time-reversal''
scenario, in which the problem of baryon pollution can be avoided, with
additional important observational consequences (see
\cite{Rezzolla2015} for an alternative proposal). In order to explore the SGRB-merger 
connection for the BNS case, more simulations of systems with different properties 
are required, to examine in detail the merger and early post-merger dynamics and to better
quantify the amount of baryon pollution and thus the potential for
generating relativistic outflows.
Furthermore, magnetic fields are likely to play a key role in the formation
of a jet and therefore investigating the nature of SGRBs demands GRMHD 
simulations.

Kilonovae or macronovae represent another important EM
counterpart to the GW signal of BNS and NS-BH mergers
\cite{Li:1998:L59,Kulkarni:2005:macronova-term,Rosswog2005,Metzger2010,Roberts2011,Barnes2013,Piran2013,Tanvir2013,Berger2013,Yang2015}. 
These thermal transients at optical and infrared wavelengths
and timescales of days to weeks are powered by heating from
radioactive decay of r-process elements produced in the expanding
sub-relativistic ejecta. The amount of r-process material synthesized
in the dynamical ejecta (e.g., 
\cite{Hotokezaka2013a,Oechslin2007,Bauswein:2013:78,Kastaun:2015:064027,Radice2016}) 
and in winds from the remnant object \cite{Dessart:2009:1681,Siegel:2014:6}, 
or from a remnant accretion
disk/torus \cite{Fernandez2013,Just2015a} depend sensitively on
properties of the matter outflows at launch, such as the distributions in
mass, velocity, entropy, and electron fraction. Numerical simulations
are necessary to investigate these properties in detail. 

\begin{table*}
  \caption{Initial data parameters: mass ratio ($q=M_\mathrm{g}^1/M_\mathrm{g}^2$),
    total baryonic mass of the system ($M_\mathrm{b}^\mathrm{tot}$), baryonic and 
    gravitational masses of each star at infinite separation 
    ($M_\mathrm{b}$ and $M_\mathrm{g}$), compactness ($M_\mathrm{g}/R_c$,
    dimensionless), initial orbital frequency
    and proper separation ($f_0$ and $d$), initial
    magnetic energy ($E_\mathrm{mag}$), initial maximum value of
    magnetic field strength ($B_\mathrm{max}$), and $A_b$, 
    the value in geometric units used in equation~(\ref{eq:Avec}) 
    in order to fix $B_\mathrm{max}$.}
  \begin{ruledtabular}\begin{tabular}{lllllll}
      Model & APR4 equal & APR4 unequal & MS1 equal & MS1 unequal & H4 equal & H4 unequal \\
      \hline
      $q$                        & $1$     & $0.90$        & $1$     & $0.91$      & $1$     & $0.91$ \\
      $M_\mathrm{b}^\mathrm{tot}$ [$M_\odot$]  & $2.98$  & $2.98$    & $2.91$  & $2.91$ & $2.92$  & $2.92$ \\
      $M_\mathrm{b}$ [$M_\odot$]        & $1.49$  & $1.58,1.41$ & $1.45$  & $1.53,1.38$ & $1.46$  & $1.54,1.38$ \\
      $M_\mathrm{g}$ [$M_\odot$]        & $1.35$  & $1.42,1.28$ & $1.35$  & $1.41,1.28$ & $1.35$  & $1.42,1.29$ \\
      $M_\mathrm{g}/R_c$                  & $0.176$  & $0.185,0.167$ & $0.134$  & $0.140,0.127$ & $0.143$  & $0.150,0.135$ \\
      $f_0$ [Hz]                 & $283$  & $284$ & $287$  & $287$ & $287$  & $286$ \\
      $d$ [km]                   & $59$  & $59$ & $57$  & $57$ & $58$  & $58$ \\
      $E_\mathrm{mag}$ [$10^{47}$erg]     & $2.42$  & $2.42$ & $2.42$  & $2.42$ & $2.42$  & $2.42$ \\
      $B_\mathrm{max}$ [$10^{15}$G]     & $3.00$  & $3.51,2.37$ & $2.05$  & $2.36,1.70$ & $2.42$  & $2.91,1.89$ \\
      $A_b$                     & $776$  & $748$ & $4714$  & $4609$ & $2816$  & $2720$
  \end{tabular}\end{ruledtabular}
   \label{tab:init_param}
\end{table*}

While NS-BH mergers inevitably end up in a BH possibly surrounded 
by a massive accretion disk, BNS mergers can lead to qualitatively different
remnants. 
Depending on the EOS and the component masses, the BNS can form a BH
(prompt collapse), a hypermassive NS (HMNS; NS with mass above the maximum
mass for uniformly rotating configurations), or a long-lived NS, which
we assume to be either supramassive (SMNS; NS with mass above the maximum
mass $M_\mathrm{TOV}$ for non-rotating configurations) or indefinitely stable.
HMNSs typically collapse to a BH on a timescale of
$\sim\!\text{ms}$ to $\sim\!100\,\text{ms}$, while SMNSs  
can typically survive for minutes or even much longer.
It is commonly believed that HMNSs are supported against collapse by rapid 
rotation of the core (see \cite{Baumgarte:2000:29} for such HMNS models) and 
consequently collapse when enough differential rotation is removed (via GW 
emission or electromagnetic torques \cite{Shibata2006a,Duez2006b,Siegel:2013:121302}).
SMNS are thought to be supported by uniform rotation and to collapse when enough angular 
momentum is carried away via magnetic dipole radiation and GWs. In contrast, a growing 
number of simulations \cite{Kastaun:2015:064027, Endrizzi:2016:164001, Kastaun:2016, 
Kastaun:2016b:arxiv, Hanauske2016arXiv} indicate that both HMNSs and SMNSs typically 
have slowly rotating cores, and that collapse is rather avoided because a significant
amount of matter in the \textit{outer} layers approaches Kepler velocity. This implies 
that the exact mechanism leading to collapse is still poorly understood, which has 
important consequences when interpreting the lifetimes of HMNSs and SMNSs. 
Therefore, special attention should be paid to the rearrangement of the radial remnant 
structure preceding collapse.

BNS mergers leading to a hypermassive, supramassive or stable 
NS are characterized by a post-merger phase in which GW emission can still be significant
for several tens of ms (or more) and in general much stronger than the short and weak 
BH ringdown signal. 
This post-merger GW emission carries the signature of the remnant structure and 
represents a promising way to constrain the NS EOS. In particular, the spectrum 
always shows a dominant peak at a frequency that strongly depends on the EOS
(e.g., \cite{Bauswein:2012:11101,Bauswein:2014:23002,Takami:2014:91104}). 

In this paper, we perform a set of GRMHD simulations of BNS mergers with 
different EOS and mass ratios, focusing most of the attention on systems leading to 
the formation of a long-lived remnant NS (i.e.~supramassive or stable). For comparison, 
we also consider two BNS mergers forming a HMNS that collapses to a BH by the end 
of the simulation. With $M_\mathrm{TOV}\gtrsim 2\,M_\odot$
\cite{Demorest:2010:1081,Antoniadis:2013:448}, the maximum mass of 
uniformly rotating configurations $\sim\!20\%$ larger, i.e.~$M_\mathrm{supra}\approx
1.2\,M_\mathrm{TOV}\gtrsim 2.4\,M_\odot$ \cite{Lasota1996}, 
and a typical remnant mass between $2.3-2.5\,M_\odot$ 
when accounting for mass loss and neutrino and GW emission \cite{Belczynski2008}, 
we expect that an important (order unity) fraction of BNS merger events should lead 
to the formation of a long-lived NS.
Despite being very likely, this case remains poorly studied in numerical relativity,
and only a few simulations of such systems were performed including magnetic fields 
(i.e.~in GRMHD) \cite{Giacomazzo2013ApJ...771L..26G,Palenzuela2015,Endrizzi:2016:164001}.

The presence of a long-lived remnant has important
consequences. First, neutrino and/or magnetically driven outflows can
provide an additional source of ejecta material for r-process
nucleosynthesis on secular timescales ($\sim\!1\,\mathrm{s}$) \cite{Dessart:2009:1681,Siegel:2014:6}. 
Second, the spindown radiation from the magnetized remnant NS represents 
an additional source of energy that can power nearly isotropic EM transients.
This emission provides a possible explanation for the long-lasting ($\sim$minutes to hours) 
X-ray afterglows observed by Swift \cite{Gehrels2004-Swift} in association with a substantial 
fraction of SGRB events \cite{Rowlinson2013,Lu2015}. 
At the same time, long-lasting afterglows are hardly explained within the popular BH-disk 
scenario of SGRBs, due to the short accretion timescale of the disk onto the BH ($\sim$seconds).
If the above interpretation is correct, this provides additional evidence that the product of BNS 
mergers is very often a long-lived NS. 
Moreover, independently from SGRBs, spindown-powered EM 
transients represent an additional and potentially very promising EM counterpart 
for multimessenger astronomy with BNS mergers 
\cite{Yu2013,Metzger2014a,Siegel:2016a,Siegel:2016b}. In addition, they may be 
connected with other astrophysical phenomena, such as fast radio bursts~\cite{Dai2016}.

Here, we initiate a systematic investigation on BNS mergers ending up in a long-lived
NS, aimed at covering all of the key aspects mentioned above. 
The paper is organized as follows. Section~\ref{sec:setup}
describes the physical models, the numerical setup and the generation 
of initial data. In Section~\ref{sec:dynamics} we discuss in detail the 
evolution from the inspiral to the post-merger phase for the different 
models. The following sections
provide a more detailed analysis of individual aspects, such as the
rotation profile of the remnant, its structure and its stability against collapse 
(Section~\ref{sec:rot}), the 
evolution of magnetic fields (Section~\ref{sec:mag}), and the implications
for SGRBs (Section~\ref{sec:sgrb}). In Section~\ref{sec:ej}
we investigate mass ejection, while Section~\ref{sec:gw} is devoted to 
the analysis of the GW emission, with 
particular emphasis on the post-merger signal. Conclusions are presented in
Section~\ref{sec:conclusion} and an appendix is added to discuss aspects
of numerical convergence.


\section{Physical models and numerical setup}
\label{sec:setup}

In this work, we study a set of magnetized BNS systems with a mass 
ratio of either $q=1$ (equal mass) or $q=0.9$ (unequal mass). The 
most relevant initial parameters of our models are 
summarized in Table~\ref{tab:init_param}. 
In the equal-mass case, each NS has a gravitational mass at infinite 
separation of $1.35$ M$_\odot$, which appears to be the most likely 
mass for NSs in a merging BNS system according to current models 
and observations (e.g., \cite{Belczynski2008,Lattimer:2012:485,Oezel2016}). 
For the unequal-mass case ($q=0.9$), we impose the same total gravitational 
mass at infinite separation. 
Both the individual masses and the mass ratios we consider span roughly 
the same range as the available BNS observations with well constrained 
masses \cite{Lattimer:2012:485,Oezel2016}.
We consider three 
different EOS to describe NS matter: APR4 \cite{Akmal:1998:1804}, 
MS1 \cite{Mueller1996}, H4 \cite{Glendenning1991}. 
These are chosen to cover a relatively wide range of compactness 
($M_\mathrm{g}/R_c\simeq0.134-0.176$ for a canonical 
$1.35$ M$_\odot$ NS). 
With the chosen masses, the final product of the merger is a SMNS for 
the APR4 EOS, a stable NS for the MS1 EOS, and a HMNS for the H4 EOS.
The latter collapses to a BH within the physical time covered by the simulations.
In order to assess the effect of magnetic fields, we also consider the equal-mass 
APR4 model without magnetic field (labelled ``B0" in the figure legends).

\begin{figure}[!ht]
  \centering
  \includegraphics[width=0.99\linewidth]{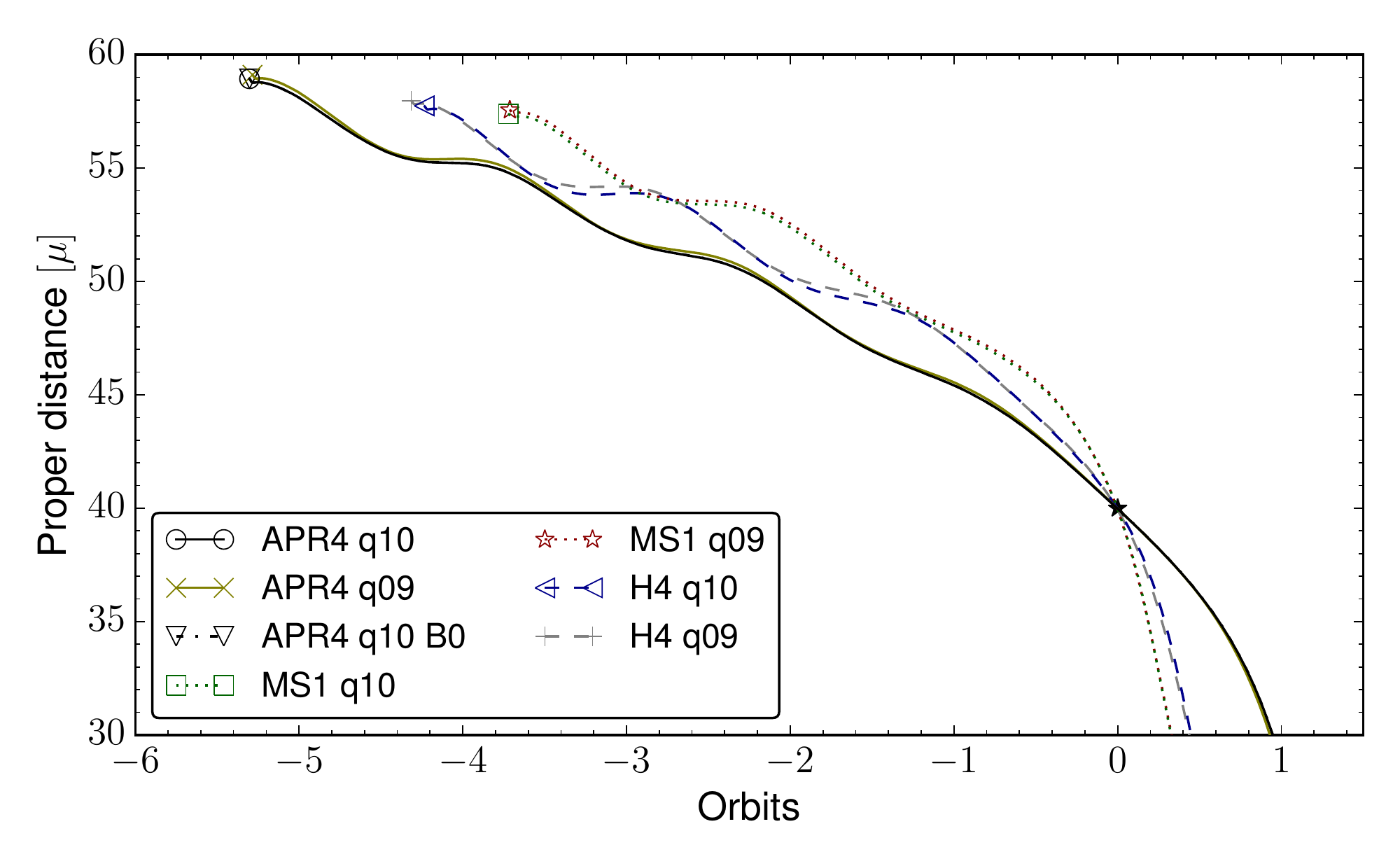}
  \caption{Proper separation between barycenters of the NSs versus orbital phase. 
  The separation is plotted in units of reduced mass 
  $\mu=M_\mathrm{g}^1 M_\mathrm{g}^2 / (M_\mathrm{g}^1 + M_\mathrm{g}^2)$,
  and the orbital phase is defined relative to a separation of $40\usk\mu$. Barycenter
  and orbital phase are computed with respect to simulation coordinates. } 
  \label{fig:separation}
\end{figure}

We compute the initial data using the publicly available
code \texttt{LORENE}~\cite{Gourgoulhon:2001:64029, Taniguchi2002}. 
Our initial binary systems are computed as irrotational and on a circular
orbit. Because of the lack of an initial radial component of the
velocity, the orbits have some minor residual eccentricity, as shown 
in \Fref{fig:separation}.
For all our models the initial 
coordinate separation is $45$ km, corresponding to a proper separation 
of $\simeq 57-59$ km.
Each EOS used in this paper has been implemented employing a
piecewise-polytropic approximation of the corresponding nuclear physics
(tabulated) EOS, taken from \cite{Read:2009:124032} for the H4 and MS1 
EOS and form \cite{Endrizzi:2016:164001} for the APR4 EOS.
In particular, H4 and MS1 are approximated with three pieces for the 
core/high density part and with four pieces for the low density part.
For APR4, we have two additional pieces at very high densities 
(see \cite{Endrizzi:2016:164001}).
During the evolution, a thermal component is added via an ideal-fluid 
EOS with adiabatic index of $\Gamma=1.8$ (same as in 
\cite{Kiuchi:2014:41502}). 

Since \texttt{LORENE}
cannot compute equilibrium configurations for magnetized BNS systems,
we add the magnetic field to \texttt{LORENE} initial configurations 
manually. Since the field geometry in actual NSs is unknown, we use 
the following analytic prescription for the vector potential $A_{\phi}$:
\begin{equation}
\label{eq:Avec}
A_{\phi} \equiv \varpi^2 A_b\, {\rm max}\,(p-p_{\rm cut},0)^{n_{\rm s}} \,,
\end{equation}
where $\varpi$ is the coordinate distance from the NS spin axis,
$p_{\rm cut}=0.04 \,\mathrm{max}(p)$ is a cutoff that determines where
the magnetic field goes to zero inside the NS, $\mathrm{max}(p)$ is
the initial maximum pressure in each star, and $n_{\rm s}=2$ is the
degree of differentiability of the magnetic field
strength~\cite{Giacomazzo2011PhRvD..83d4014G}. The resulting field
is dipole-like in the interior of the NSs and zero outside.
The value of $A_b$ is
chosen such that for the equal-mass APR4 model, the maximum of the 
initial magnetic field strength is $\approx 3 \times 10^{15}$ G. This corresponds 
to a magnetic energy of $\simeq 1.21\times 10^{47}$ erg for each NS. 
The values of $A_b$ in the other models are adjusted in order to maintain 
the same total magnetic energy.
With this choice, all models have the same energy budget at infinite separation
in terms of both the total gravitational mass and the total magnetic energy.

\begin{table*}
\caption{Outcome of our BNS mergers. $M_\mathrm{BH}$ and $J_\mathrm{BH}$ are black hole
  mass and angular momentum $3.4\msec$ after formation (only 
  for collapsing models). 
  $M_\mathrm{blk}$ and $R_\mathrm{blk}$ are bulk mass and bulk 
  radius (see text for definitions), while $\nu_\mathrm{c}$ and 
  $\nu_\mathrm{max}$ denote the remnants central and maximum rotation 
  rates, all computed $20\usk\milli\second$ after merger.
  $f_\mathrm{merge}$ is the gravitational wave instantaneous frequency at the time of 
  merger, $f_\mathrm{pm}$ is the frequency of the maximum 
  in the post-merger part of the gravitational wave power spectrum, and
  $f_{10}$ is the average instantaneous frequency during the first $10\msec$ after 
  merger (see Section \ref{sec:gw}). $M_\mathrm{disk}$ is the mass outside the apparent 
  horizon, or the mass outside $r>20$ km if no black hole is formed.
  $M_\mathrm{fb}$ is the bound mass outside $r>60$ km. Both are 
  measured at $t=3.4\, \mathrm{ms}$ after black hole formation, or
  $t=20\, \mathrm{ms}$ after merger if no black hole is formed.
  Finally, $M_\mathrm{ej}$ and $v_\mathrm{esc}$ are our estimates for 
  the total ejected mass and the average escape velocity. The values 
  in brackets for the APR4 model refer to the high-resolution 
  run (the measures absent for the standard resolution run were not 
  implemented at the time).}
\begin{ruledtabular}
\begin{tabular}{lcccccc}
Model&
APR4 equal&
APR4 unequal&
MS1 equal&
MS1 unequal&
H4 equal&
H4 unequal
\\\hline 
$M_\mathrm{BH}\,[M_\odot]$&
---&
---&
---&
---&
$2.49$&
$2.42$
\\
$J_\mathrm{BH}/M^2_\mathrm{BH}$&
---&
---&
---&
---&
$0.63$&
$0.57$
\\
$M_\mathrm{blk}\,[M_\odot]$&
($2.47$)&
$2.42$&
$2.35$&
$2.25$&
$2.48$&
$2.37$
\\
$M_\mathrm{blk}/R_\mathrm{blk}$&
($0.30$)&
$0.30$&
$0.21$&
$0.21$&
$0.27$&
$0.26$
\\
$\nu_\mathrm{c}\, [\mathrm{kHz}]$&
$0.73$ ($0.69$)&
$0.64$&
$0.34$&
$0.27$&
$0.69$&
$0.52$
\\
$\nu_\mathrm{max}\, [\mathrm{kHz}]$&
$1.65$ ($1.64$)&
$1.59$&
$0.99$&
$1.01$&
$1.35$&
$1.24$
\\
$f_\mathrm{merge}\, [\mathrm{kHz}]$&
$2.12$ ($2.12$)&
$2.09$&
$1.46$&
$1.36$&
$1.54$&
$1.51$
\\
$f_\mathrm{pm}\, [\mathrm{kHz}]$&
$3.35$ ($3.33$)&
$3.24$&
$2.03$&
$2.09$&
$2.54$&
$2.55$
\\
$f_{10}\, [\mathrm{kHz}]$&
$3.33$ ($3.32$) &
$3.25$ &
$1.97$ &
$1.96$ &
$2.45$ & 
$2.36$ 
\\
$M_\mathrm{disk}\,[M_\odot]$&
($0.201$)&
$0.252$&
$0.387$&
$0.479$&
$0.126$&
$0.211$
\\
$M_\mathrm{fb}\,[M_\odot]$&
($0.121$)&
$0.133$&
$0.180$&
$0.191$&
$0.105$&
$0.175$
\\
$M_\mathrm{ej}\, [10^{-2} \, M_\odot]$&
$1.31$ ($1.27$)&
$0.74$&
$0.08$&
$0.09$&
$0.07$&
$0.10$
\\
$v_\mathrm{esc}\, [c]$&
($0.12$)&
$0.10$&
$0.10$&
$0.10$&
$0.11$&
$0.13$
\end{tabular}
\end{ruledtabular}
\label{tab:outcome}
\end{table*}

We note that half of the total magnetic energy corresponds to the value
for a magnetized NS with a simple purely poloidal/dipolar configuration 
and $B_\mathrm{pole}\approx 2.4\times 10^{14}$ G (as computed 
with the ``magstar" \texttt{LORENE} code). 
For more realistic configurations including also a strong toroidal magnetic field 
inside the NS, the same magnetic energy could even correspond to 
a $B_\mathrm{pole}$ as low as $\sim10^{13}$ G \cite{Ciolfi2013}.
Since NSs in binary systems are expected to have 
$B_\mathrm{pole}\sim10^{12}$ G, we are imposing magnetic energies 
a factor of $10^2-10^4$ higher than the common expectations. 
Nevertheless, GRMHD simulations of BNS mergers performed 
at very high resolution have recently confirmed that when magnetic 
field amplification mechanisms such as the Kelvin-Helmholtz instability 
are well resolved, the magnetic field can easily reach strengths of the 
order of $\sim10^{15}$ G or higher (see \cite{Kiuchi:2015:1509.09205} 
and refs.~therein). 
Since our resolution is insufficient to fully resolve these amplification 
mechanisms, a lower (and more realistic) initial magnetic energy 
would result in a post-merger magnetic field orders of magnitude 
weaker than expected. 
For the resolution that we can currently afford, our choice allows us to 
explore more realistic post-merger field strengths despite the lower
amplification factors.
We stress, however, that this is by no means equivalent to fully resolving 
the amplification of a weaker initial field up to ${\sim}10^{15}$ G or more.
We also note that the magnetic field strengths we impose are still sufficiently 
low to safely neglect deviations from hydrostatic equilibrium as well as
constraint violations (magnetic energy is ${\sim}10^{6}$ times smaller 
than the binding energy of each NS).

For the evolution we use our GRMHD
code \texttt{Whisky}~\cite{Giacomazzo:2007:235,
Giacomazzo2011PhRvD..83d4014G, Giacomazzo2013ApJ...771L..26G} coupled
with the publicly available \texttt{Einstein
Toolkit}~\cite{Loeffler:2012:115001}. The \texttt{Einstein Toolkit} is
a collection of publicly available codes, including
the \texttt{Cactus} computational framework, the \texttt{Carpet}
driver, and the \texttt{McLachlan} code. In particular we use
the \texttt{McLachlan} code to evolve Einstein's equations using the
BSSNOK formulation for the
spacetime~\cite{Baumgarte:1998:24007,Shibata:1995:5428,Nakamura:1987:1}. 
The GRMHD equations are instead evolved by our \texttt{Whisky} code, which
uses high-resolution shock-capturing schemes to solve the GRMHD
equations written in a flux-conservative form via the ``Valencia''
formulation~\cite{Anton:2005gi}. The fluxes are computed with the HLLE
approximate Riemann solver~\cite{Harten:1983:35} that uses the
primitive variables reconstructed at the interfaces between the cells
via the piecewise-parabolic method~\cite{Colella:1984:174}. In order
to preserve the divergence-free character of the magnetic field, we
evolve the vector potential and compute the magnetic field from it. To
avoid spurious magnetic field amplifications at the boundaries between
refinement levels, we use the modified Lorenz
gauge~\cite{Etienne:2011re,Farris2012}. 
We also set a density floor for the rest-mass density $\rho$ 
equal to $\rho_{\rm atmo}=10^{-11}\approx
6.2 \times 10^{6}$ g cm$^{-3}$. Where $\rho$ falls below that value, we
reset it to $\rho_{\rm atmo}$ and set the velocity to zero. 

In all our simulations we use ``moving box'' mesh refinement provided by the 
\texttt{Carpet} driver. We use six refinement levels, with the grids of 
the two finest levels following each of the two NSs during the inspiral phase. 
At merger, we switch to fixed mesh refinement, with a central 
finest grid covering a radius of $30\kmeter$, large enough to contain the 
remnant object and the innermost part of the disk. 
We employ a resolution on the finest grid of
$dx\approx 220$ m. 
This fiducial resolution allows us to cover the radii of the initial NSs 
with $\approx50-70$ points, depending on the EOS. The equal-mass 
APR4 model is also evolved at higher and lower resolutions in order 
to assess the numerical accuracy (see the Appendix). 
The highest resolution employed in this work (for only one simulation) 
is $dx\approx 177$ m.
We note that recent GRMHD simulations of BNS mergers have been also 
performed with higher or much higher resolution 
\cite{Kiuchi:2014:41502,Kiuchi:2015:1509.09205}.
The outer boundary of our computational domain is located 
at $\approx 1250$ km. To save computational
resources we also enforce a reflection symmetry across the $z=0$
plane.


\section{Merger and postmerger dynamics}
\label{sec:dynamics}

\begin{figure*}[!ht]
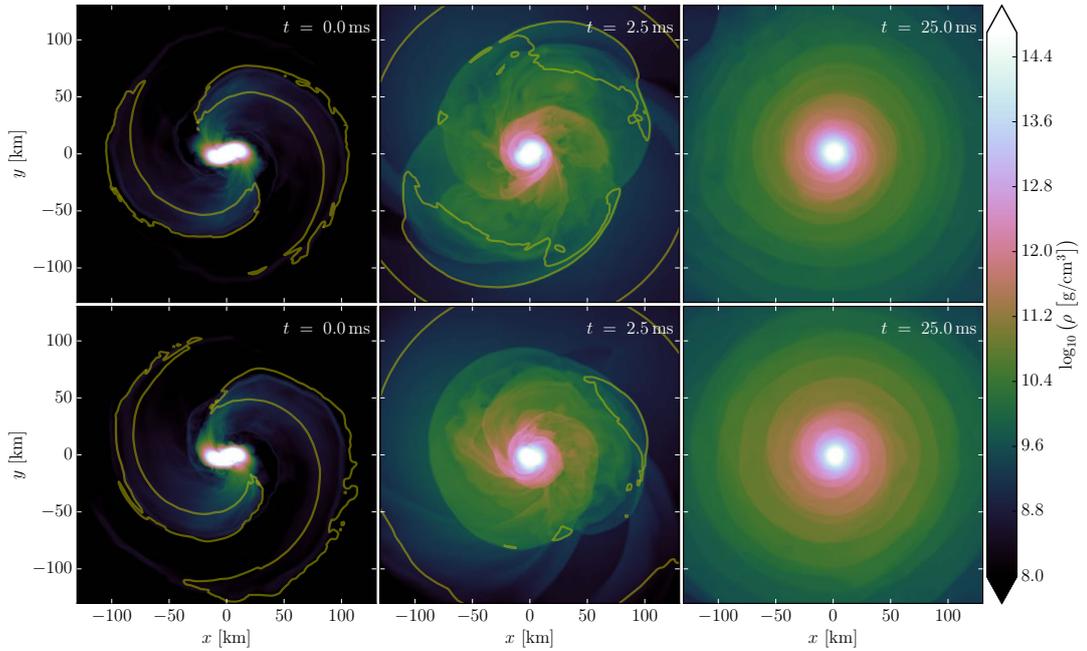

  \centering
  \includegraphics[width=0.95\linewidth]{{{fig2}}}
  \caption{Rest mass density evolution on equatorial plane for the APR4 equal-mass (top row)
  and unequal-mass (bottom row) models. The contours indicate matter ejected that is unbound 
  according to the geodesic criterion (see Section~\ref{sec:ej}). The times of the snapshots
  denote the time after merger.} 
  \label{fig:rho_xy_APR4}
\end{figure*}

\begin{figure*}[!ht]
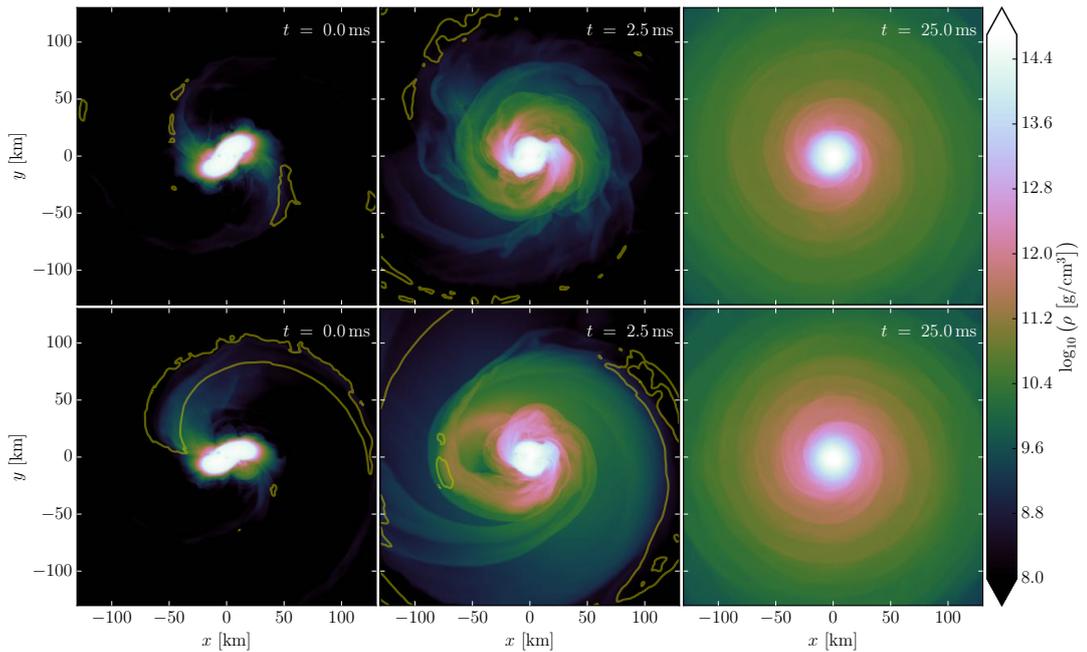

  \centering
  \includegraphics[width=0.95\linewidth]{{{fig3}}}
  \caption{Same as Fig.~\ref{fig:rho_xy_APR4} for MS1 equal-mass (top row) and unequal-mass (bottom row) 
  models. } 
  \label{fig:rho_xy_MS1}
\end{figure*}

\begin{figure*}[!ht]
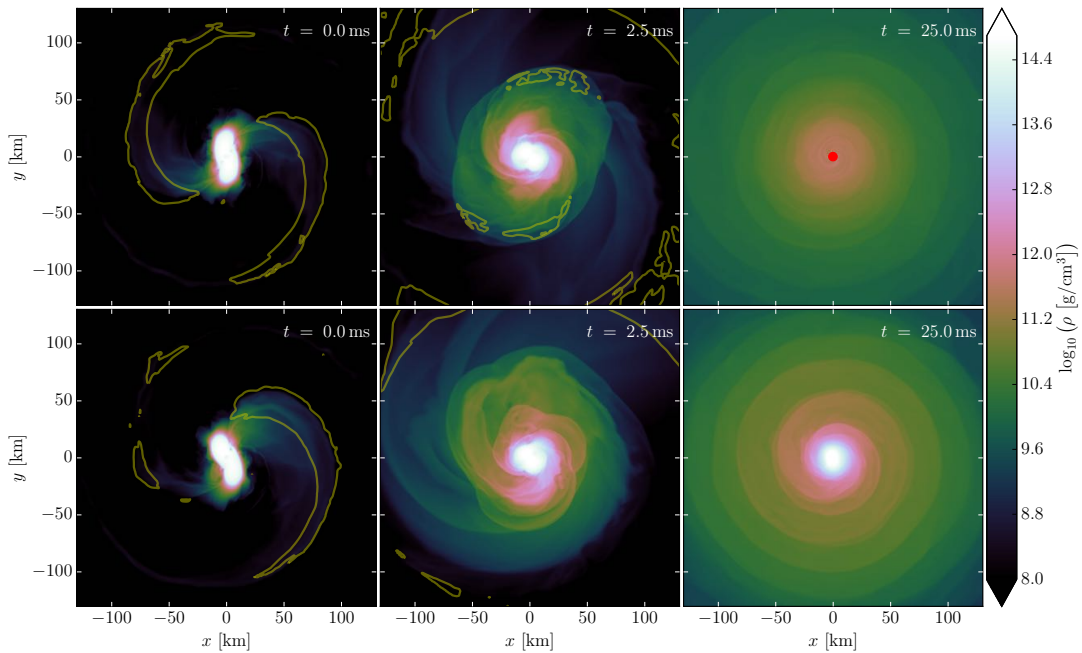

  \centering
  \includegraphics[width=0.95\linewidth]{{{fig4}}}
  \caption{Same as Fig.~\ref{fig:rho_xy_APR4} for H4 equal-mass (top row) and unequal-mass (bottom row) 
  models. For the equal-mass model at $25\msec$ (upper right panel), a black hole is already formed, 
  and the red disk indicates the apparent horizon.} 
  \label{fig:rho_xy_H4}
\end{figure*}

\begin{figure*}[!ht]
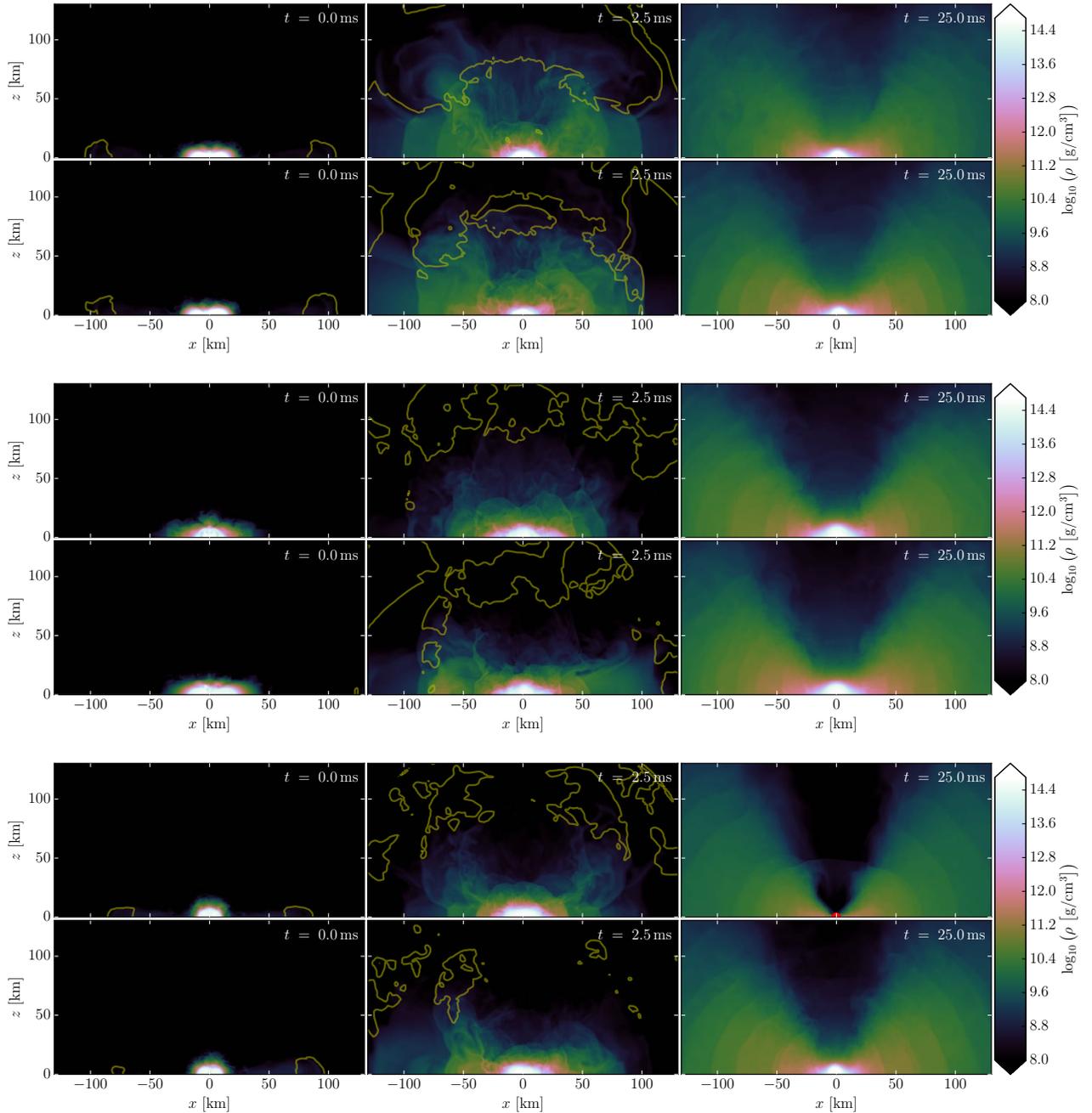

  \centering
  \includegraphics[width=0.95\linewidth,trim=0 11mm 0 11mm, clip]{{{fig5a}}}
  \includegraphics[width=0.95\linewidth,trim=0 11mm 0 11mm, clip]{{{fig5b}}}
  \includegraphics[width=0.95\linewidth,trim=0 11mm 0 11mm, clip]{{{fig5c}}}
  \caption{Same as Figures~\ref{fig:rho_xy_APR4} to \ref{fig:rho_xy_H4} but for the
  rest mass density on the meridional plane. From top to bottom: APR4, MS1, and H4 models.} 
  \label{fig:rho_xz}
\end{figure*}

In this Section, we describe basic aspects of the dynamics of the six reference models 
considered in this work. The key numeric results are listed in \Tref{tab:outcome}.  
We recall that four of these models form long-lived NSs (supramassive or stable for 
the APR4 and MS1 EOS, respectively), while the two models employing the H4 EOS 
produce a HMNS collapsing to a BH within few tens of ms.
 
The inspiral phase is shown in \Fref{fig:separation}, depicting the separation
versus orbital phase. We observe a clear trend for the impact of the EOS: the 
more compact the stars (see \Tref{tab:init_param}), the more orbits before merger. 
Note the oscillations around the overall decrease in separation correspond to the 
residual eccentricity of the initial data. Correcting the eccentricity might lead 
to some quantitative changes, but not enough to affect the general trend
(see, e.g., \cite{Maione2016} and refs.~therein for more details on 
eccentricity in BNS merger simulations).

\begin{figure*}
  \centering
  \includegraphics[width=0.9\textwidth]{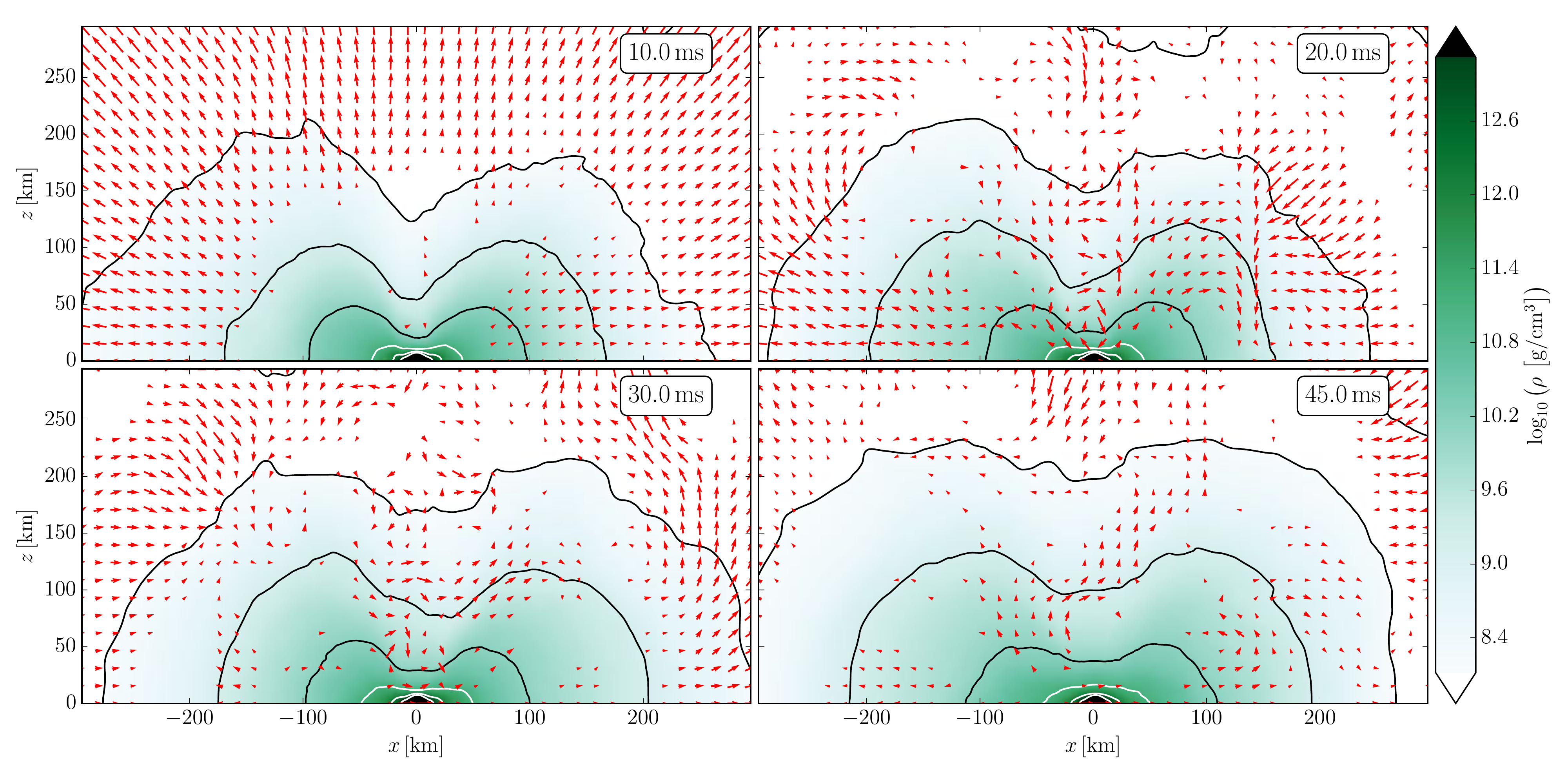}
  \caption{Velocity (arrows) and density (color and contour lines) in the meridional plane 
  for the equal-mass APR4 model, at times $10,20,30,45\msec$ after merger, 
  averaged over the time interval $\pm2\msec$ to remove the contribution of oscillations.
  The contours correspond to $10^{-n}$ of the initial central density, with $n=1\ldots 6$.
  Note that the scale of the arrows is not the same in all panels, with the maximum 
  velocities being $v/c = 0.05\,(t=10\msec), \,0.09 (20\msec), \, 0.05 (30\msec), \,  
  0.07 (45\msec)$.
  } 
  \label{fig:apr4q10_mflux_xz}
\end{figure*}

\begin{figure*}
  \centering
  \includegraphics[width=0.9\textwidth]{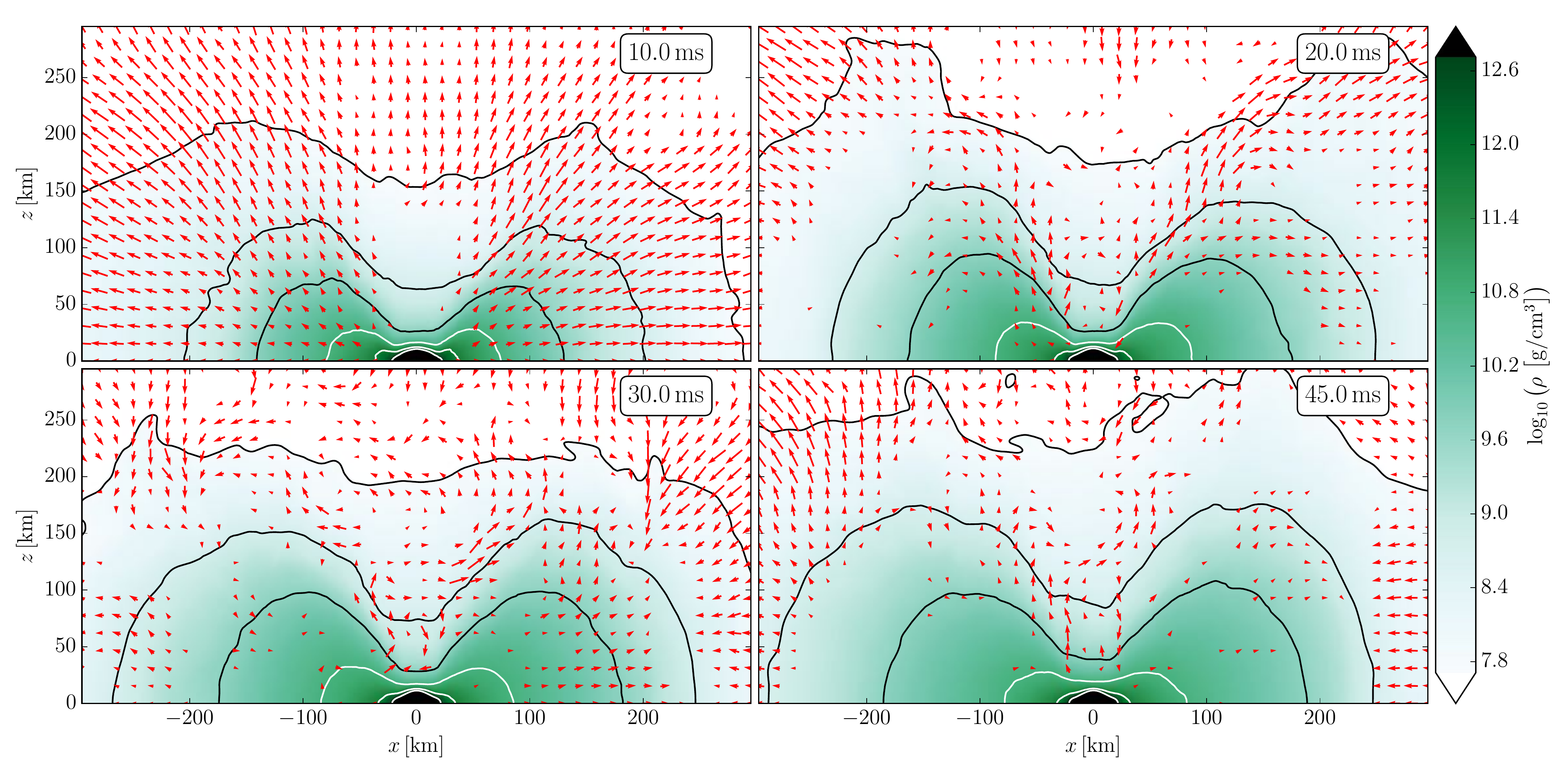}
  \caption{Like \Fref{fig:apr4q10_mflux_xz}, but for the equal-mass MS1 model.
  Note that the scale of the arrows is not the same in all panels, with the maximum 
  velocities being
  $v/c = 0.08\,(t=10\msec), 0.05\, (20\msec),0.05 \, (30\msec), \, 0.06 (45\msec)$.
  } 
  \label{fig:ms1q10_mflux_xz}
\end{figure*}

\begin{figure*}
  \centering
  \includegraphics[width=0.9\textwidth]{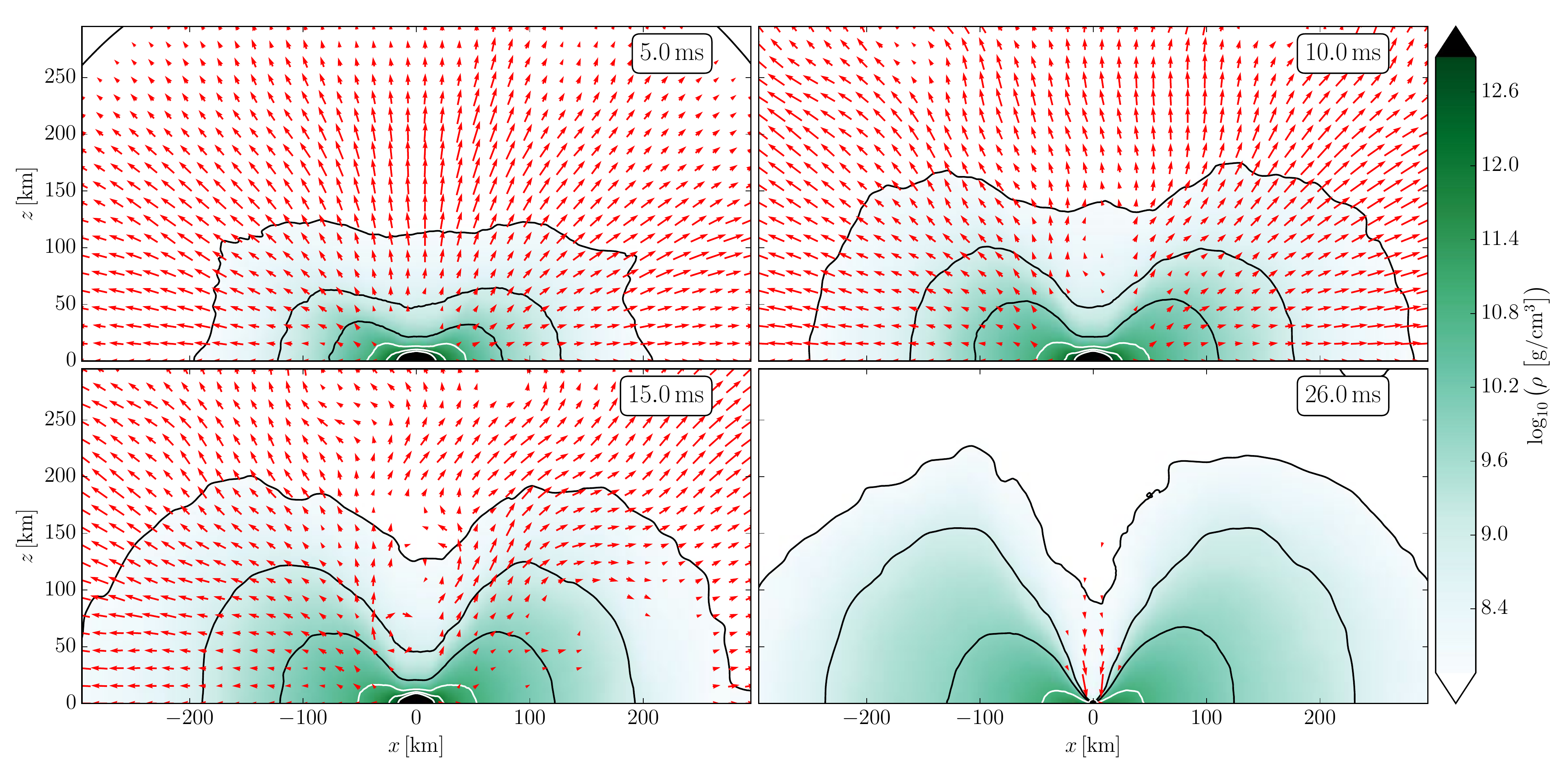}
  \caption{Like \Fref{fig:apr4q10_mflux_xz}, but for the equal-mass H4 model.
  Note that the scale of the arrows is not the same in all panels, with the maximum 
  velocities being
  $v/c = 0.13\,(t=5\msec), 0.07\, (10\msec), 0.05\, (15\msec), 0.40\, (26\msec)$.
  The black disk in the lower right panel marks the apparent horizon.
  } 
  \label{fig:h4q10_mflux_xz}
\end{figure*}

\begin{figure}
  \centering
  \includegraphics[width=0.49\textwidth]{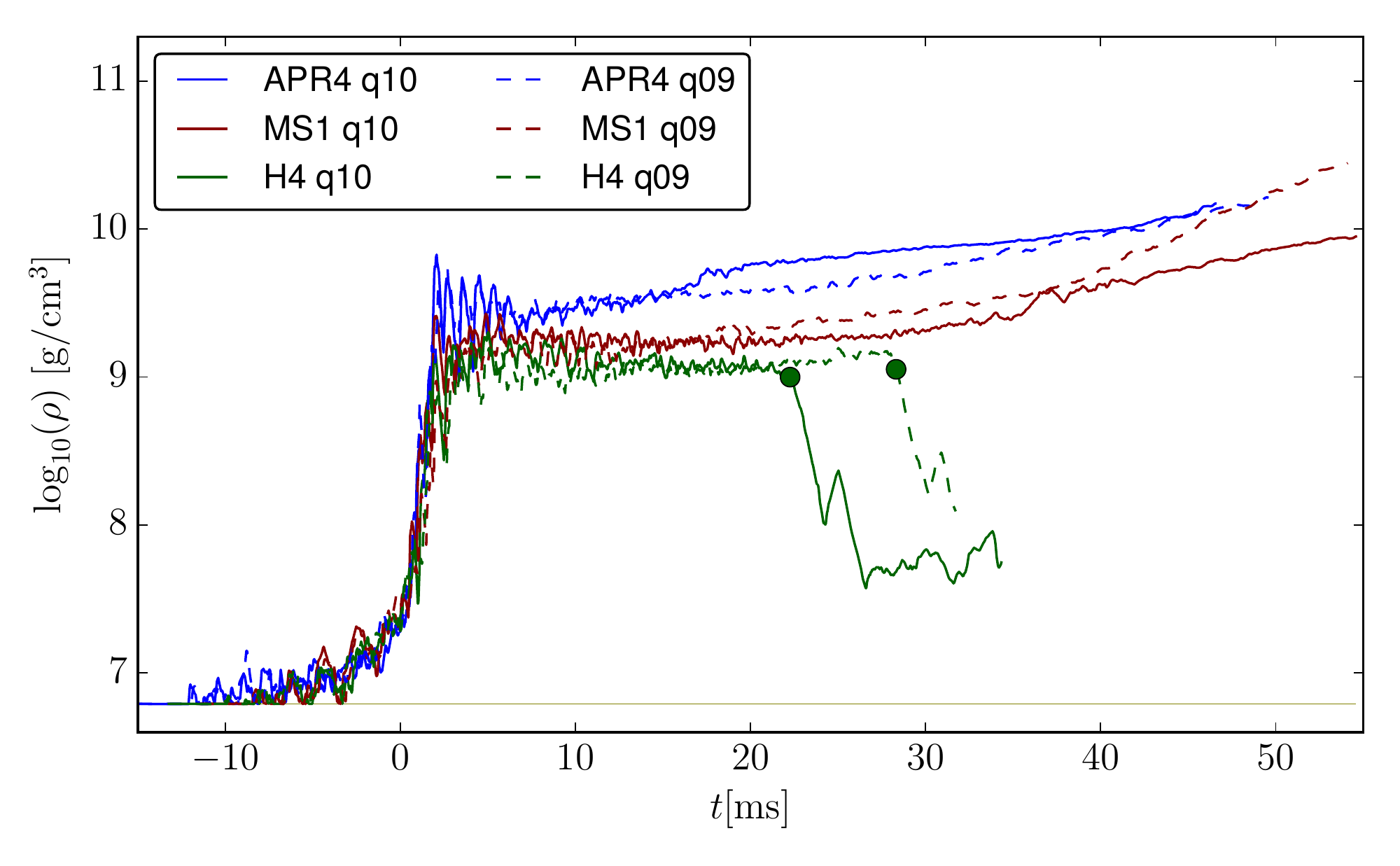}
  \caption{Evolution of rest-mass density along the $z$ axis, averaged 
  between $z=30\kmeter$ and $z=50\kmeter$. The horizontal line marks the density of the 
  artificial atmosphere.
  } 
  \label{fig:rho_funnel}
\end{figure}

Differences between mass ratios $1$ and $0.9$ are instead very small. 
Comparing the equal-mass APR4 model to the corresponding unmagnetized case, 
we also find that the magnetic field of the given strength has no impact on the inspiral phase.

An overview of the merger and post-merger evolution is given in  
Figures~\ref{fig:rho_xy_APR4} to~\ref{fig:rho_xy_H4}, showing
snapshots of the rest-mass density in the orbital plane at times 
$0, 2.5, 25\msec$ after merger. 
We define the time of merger, $t_\mathrm{merge}$, as the retarded time at which 
the GW signal reaches its maximum amplitude. Throughout this article, 
all times are given relative to $t_\mathrm{merge}$, i.e. times generally refer
to the time after merger. 
Fig.~\ref{fig:rho_xz} shows the same evolution as seen on the meridional 
plane. From those figures it is clear that, after a highly dynamic merger 
phase, the system settles within $\sim\!20\msec$ to a quasi-stationary state 
composed of a massive NS surrounded by an accretion disk. 
For the APR4 and MS1 models, such a configuration remains almost unchanged 
until the end of the simulations (more than $45\msec$ after merger), while 
for the H4 models a BH is formed respectively $22$ and $28\msec$ after merger 
for the equal and unequal-mass cases.

In order to quantify the disk mass, we provide in \Tref{tab:outcome} the total 
mass either outside the apparent horizon or outside a radius $r>20 \kmeter$ if no
BH is formed. In the equal-mass case, this estimate gives $M_\mathrm{disk}\!\approx\!0.2\, M_{\odot}$
for APR4 and almost $0.4\, M_{\odot}$ for MS1.
Going from equal to unequal mass, both models 
result in a $\sim\!25\%$ higher disk mass. 
For the H4 models, a few ms after collapse the BH is surrounded by a disk of 
$\sim\!0.13\,$($0.21$)~$M_{\odot}$ for equal (unequal) mass. In this case, the 
mass ratio has a much larger impact on $M_\mathrm{disk}$. 
We note that the HMNS lifetime is also longer for the unequal-mass case.
Part of the increased disk mass could be due to a higher amount of matter 
expelled from the remnant via oscillations or shocks.
The properties of the BHs shortly after formation are very similar for 
equal- and unequal-mass case, 
with BH masses of $2.50 \,M_{\odot}$ and $2.42 \,M_{\odot}$,
and spins of $0.62$ and $0.57$, 
respectively.
Since the disk smoothly transitions into a fallback component 
on non-circular orbits, we also provide the mass outside $r>60\kmeter$,
$M_\mathrm{fb}$, as a ballpark figure for the outer disk/fallback 
component. We find that $40$--$80\%$ of the total disk mass is outside 
$60\kmeter$.

Figures~\ref{fig:rho_xy_APR4} to~\ref{fig:rho_xz}
also show regions of matter that is unbound according to the geodesic criterion.
In all models, we can distinguish a tidal contribution to the ejected matter confined 
to the equatorial plane (clearly visible at $t\!=\!0$) and a later, more isotropic ejection 
which we attribute to breakout shocks. 
Our best estimate for the amount of unbound matter is given in \Tref{tab:outcome}. 
A more detailed discussion on mass ejection  
is given in \Sref{sec:ej}.

For the long-lived remnant cases (APR4 and MS1 models), the above dynamical 
ejecta are followed by a slower outflow of material that is bound according to the geodesic
criterion, and that might fall back onto the remnant at later times. 
As will be discussed in \Sref{sec:ej}, it is also possible that some of this matter
will become unbound as a result of the magnetic pressure, and constitute a baryon-loaded 
wind. Note that such winds are likely to play an important role for the long-term EM 
emission from the supramassive or stable NS 
(e.g.~\cite{Yu2013,Metzger2014a,Ciolfi:2015:36,Siegel:2016a,Siegel:2016b}).
Density and velocity of the outflow in the meridional plane are shown at different times
in Figures~\ref{fig:apr4q10_mflux_xz} and \ref{fig:ms1q10_mflux_xz}.
Since we are not interested in fluctuations, we averaged over a duration of 
$4\msec$. We find a relatively isotropic radial outflow with maximum velocities 
around $0.05$--$0.08\usk c$ at $t=10\msec$. At $t>20\msec$ however, the flow 
patterns consist mainly of large eddies, with a smaller net flux. 
These post-merger matter outflows will be discussed further in \Sref{sec:ej}.

For the H4 models, we find a strong influx of matter along the $z$ axis after
the BH is formed, as shown in \Fref{fig:h4q10_mflux_xz}. This is expected 
since matter along the $z$ axis could only be supported by the vertical pressure 
gradient, which can only be sustained by a NS remnant, not a BH.
The inflow after the collapse quickly leads to a funnel of reduced density.

The baryon pollution along the orbital axis has important consequences for the possibility 
of launching relativistic jets which could give rise to SGRBs 
\cite{Murguia-Berthier2014,Nagakura2014,Murguia-Berthier2017a} (see Section \ref{sec:sgrb}).
Fig.~\ref{fig:rho_funnel} shows the density averaged along the $z$ axis
between $z=30$ and $z=50\kmeter$.
For all models, we find densities of the order $10^{9}\usk\gram\per\centi\meter\cubed$
few ms after merger, and subsequently a slow and persistent increase.
For the H4 models, the density drops sharply by almost two orders of magnitude 
when a BH is formed. At least for the equal-mass model, the density seems to stabilize 
at this level or even increase slightly. The unequal-mass simulation ends shortly after BH
formation, but we expect a similar behavior.


\section{Rotation profile and remnant structure}
\label{sec:rot}

\begin{figure*}
  \centering
  \includegraphics[width=0.99\linewidth]{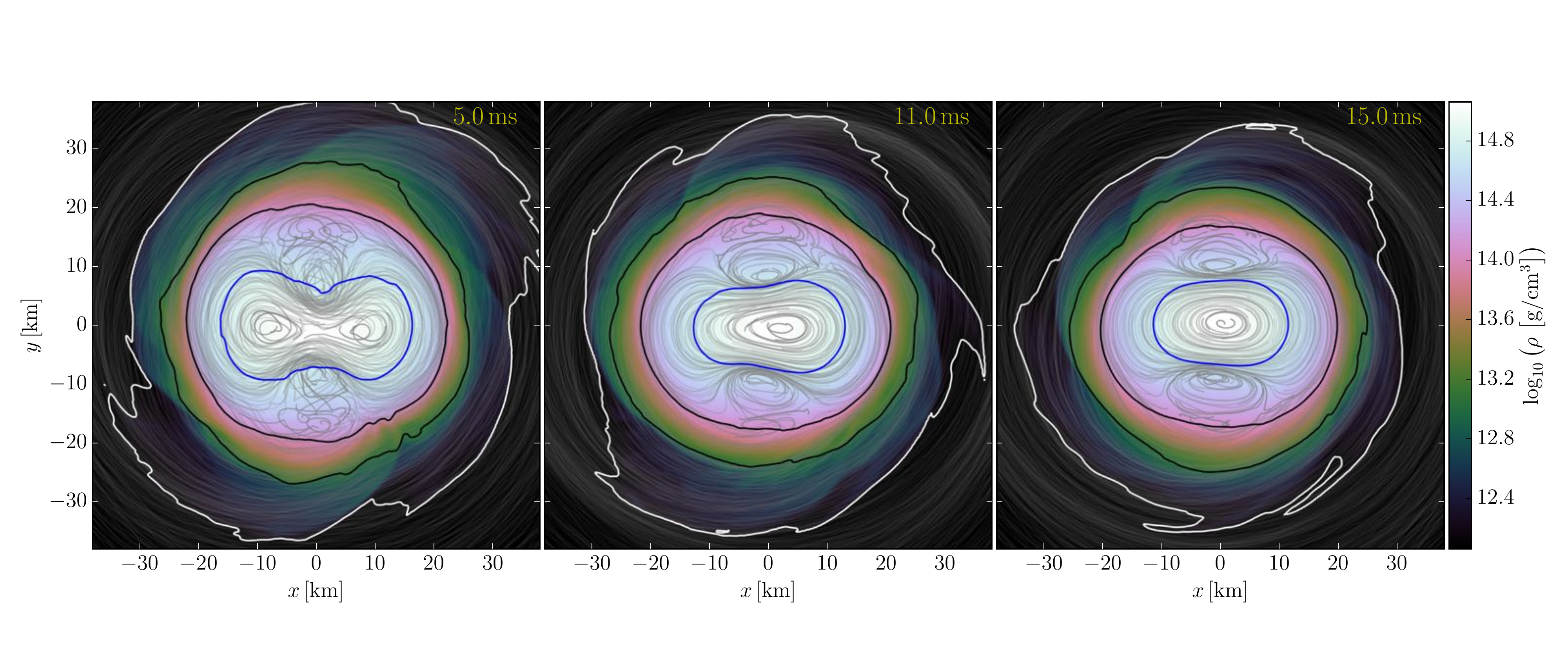}
  \caption{Remnant structure in the equatorial plane at different times for the equal 
  mass H4 model. The density is shown as color plot. The thick lines are isodensity contours
  at $0.5$, $0.1$, $0.01$, $0.001$ times the maximum density. The thin lines are fluid
  trajectories in a frame corotating with the $m=2$ density deformation, during
  a time interval $\pm 1 \msec$ around the time of each snapshot.} 
  \label{fig:h4q10_traj_xy}
\end{figure*}

\begin{figure*}
  \centering
  \includegraphics[width=0.99\linewidth]{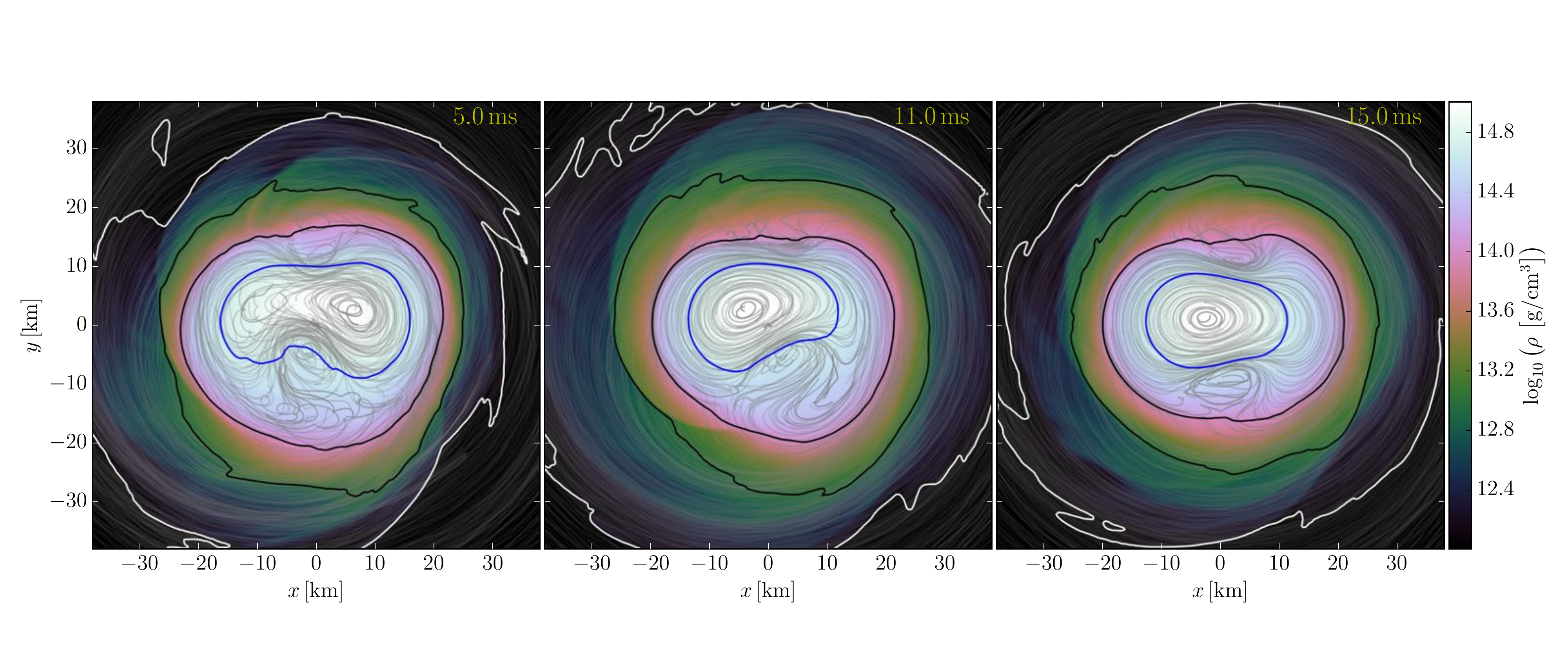}
  \caption{Like \Fref{fig:h4q10_traj_xy}, but showing the unequal-mass H4 model.} 
  \label{fig:h4q09_traj_xy}
\end{figure*}

In the following, we investigate the structure of the fluid flow inside the remnant
in the equatorial plane, focusing at first on the H4 models. 
As in \cite{Kastaun:2016, Kastaun:2015:064027}, we 
track fluid elements in a frame corotating with the $m=2$ component of the density 
deformation.
The fluid flow together with the density distribution at different times are shown 
in \Fref{fig:h4q10_traj_xy} for the H4 equal-mass case and in \Fref{fig:h4q09_traj_xy} 
for the H4 unequal-mass case.
As one can see, the remnants are still strongly deformed at $15\msec$ after merger.
We also find that the fluid flow does not correspond to simple differential rotation. 
Instead, we observe secondary vortices. Those vortices are related to the density 
deformation, although it is unclear if they are causing it or are caused by it.
Most likely, both density deformation and vortices influence each other.
In any case, the vortices remain stationary with respect to the deformation most of 
the time, although there is some gradual evolution towards a more axisymmetric state.
For the unequal-mass model however, a more rapid rearrangement seems to happen between
$5$--$11\msec$ after merger. Also, not surprisingly, the structure shortly after merger 
is decidedly less symmetric for the unequal-mass case. We note that a similar 
rearrangement of vortices and deformation pattern has been found in 
\cite{Kastaun:2015:064027} for a binary model with equal mass, but unequal NS spin.
It is also worth pointing out that, roughly speaking, the deformation of the outer layers 
is rotated 90 degrees with respect to the core, which implies contributions to the 
quadrupole moment with opposite signs. The impact on the GW signal will be discussed 
further in \Sref{sec:gw}.

For the long-lived remnant cases (APR4 and MS1), similar structures appear in the early 
post-merger phase. Nevertheless, within $15-20$~ms the system settles to a more ordered 
quasi-stationary structure characterized by simple differential rotation (see, e.g., Fig.~9 of
\cite{Kastaun:2016b:arxiv}, showing the same as \Fref{fig:h4q10_traj_xy} for our 
unmagnetized APR4 equal-mass model).

We now turn to discuss the rotation profiles of the remnants.
For this, we employ the methods described in \cite{Kastaun:2015:064027}.
In particular, we use a coordinate system that is defined independent of the 
spatial gauge conditions and prevents non-axisymetric as well as spiral distortions,
given that the spacetime is axisymmetric (see \cite{Kastaun:2015:064027} for details).
We restrict the analysis to the 
equatorial plane, because the new coordinate system is only defined there and the 
required data is saved only on coordinate planes.

\begin{figure}
  \centering
  \includegraphics[width=0.99\linewidth]{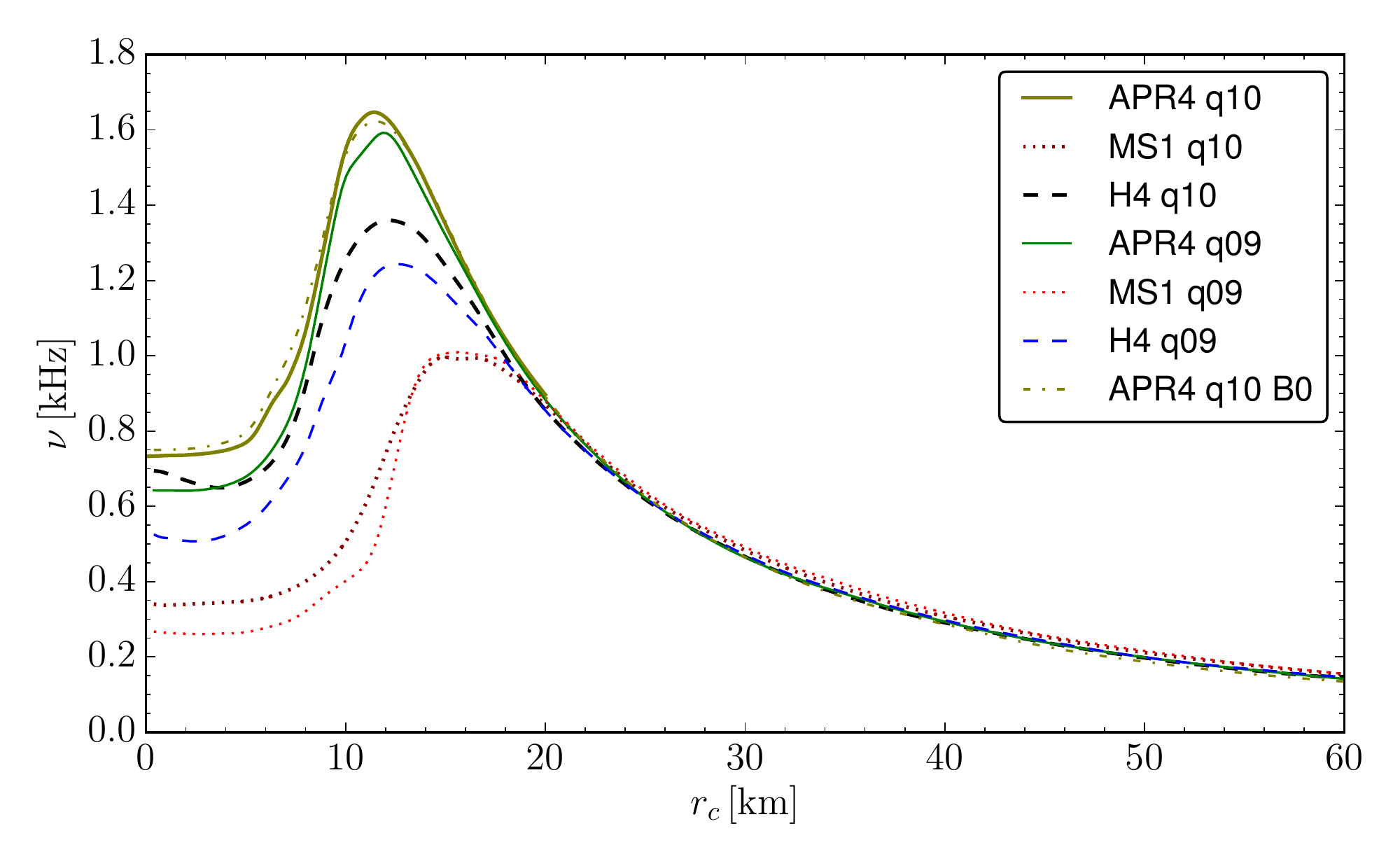}
  \caption{Rotation rate (as seen from infinity) in the equatorial plane $20\msec$ after merger,
  averaged in $\phi$ and over a time window of $\pm 1 \msec$, versus
  circumferential radius.} 
  \label{fig:rot_profs}
\end{figure}

\Fref{fig:rot_profs} shows the rotation rate for all models at a time $20\msec$ after
merger. To reduce the influence of residual oscillations, we average over the time interval
$20 \pm 1 \msec$.
All the rotation profiles show a clear maximum away from the center,
and a slow central rotation rate below $0.8 \kHz$. 
\Fref{fig:rot_profs} also shows part of the disk, which smoothly joins the remnant.
For $r > 20 \kmeter$, the rotation rates are given approximately by the Kepler velocity, 
which depends almost exclusively on the remnant mass.
Our findings are similar to the results obtained
for different models in \cite{Kastaun:2015:064027, Endrizzi:2016:164001, Kastaun:2016, Kastaun:2016b:arxiv,
Hanauske2016arXiv}. 
The models in those publications together with the present one include hypermassive, supramassive,
and stable remnants, different mass ratios, and even binaries with initial aligned spin.
The general shape of the rotation profiles shown in \Fref{fig:rot_profs} seem to be a generic 
property of merger remnants.

Since the rotation profiles show that the cores are rotating slowly, we expect that the 
inner core can be approximated by a spherical TOV solution. 
In order to judge the importance of centrifugal forces in the core, we computed the ratio 
of rotation rate and orbital frequency of a test mass in circular orbit (both measured by zero 
angular momentum observers, i.e.~removing the frame dragging) at the center of the remnants 
$20\msec$ after merger. We found values ranging between $0.02$ (APR4 unequal-mass model) 
and $0.06$ (H4 equal-mass model). This indeed strongly suggests TOV-like cores.
In order
to quantify the radial mass distribution in an unambiguous way, we use the measures 
described in \cite{Kastaun:2016}.
These replace the density versus radius measures used for spherical stars with the  
baryonic mass as function of proper volume contained in isosurfaces of constant rest mass density.
Further, to express compactness of the remnant in absence of a clear surface, we define 
the compactness of each isodensity surface as the ratio between the contained baryonic mass 
and the radius of an Euclidean sphere with the same proper volume. This compactness has a maximum,
which we use to define the bulk isodensity surface, and the corresponding bulk compactness, 
bulk mass, and bulk volume.

\begin{figure}
  \centering
  \includegraphics[width=0.99\linewidth]{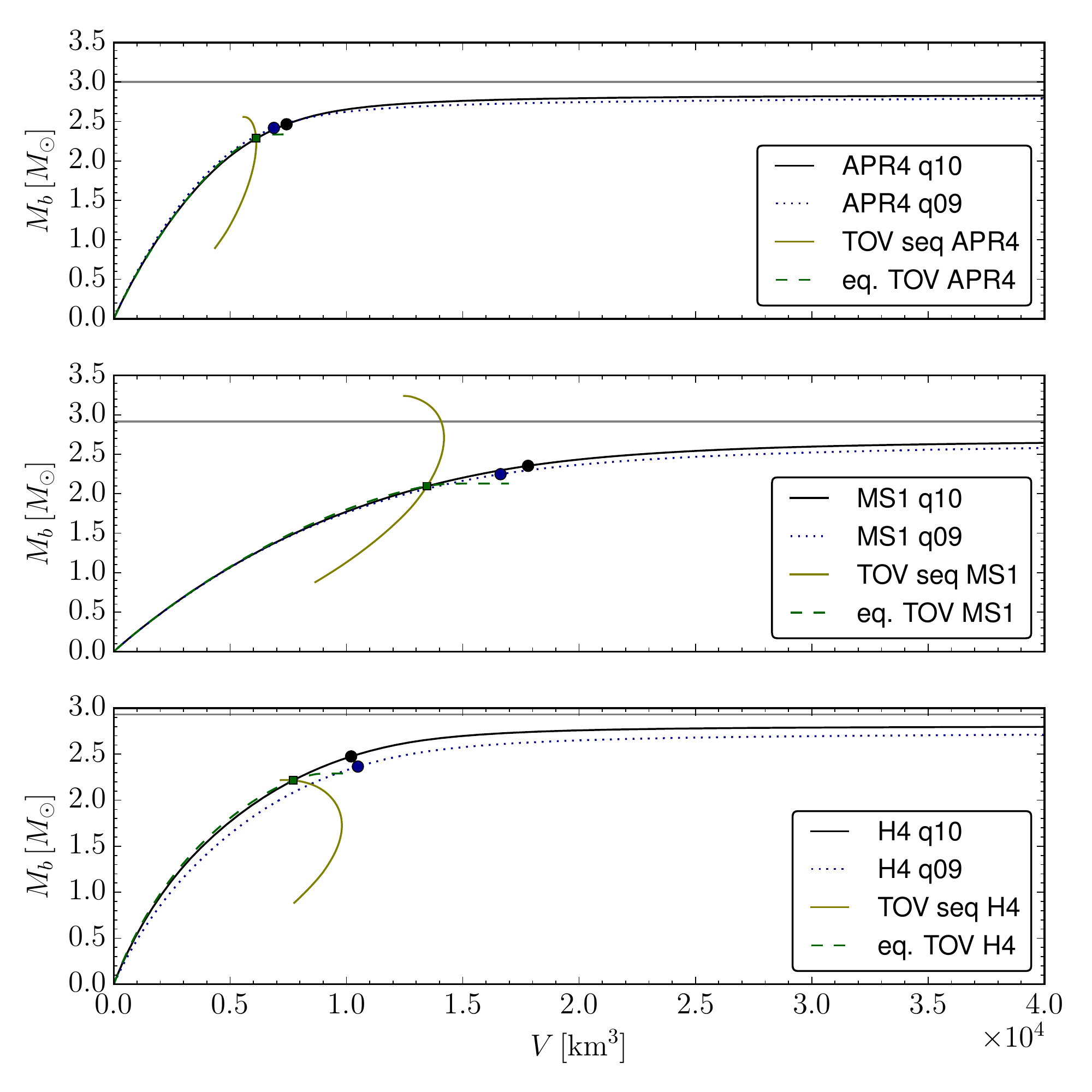}
  \caption{Mass versus volume relations of the merger remnant isodensity surfaces, 
  $20\msec$ after merger. The maximum compactness shell (the bulk) is marked by circle symbols.  
  For comparison, we show bulk mass and bulk volume (see text) 
  of TOV sequences obtained with the same EOS (continuous yellow line), and the mass-volume relation
  of the TOV core equivalent approximating the inner remnant core (dashed green line, bulk marked by 
  square symbol). Horizontal grey lines mark the total baryon mass of the system.} 
  \label{fig:mass_volume}
\end{figure}

The mass-versus-volume relations for the merger remnants are shown in \Fref{fig:mass_volume},
while the bulk properties of the remnants are given in \Tref{tab:outcome}.
For the models at hand, the radial mass distribution and the bulk compactness are 
mainly determined by the EOS, while the mass ratio has a minor impact (at fixed total 
gravitational mass). \Fref{fig:mass_volume} also shows the 
relation of bulk mass versus bulk volume for sequences of TOV solutions with the EOS used in 
this work. We use the intersection with the remnant profile to find a TOV model approximating 
the inner core of the remnant, called TOV core equivalent in the following. 
By comparing the mass-versus-volume relation of the TOV core equivalent and the remnant, we
find that {\it the structure of the core of the remnants is very well approximated by TOV core 
equivalent solutions}. 
\Fref{fig:mass_volume} also shows that the differences between TOV equivalent and actual 
remnant become gradually larger between the bulk of the TOV core equivalent (square symbol) 
and its surface. This is due to the fact that for the remnant, centrifugal forces become 
important in the outer envelope.

\begin{figure}
  \centering
  \includegraphics[width=0.99\linewidth]{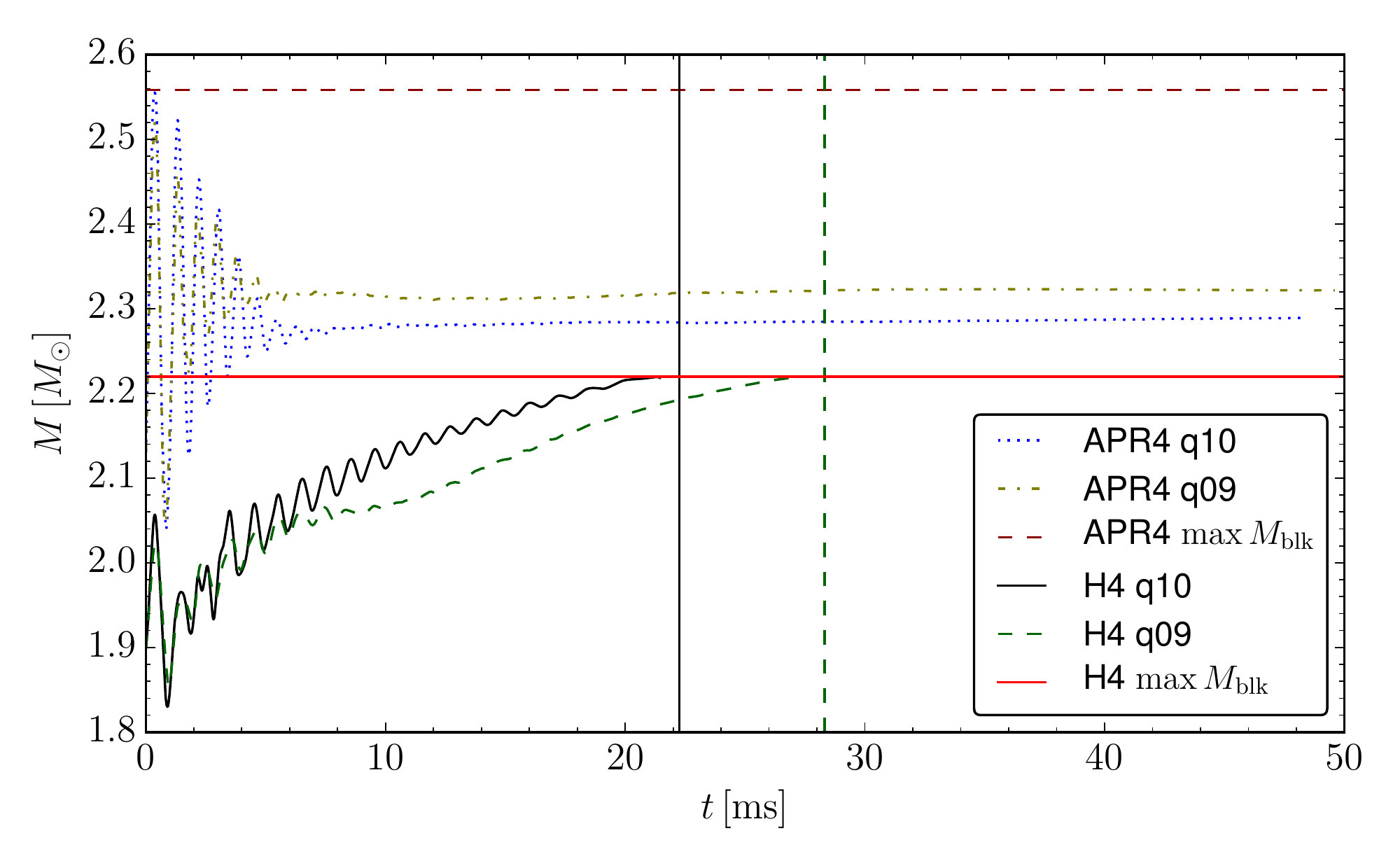}
  \caption{Evolution of the bulk mass of the TOV core equivalent approximating the
  merger remnant for the H4 and APR4 models. The horizontal lines mark the maximum bulk mass
  of stable TOV solutions for each EOS, while the vertical lines mark black hole formation. } 
  \label{fig:eqtov_evol}
\end{figure}

It is reasonable to assume that if there is no stable TOV solution approximating the inner core, it 
either has to rotate more rapidly or collapse. This gives us another critical mass, namely
the bulk mass of the maximum (gravitational) mass TOV star. This mass is $2.56\usk M_\odot$ for the APR4 EOS,
$2.22\usk M_\odot$ for the H4 EOS, and $3.24\usk M_\odot$ for the MS1 EOS.
Note that for the H4 simulations, the bulk mass of the TOV core equivalent is very close to 
the maximum value allowed for a stable star, while for the other models it is much lower. At the 
same time, only the H4 models collapsed to a BH on the timescale of the evolution.
To investigate this aspect further, we computed the evolution of the TOV core equivalent bulk mass 
for the H4 models, which is shown in \Fref{fig:eqtov_evol}. For those two runs, the core mass 
slowly approaches the critical one, and the collapse occurs as soon as the latter is reached.
We therefore propose a new conjecture: {\it merger remnants that do not admit a TOV core equivalent
promptly collapse to a BH}. 
This differs from the classification into supra- and hypermassive stars because it is a constraint
on the mass of the inner core, not the total mass. An important consequence is that, given the EOS, 
the presence of a post-merger phase (e.g., observed via the post-merger GW signal) 
would put a constraint on the mass, volume, and compactness of the inner core. 

Of course, our conjecture needs to be validated for more models. Also, it is meant for the phase when 
the remnant has settled down and can be regarded as stationary, not for the strongly oscillating phase
directly after merger. If it is also relevant for this phase is however an interesting question. 
For example, \Fref{fig:eqtov_evol} also shows the APR4 models, for which the core equivalent of the 
late remnant is well below the critical mass. During the early post-merger phase, however, it 
comes very close to the critical 
value for a short time. According to \cite{Hotokezaka:2011:124008}, the threshold for prompt
collapse of equal mass binaries with the APR4 EOS is reached at a single star ADM mass around 
$1.4\usk M_\odot$. Our APR4 equal-mass model with $M_g=1.35\usk M_\odot$ is indeed close to this threshold.

Another noteworthy observation is that the TOV core equivalents of the equal- and unequal-mass 
H4 models are very similar for the first $6\msec$ after merger and then suddenly start to differ.
This might be caused by the aforementioned change in the fluid flow happening around the same time.
If the two events are in fact related, it would imply that the vortex structure also has a direct 
impact on HMNS lifetimes.

\begin{figure}
  \centering
  \includegraphics[width=0.99\linewidth]{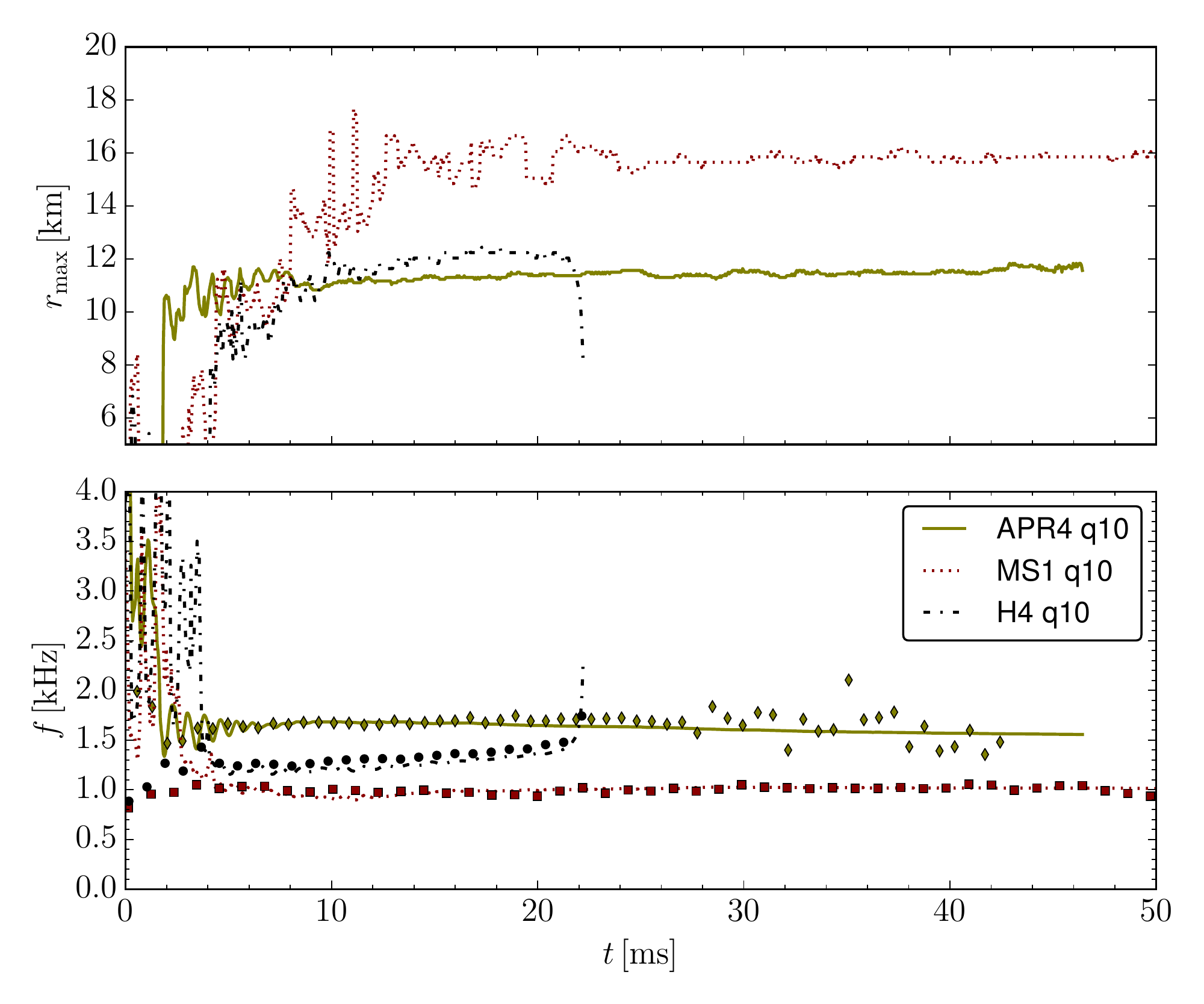}
  \caption{Top panel: location of the maximum of the rotation profile shown in \Fref{fig:rot_profs} 
  as function of time. Bottom panel: maximum rotation frequency (lines) compared to half of the 
  gravitational wave  instantaneous frequency (markers).
  } 
  \label{fig:rot_max}
\end{figure}

Returning to the rotation rate of our models shown in \Fref{fig:rot_profs}, 
we find that (for the given total mass) 
the EOS has a much stronger influence on the maximum rotation rate than the mass ratio. 
The APR4 EOS results in the highest rotation rate, followed by the H4 EOS, and then  
the MS1 EOS. We also notice a correlation between 
the maximum rotation rate and the position of the maximum, which is located further 
out for the models with smaller maximum rotation rate.
Intuitively, one might expect more compact models to rotate faster.
We find indeed that the bulk compactness of the remnants  
follows the same ordering as the maximum rotation rate, 
with the most compact remnant obtained for the APR4 case (see \Tref{tab:outcome}). 
For the equal-mass APR4 model, \Fref{fig:rot_profs} also shows the profile for the 
non-magnetized case, which is almost identical to the magnetized one.

The time evolution and radial location of the maximum rotation rate are shown in \Fref{fig:rot_max}.
Directly after merger, the maximum is located near the origin, although this measure is 
not meaningful during this phase because the fluid flow cannot be described as simple differential rotation. 
After around $5\msec$, however,
the remnant has settled down to a state similar to \Fref{fig:rot_profs}, with the maximum at the 
outer layers. Subsequently, the APR4 and MS1 models show only minor drifts of the maximum 
rotation rate on the timescale of the simulation. Also the location of the maximum varies only slightly. 
This indicates that the remnant will change on timescales much longer than the time window covered 
by our simulation.
The H4 model on the other hand exhibits a moderate increase of the rotation 
rate until the collapse to a BH occurs. 
\Fref{fig:rot_max} also shows the instantaneous GW frequency. The GW signal will be discussed in detail 
in \Sref{sec:gw}. Here we point out that the angular velocity of the $m=2$ GW pattern (i.e. half the GW 
frequency) closely follows the maximum rotation rate. This is not surprising if the maximum rotation rate 
is tied to the main $m=2$ deformation of the remnant, which seems to be the case.
This relation seems robust, as it was also found for different models in 
\cite{Kastaun:2015:064027, Endrizzi:2016:164001, Kastaun:2016, Hanauske2016arXiv} and we are not aware of 
a single counter-example.


\section{Magnetic fields}
\label{sec:mag}

\begin{figure*}
  \centering
  \includegraphics[width=0.49\textwidth]{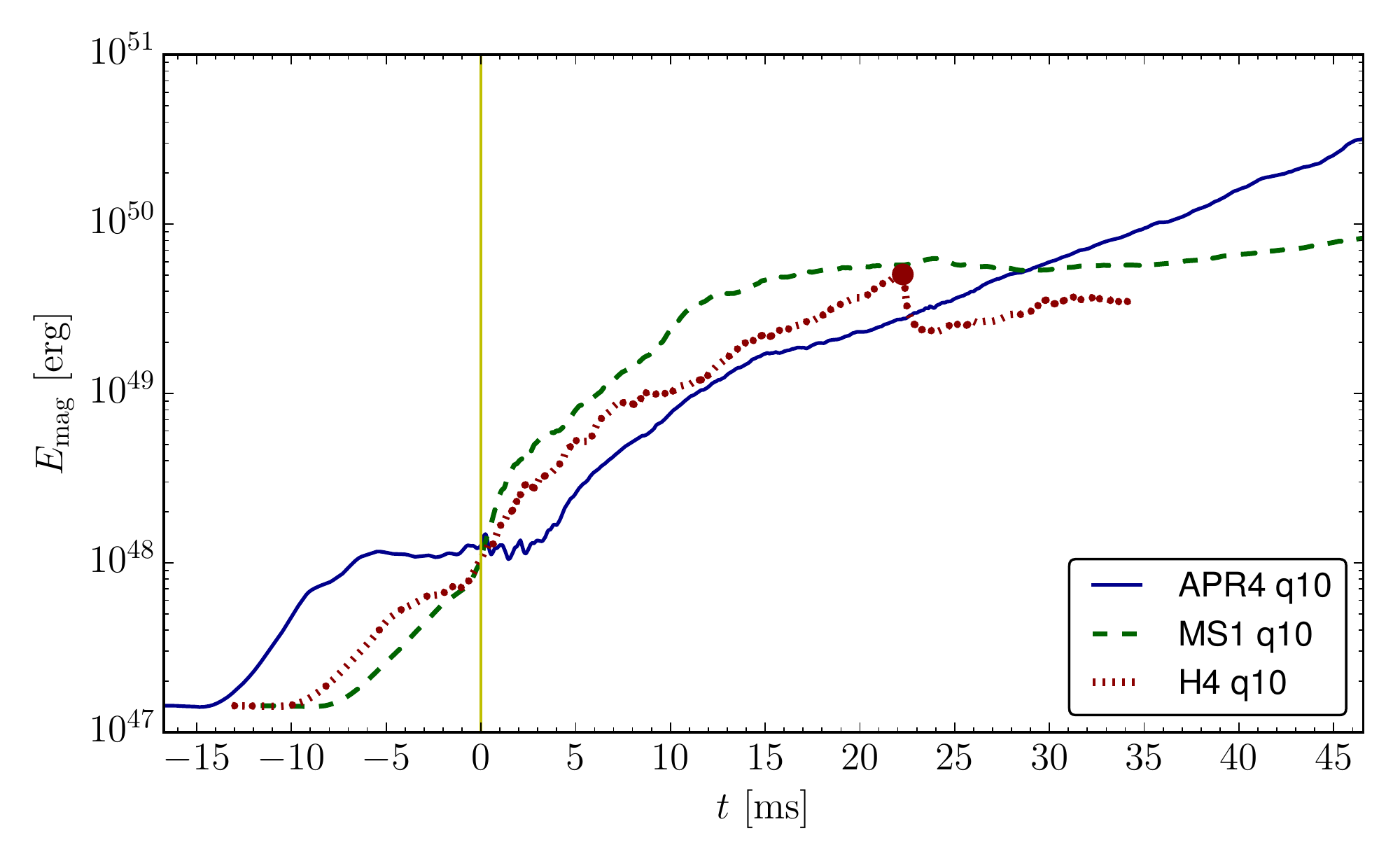}
  \includegraphics[width=0.49\textwidth]{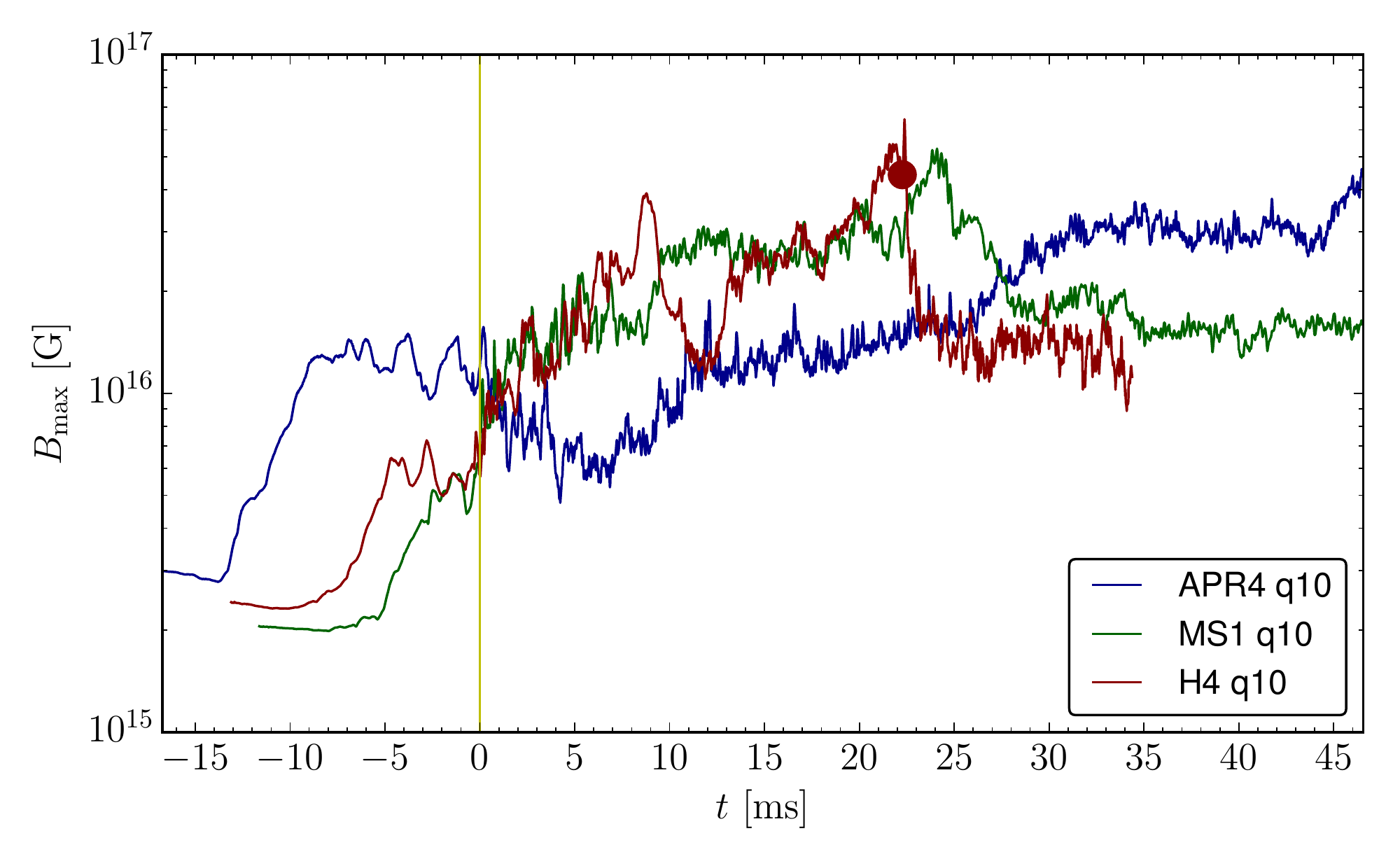}
  \includegraphics[width=0.49\textwidth]{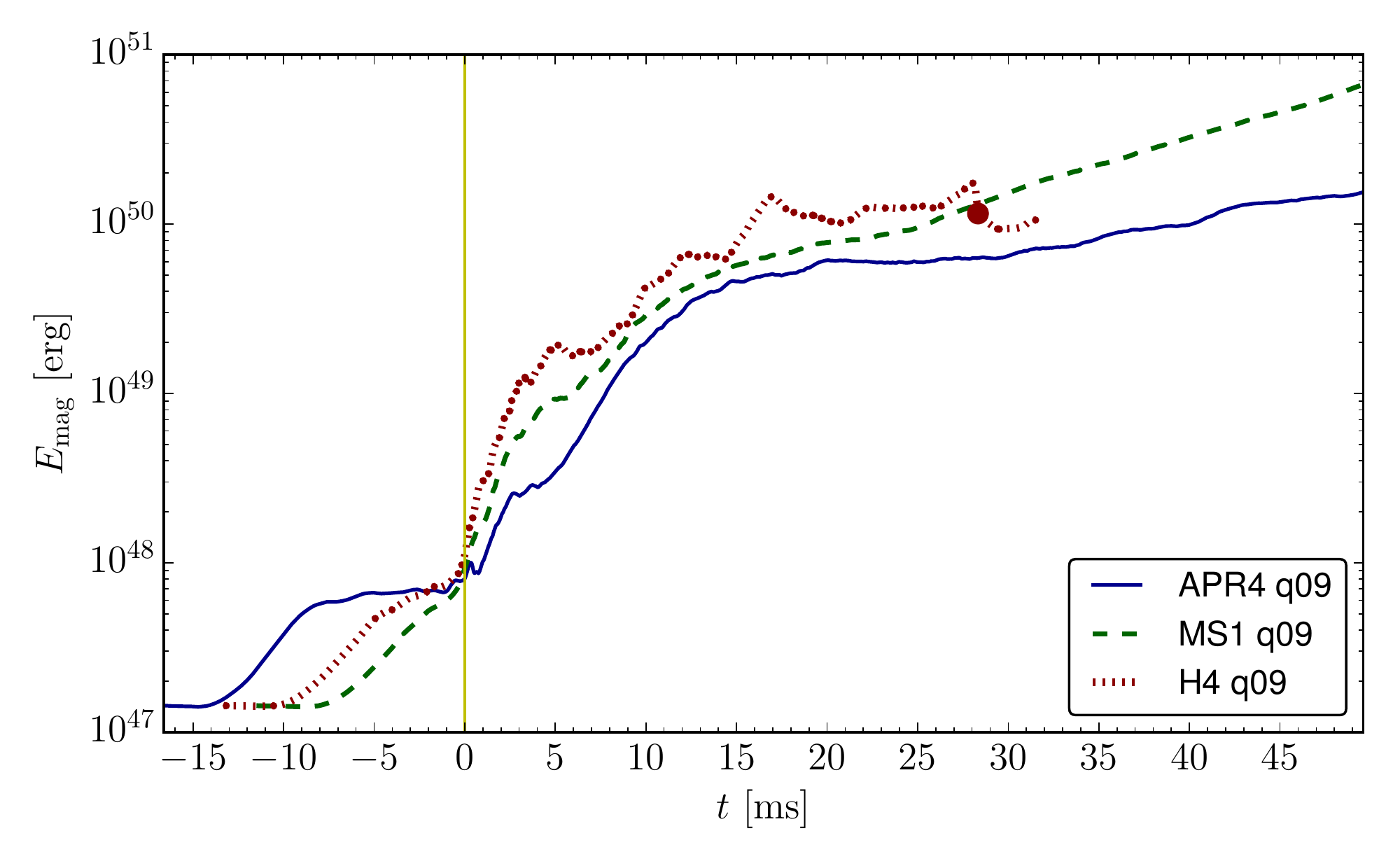}
  \includegraphics[width=0.49\textwidth]{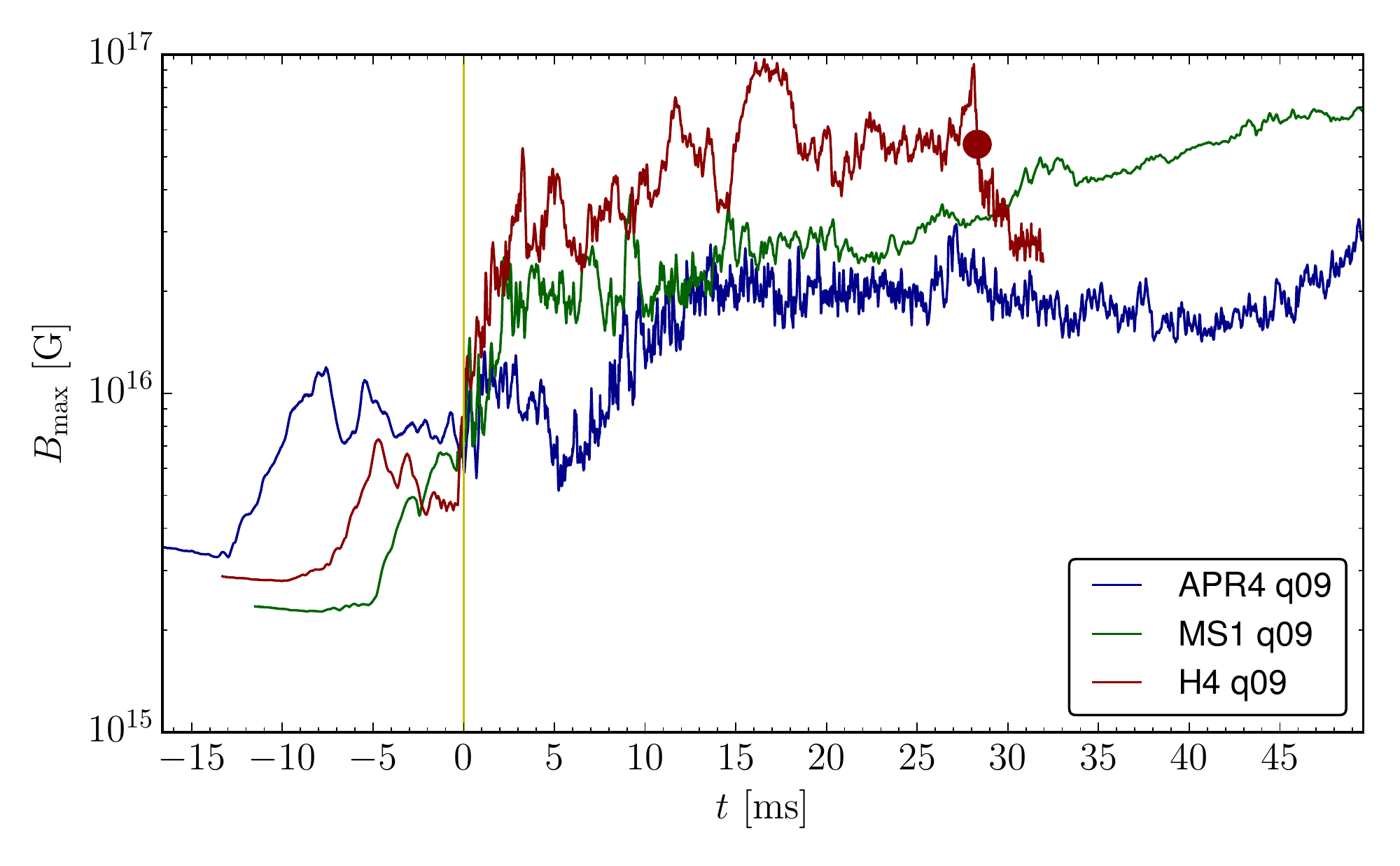}
  \caption{Evolution of magnetic energy (left) and maximum magnetic field strength (right) for the equal-mass (top) and unequal-mass (bottom) models. The vertical line marks the time of merger. The red circle (H4 EOS) marks the time of collapse to black hole.
  } 
  \label{fig:mag_1D}
\end{figure*}

In this Section we discuss the evolution of magnetic fields. \Fref{fig:mag_1D} (left panels) shows the magnetic energy evolution for the different EOS and mass ratios considered in this work.
All of our BNS models experience magnetic field amplification prior to merger, starting when the two NSs are at a proper distance of $\sim\!54$~km. After a few ms and about one order of magnitude increase in magnetic energy, the amplification can stall for some time and continue when the two NS cores effectively merge. Depending on the EOS, this stalling can last up to $\sim\!10$~ms (APR4 case) or be absent (MS1 case), with a duration that increases with compactness. 
This might simply be due to the different duration of the inspiral phase.

The cause of the pre-merger amplification and its saturation is still unclear. 
As discussed in the Appendix, the initial growth does not seem to be a result of insufficient resolution,
although we cannot rule out that it is caused by interaction of the NSs with the  artificial atmosphere.
The violations of the GR constraint equations introduced by adding the magnetic field can safely be neglected,
and also the deviation from hydrostatic equilibrium due to the additional magnetic pressure is too small and will only
lead to small oscillations. 
By looking at the magnetic field strength at the boundaries of 
the moving grids during inspiral, we find no evidence of spurious 
magnetic field amplification.
However, we cannot exclude that generic imperfections of the initial data lead to fluid 
flows that amplify the magnetic field. 
For the saturation phase, the resolution has a larger influence and it is not clear if the saturation is a 
purely numerical artifact or if the saturation mechanism is physical, but harder to resolve. 
Another effect that can be excluded is the development of a hydromagnetic instability
such as the Tayler instability of purely poloidal magnetic fields \cite{Markey1973}, since the Alfv\'en 
timescale inside the NSs before merger is at least one order of magnitude larger than the observed 
amplification timescale (see e.g. \cite{Ciolfi2011}).  

The only remaining physical explanation seems to be the time-changing tidal deformation during the inspiral. 
Although it is by no means clear how it would amplify the field, we note that a recent 
study \cite{Osso:2013:428518} suggested that the tidal forces can drive significant fluid 
flows inside the NSs in the late inspiral. 
If the observed amplification was indeed a physical effect, it would be very interesting. 
In particular, we note that all models end up with the same magnetic energy at the time of merger, independent of mass ratio and EOS. 
This would indicate that the magnetic energy at merger might be determined by the saturation scale
of the mechanism responsible for the amplification.
We stress again that our findings are not conclusive since we cannot rule out unphysical causes. 
In any case, the topic deserves further investigation.

We also note that magnetic field amplification prior to merger was already reported in other 
studies. For instance, Kiuchi et al. \cite{Kiuchi:2014:41502} evolved an equal-mass H4 model 
with different initial magnetic field strengths and resolutions and obtained in all runs a factor 
$\sim2$ amplification in magnetic energy in the last $5$~ms of pre-merger evolution 
(see Fig.~2 of \cite{Kiuchi:2014:41502}). 
Within the same $5$~ms time window, the above behavior is very similar to what we obtain for 
our equal-mass H4 model (see top left panel of Fig.~\ref{fig:mag_1D}).

When the two NSs merge, magnetic fields are strongly amplified 
by about one order of magnitude or more (factor $\sim50-500$ in magnetic energy). 
One key mechanism that is known to strongly amplify the toroidal component of the field is the Kelvin-Helmholtz (KH) instability, which develops in the shear layer separating the two NS cores when they come into contact. This effect is most likely responsible for the particularly steep increase of magnetic energy during the first $5\msec$ observed in all our simulations. For unknown reasons, the onset of the amplification is slightly delayed for the APR4 models (respectively by ${\sim}4$ and ${\sim}1\msec$ for the equal- and unequal-mass cases). 
Judging by the initial growth of the magnetic energy, both EOS and mass ratio have an influence on the KH instability. 
The effect of the KH instability in the early post-merger phase is also evident in terms of maximum magnetic field strength \Fref{fig:mag_1D} (right panels). After merger, this maximum is achieved in the equatorial region and corresponds to a magnetic field that is essentially toroidal.  
As a note of caution, we stress that the resolution employed in our simulations determines the smallest scale at which the KH instability is effective. As will be shown in the Appendix, the magnetic field after merger is not converging and higher resolution results in a steeper growth \footnote{Recent GRMHD simulations performed by 
Kiuchi et al.~\cite{Kiuchi:2015:1509.09205} at much higher resolution (up to a finest grid spacing of $\approx17$ m) 
show clearly that the finest grid spacing we employ is insufficient to 
properly resolve the KH instability.}. 
Nevertheless, with higher resolution the magnetic fields experience a faster amplification but also an earlier saturation, and the magnetic energy achieved in the end does not differ by more than a factor of two (when comparing medium and high resolutions).

\begin{figure}
  \centering
  \includegraphics[width=0.49\textwidth]{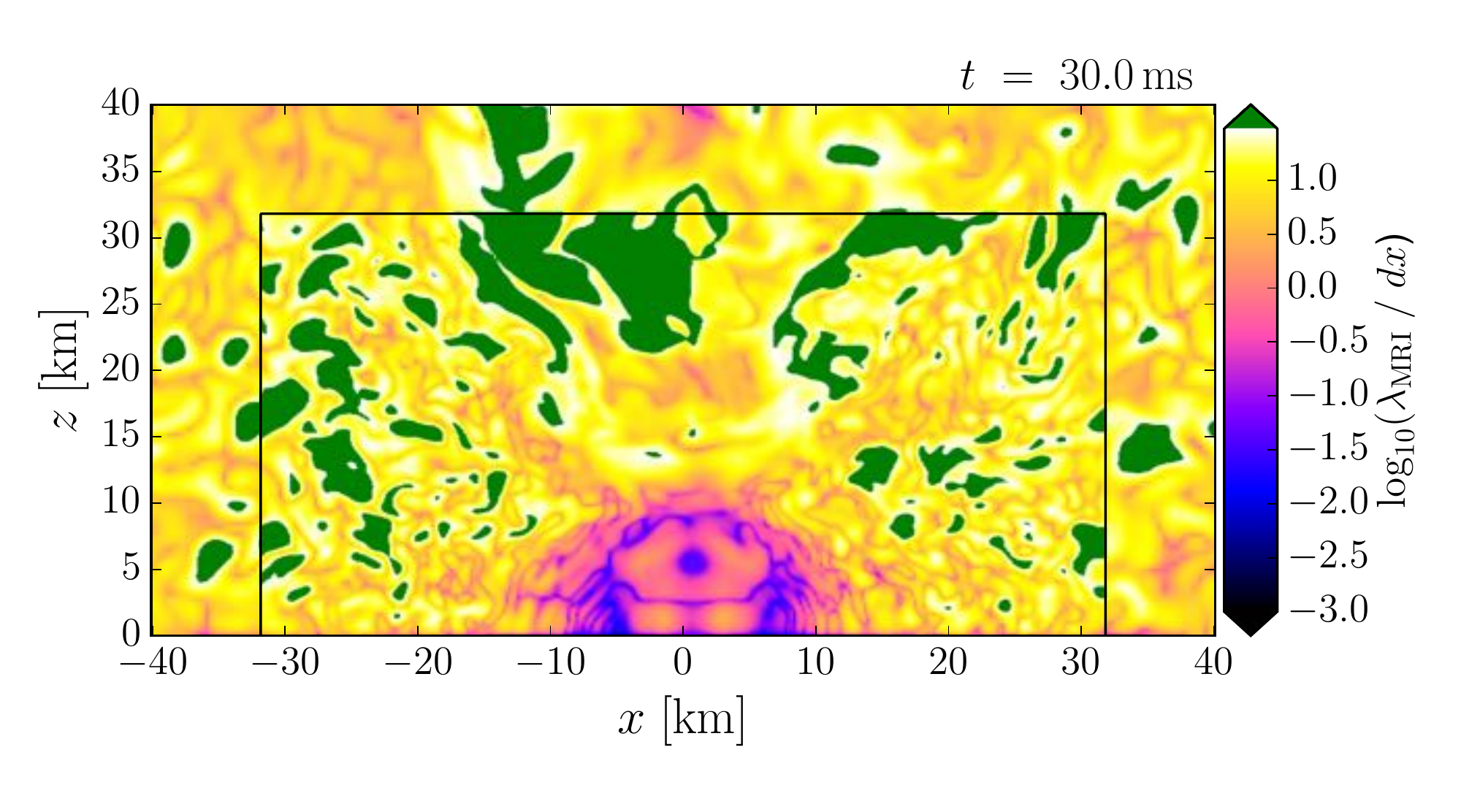}
  \includegraphics[width=0.49\textwidth]{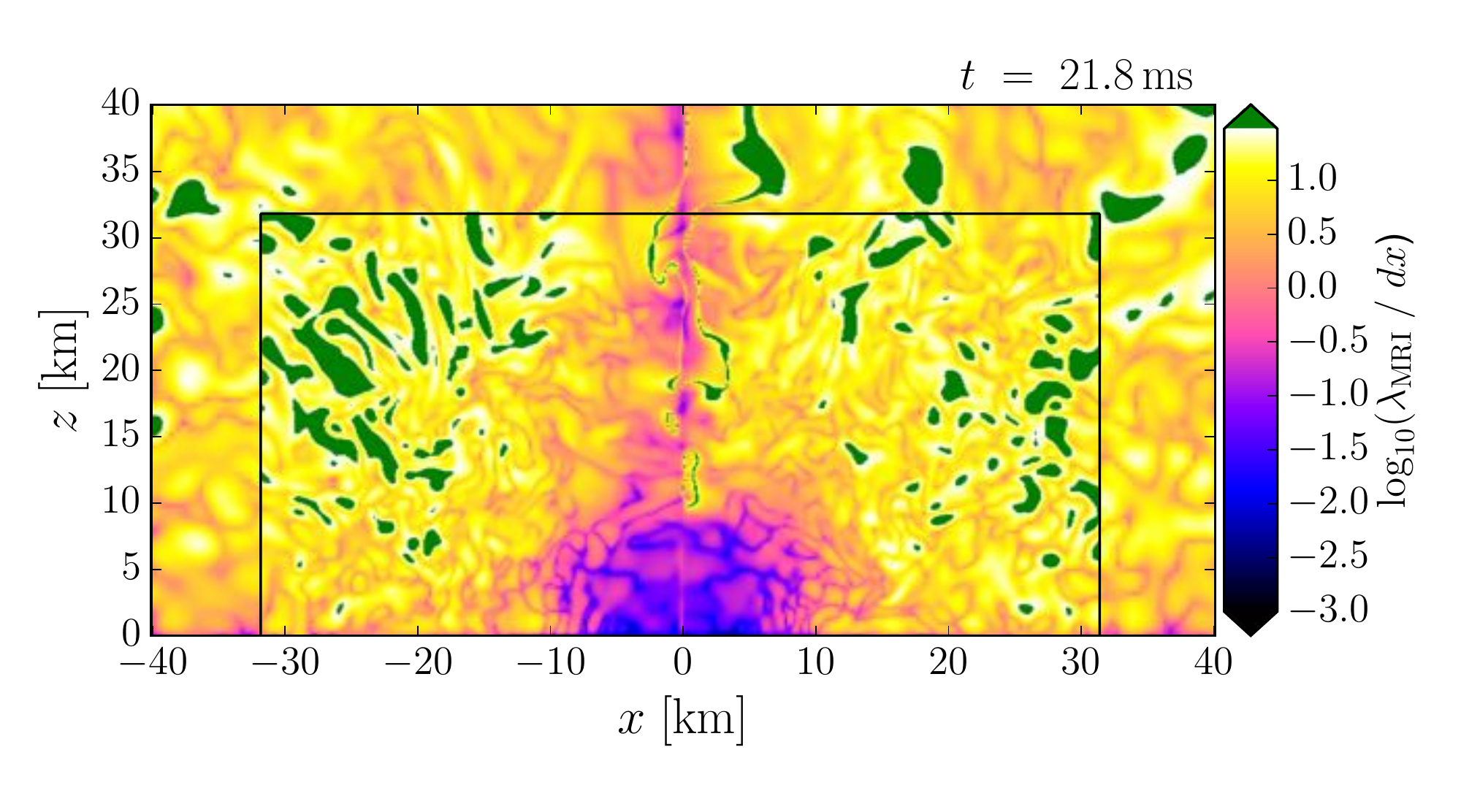}
  \caption{Meridional view of $\lambda_\mathrm{MRI}/dx$. This quantity gives an estimate of the number of grid points to resolve the wavelength of the fastest growing MRI mode (see text). Top: APR4 equal-mass case 30 ms after merger. Bottom: H4 equal-mass case, 0.5 ms prior to collapse.
  } 
  \label{fig:MRI}
\end{figure}

For all models, the rapid growth attributed to the KH instability only lasts for a few ms, after which the magnetic energy can still grow by more than one order of magnitude. At this later stage, we assume that the KH instability is gradually substituted by other amplification mechanisms, associated with turbulence and/or differential rotation. Apart from magnetic winding, which is well resolved and contributes in part to the growth of the toroidal field, the amplification mechanisms at play are limited by the smallest scales we can resolve. A potentially powerful mechanism is the magnetorotational instability (MRI) \cite{Balbus:1991,Duez2006a,Siegel:2013:121302}. In order to assess whether the MRI is contributing to the observed amplification we estimated the wavelength of the fastest growing MRI mode as $\lambda_\mathrm{MRI} \approx (2\pi/\Omega) \times B/\sqrt{4\pi\rho}$, where $\Omega$ is the angular velocity and $B$ is the magnetic field strength \footnote{A more accurate definition would require to compute the magnetic field strength along the direction of the wave vector under consideration. Therefore, by considering the total magnetic field strength, we are possibly overestimating $\lambda_\mathrm{MRI}$.}. 
Typically, the MRI is effective in numerical simulations when $\lambda_\mathrm{MRI}$ is resolved with at least 10 grid points (see, e.g., \cite{Siegel:2013:121302}). 
As shown in \Fref{fig:MRI}, this requirement is satisfied in most of the region outside the remnant, for both long-lived (APR4 and MS1) and short-lived (H4) remnants. We conclude that MRI is likely playing an active role in our simulations. 

\begin{figure*}
  \centering
  \includegraphics[width=0.99\textwidth]{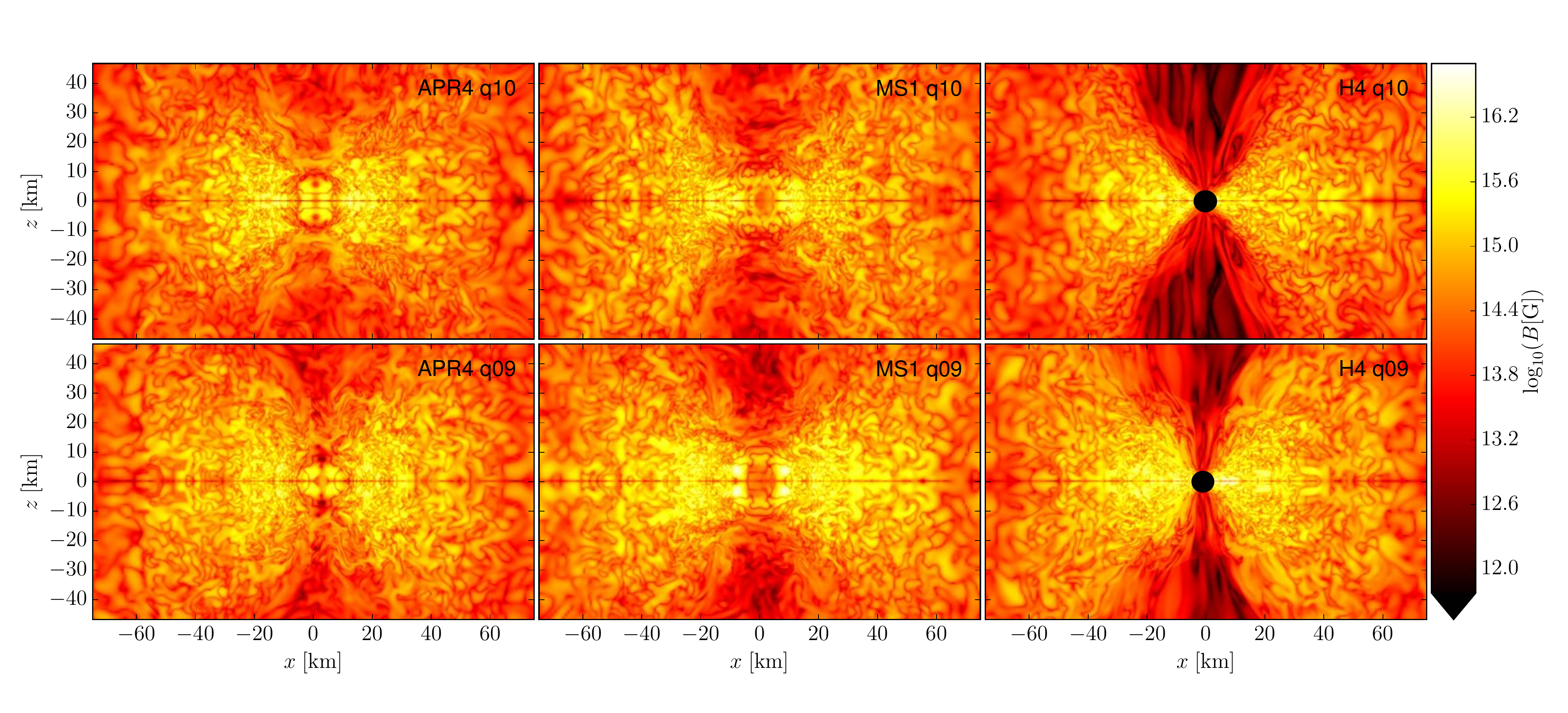}
  \caption{Meridional view of the magnetic field strength 30 ms after merger for different EOS (left to right: APR4, MS1, H4) and mass ratios (top row $q=1$, bottom row $q=0.9$). The region inside the apparent horizon is indicated in black (H4 case).
  } 
  \label{fig:mag_2D}
\end{figure*}

\Fref{fig:mag_2D} shows the magnetic field strength in the meridional plane 30 ms after merger. For all six models, the maximum field strength is in excess of $10^{16}$~G (cf.~right panels of \Fref{fig:mag_1D}) and is achieved in the inner equatorial region. After this time, the hydrodynamic evolution of both the APR4 and MS1 models has reached a quasi-stationary state, while the magnetic energy keeps growing more or less exponentially. For APR4 models, the unequal-mass case shows a lower amplification rate at this later stage, while for MS1 models it is the other way around. This suggests that the magnetic field evolution depends on EOS and mass ratio in a complex way and that the effect of the two cannot be easily disentangled.

The overall amplification of magnetic energy $45\msec$ after merger is between two and three orders of magnitude with respect to the energy at merger time. 
Note that the initial amplification attributed to the KH mechanism only accounts for a small fraction of the final energy, 
and hence the final amplification factor is dominated by the late amplification via MRI and magnetic winding. 
From \Fref{fig:MRI}, we expect to resolve the MRI, at least outside the remnant.
Our resolution study (see Appendix for details) indeed indicates that the final magnetic energy is starting to converge, 
in contrast to the early growth. 
We expect the final total magnetic energy to be accurate within an order of magnitude.
With the MRI in the disk dominating the final magnetic energy, we would also not expect substantial changes when using 
a subgrid model \cite{Giacomazzo:2015}.
Note that the numerical accuracy of the amplification strongly depends on the initial field strength, 
because the scale of the fastest growing MRI mode is proportional to the magnetic field. In a previous study 
\cite{Endrizzi:2016:164001} employing similar resolutions, we started with a much lower initial magnetic energy and
found no indication for numerical convergence in the final value of $E_\mathrm{mag}$ (which was still smaller than 
the ones reached in this work).
When taking our simulations as indication for real mergers with similar initial magnetic energy, we believe that
the main uncertainty is not the numerical accuracy but the geometry of the initial magnetic field, 
in particular the field outside the stars. This aspect will be further investigated in future studies.

\begin{figure*}
  \centering
  \includegraphics[width=0.9\textwidth]{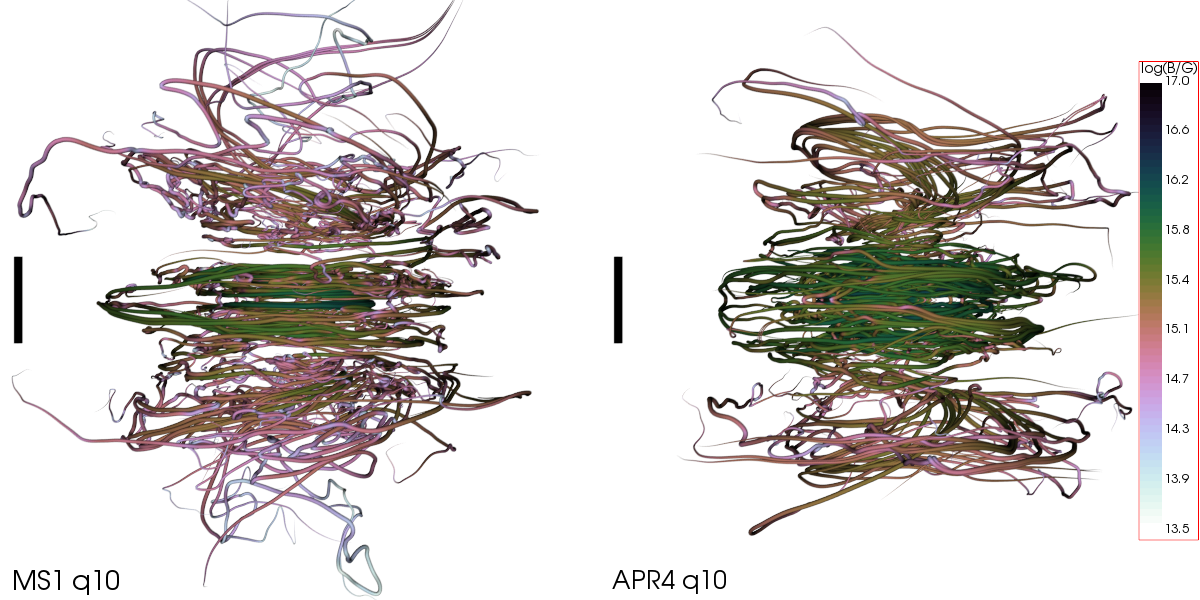}
  \caption{Structure of magnetic field $45\msec$ after merger for the equal-mass
  MS1 (left) and APR4 (right) models.
  The coloring indicates the magnetic field strength ($\log_{10}(B[\mathrm{G}])$, 
  same color scale for both models). For more quantitative results 
  see \Fref{fig:bhist_APR4_MS1}. The black bars provide a length scale of 
  $20 \kmeter$.
  } 
  \label{fig:field3d_apr4_ms1}
\end{figure*}

We now turn our attention to the geometrical structure of the magnetic field obtained 
towards the end of the simulations. For a qualitative description, we visualize the 
field lines using the same method as in \cite{Kawamura:2016:064012}. In short, the method
tries to show only the field lines with the largest average ratio of magnetic field 
strength to the maximum field strength at same $\theta$ coordinate. This is adapted
to more or less axisymmetric configurations where the field strength varies strongly 
between the pole and the equatorial plane. For details, see \cite{Kawamura:2016:064012}.
\Fref{fig:field3d_apr4_ms1} shows the field lines in 3D for the equal-mass MS1 and APR4 models 
$45\msec$ after merger. As already pointed out, the field is
largest on the equatorial plane, where it is predominantly toroidal. The field around the axis
is weaker and mostly unordered, with a slight tendency to helical structures.
This is in contrast to the cases described in \cite{Kawamura:2016:064012}, where the 
central object was a BH and where more ordered twister-like structures were found 
along a cone around the orbital axis. We expect that our H4 models, if evolved for long 
enough after collapse to a BH, would also develop a similar geometry.
In the case of a long-lived remnant (APR4 and MS1 models), however, the 
formation of analogous structures on longer timescales cannot be excluded. A notable difference 
is that magnetic fields along the orbital axis, although disordered, can exceed $10^{14}$~G 
while in the cases where a BH is formed (H4 models) they hardly exceed $10^{13}$~G. 

\begin{figure}
  \centering
  \includegraphics[width=0.99\columnwidth]{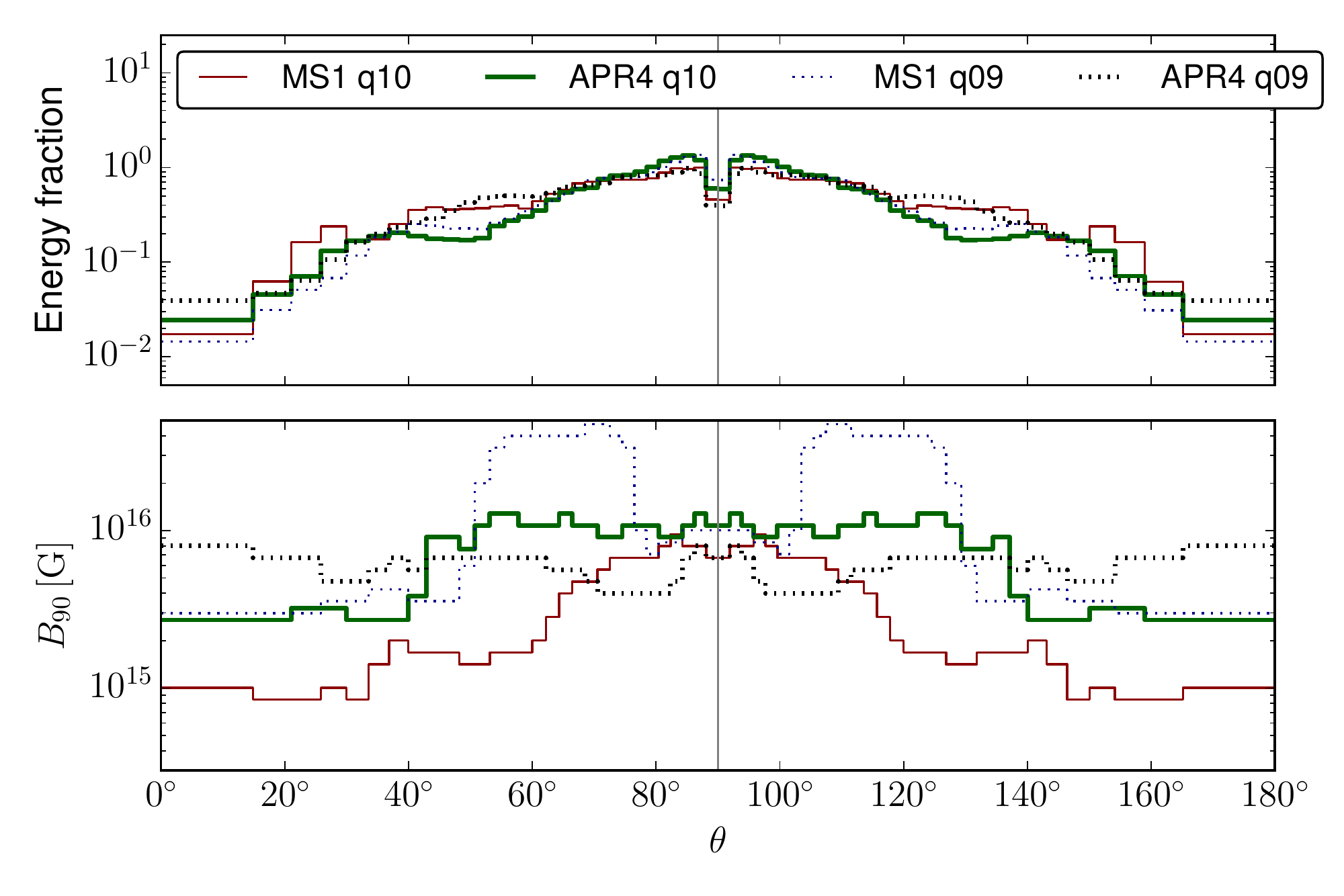}
  \caption{Distribution of magnetic field with respect to $\theta$-coordinate, for APR4 
  and MS1 models $45\msec$ after merger. 
  Top: histogram of magnetic energy employing bins regularly spaced in $\cos(\theta)$, 
  where $\theta = 0$ on the positive $z$-axis and $\theta = 90\degree$ at the equator. 
  Each curve is normalized to the total magnetic energy. 
  Bottom: characteristic field strength $B_{90}$ defined as the value for which $90\%$ of 
  the magnetic energy inside a given $\cos(\theta)$ bin is contributed by regions with  
  field strengths below $B_{90}$.} 
  \label{fig:bhist_APR4_MS1}
\end{figure}

For a quantitative description of the field distribution in the polar angle $\theta$, we use the same measures as in
\cite{Kawamura:2016:064012}: we sum up the magnetic energy in a 2D histogram binned
by $\cos(\theta)$ and magnetic field strength. For each bin in $\theta$, we define 
the field strength $B_{90}$ such that $90\%$ of the magnetic energy in the same bin is contained 
in regions with lower field strength. This measure is in-between average and 
maximum norm, but less sensitive to single points than the latter.
We also compute the total energy in each $\cos(\theta)$ bin (regardless of field strength).
The result is shown in \Fref{fig:bhist_APR4_MS1}. 
We find that for all models most of the magnetic energy is in the equatorial region. 
The characteristic field strength $B_{90}$, on the other hand, shows a different behavior
for different models. The APR4 equal-mass case has a rather flat value around $10^{16}$~G
between $\theta{\approx}40\degree$ and $140\degree$ (equatorial region) 
and around $3\times10^{15}$~G near the axis.
The APR4 unequal-mass has similar values except along a cone  
of half-opening angle of ${\approx}60\degree$--$70\degree$ around the spin axis, where 
$B_{90}$ is as strong as $4\times10^{16}$~G. The MS1 equal-mass model has the lowest 
$B_{90}$ of ${\approx}10^{15}$~G along the axis and almost $10^{16}$~G in the equatorial region 
($70\degree$--$110\degree$). Finally, the MS1 unequal-mass model has a rather flat value of 
$B_{90}{\approx}10^{16}$~G at all angles.
These results show that there is no unique behavior at this stage of the evolution. In order to assess 
whether a common ordered structure would emerge at a later time (e.g., a structure favorable for 
jet formation), long-term simulations extending far beyond the timescales covered in this work
are needed. 
 
\section{Short gamma-ray bursts}
\label{sec:sgrb}

\begin{figure*}
  \centering
  \includegraphics[width=0.49\textwidth]{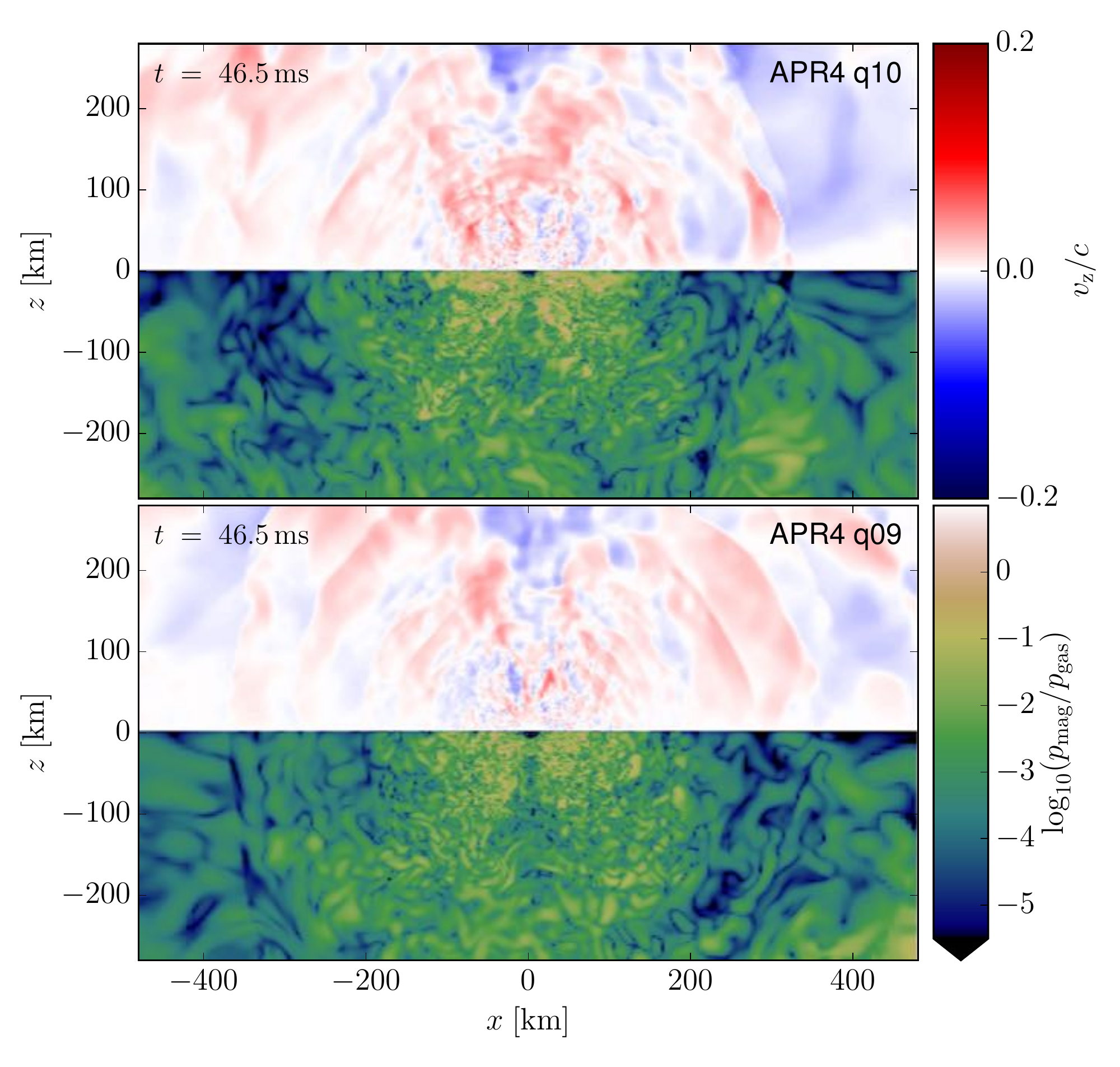}
  \includegraphics[width=0.485\textwidth]{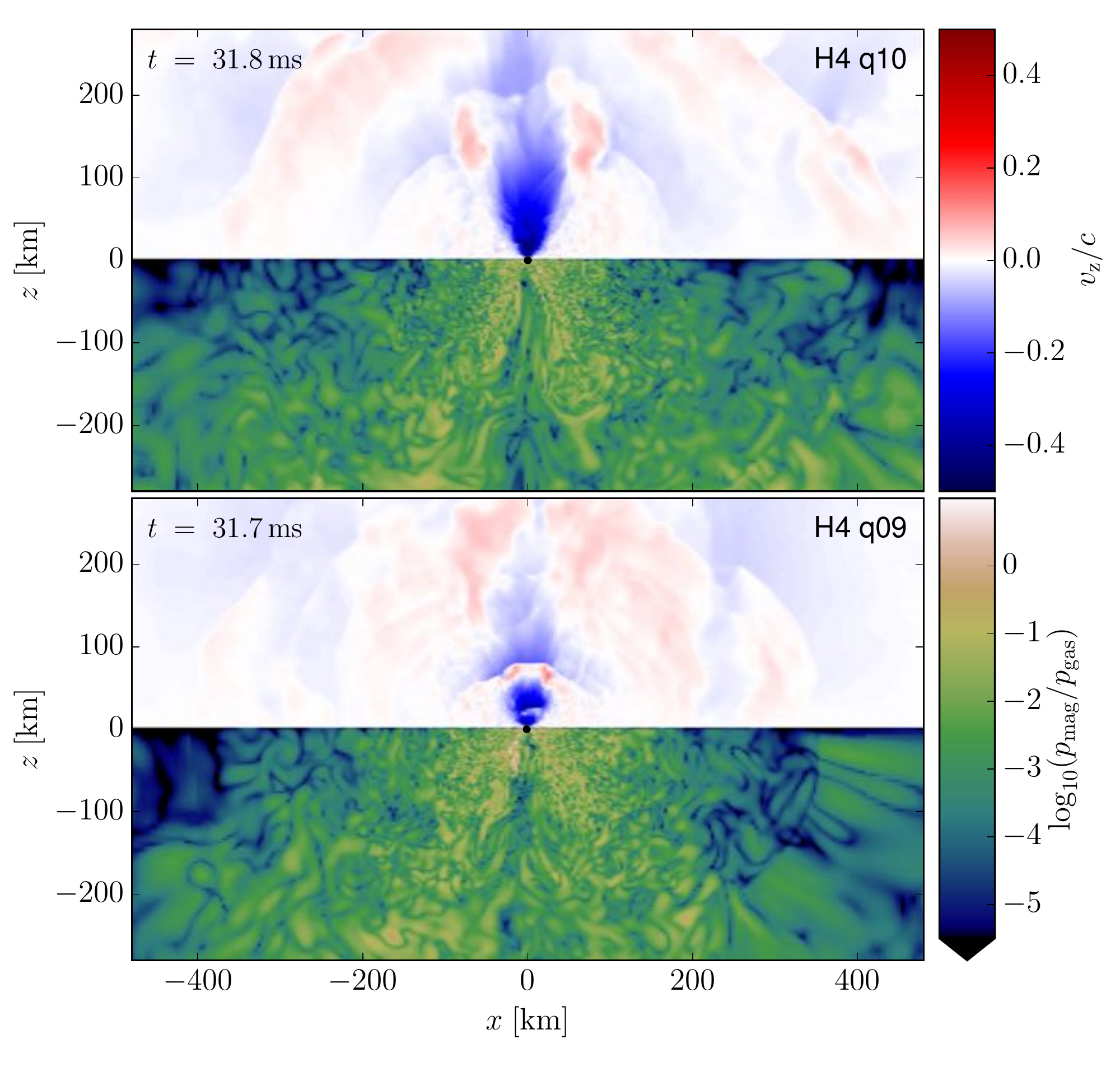}
  \caption{Meridional view of the fluid velocity perpendicular to the orbital plane (i.e. the $z$-component) and of the magnetic-to-fluid pressure ratio (on the top and bottom half of each panel, respectively), towards the end of our simulations. Left: APR4 models with equal mass (top) and unequal mass (bottom). Right: the same for H4 models (region in black is inside the apparent horizon). 
  } 
  \label{fig:vz_MHDratio}
\end{figure*}

In what follows, we discuss the results of our simulations in the context of SGRBs.
BNS and NS-BH mergers represent primary candidates as progenitors of these events
\cite{Paczynski1986,Eichler:1989:126,Narayan1992,Barthelmy2005a,Fox2005,Gehrels2005,Shibata2006b,Tanvir2013,Berger2013,Paschalidis:2015:14,Ruiz2016}. 
One main reason is that a common product of such mergers is a compact object (a massive NS or a BH) 
surrounded by an accretion disk of mass $\gtrsim\!0.1 M_\odot$, and the 
corresponding accretion timescale ($\sim\!1$~s) matches the duration 
of the SGRB prompt emission ($<\!2$~s).
In addition, the lack of supernova associations, the diverse types of host galaxies
(which include also early-type galaxies), and the large offsets from the 
center of the host galaxy, are all in favor of a binary compact object
origin \cite{Berger2014}.

The most commonly discussed scenario is the one in which a compact 
binary merger leads to the prompt formation of a BH surrounded by a massive 
accretion disk \footnote{The same scenario also applies to the case in which the product of the merger is 
a HMNS, collapsing to a BH within a few tens of ms.}.
The accretion onto the BH is what provides the source of power.
Since the gamma-ray emission is believed
to be generated within a relativistic outflow,
an additional key ingredient is the ability of
the system to drive a jet. Two main mechanisms have been proposed as
energy sources capable of launching a jet: 
(i) the deposition of thermal energy at the poles of the BH via the 
annihilation of neutrinos and antineutrinos copiously emitted by the hot 
accretion disk \cite{Eichler:1989:126,Ruffert:1999:573}, and (ii)
the action of large scale magnetic fields threading the 
accretion disk and tapping the rotational energy of the BH via the Blandford-Znajek 
mechanism \cite{Blandford1977} (analogous to the well established case 
of AGNs/blazars \cite{McKinney2009}).
Recent simulations indicate that the neutrino mechanism, while potentially important, 
seems to be too weak to drive a powerful enough jet on its own, 
especially in the BNS merger case \cite{Just2016,Perego2017}. 
Hence, the energy requirements favor magnetic fields as the main driving 
force.  

In the last few years, GRMHD simulations of BNS or NS-BH mergers provided 
important hints on the possibility of launching a magnetically driven jet
(e.g.~\cite{Rezzolla:2011:6,Kiuchi:2014:41502,Paschalidis:2015:14,Ruiz2016,Kawamura:2016:064012}). 
In particular, Ruiz et al.~\cite{Ruiz2016} reported for the first time in a 
BNS merger simulation the emergence of a
collimated and mildly relativistic outflow along a baryon-poor and 
magnetically dominated funnel surrounding the BH spin axis 
(referred to as ``incipient jet"). A similar result was obtained earlier 
for the NS-BH case \cite{Paschalidis:2015:14}.
No other group has so far reported an analogous result. 
In the most recent paper on the subject \cite{Kawamura:2016:064012}, 
BNS merger simulations performed by our group 
showed the formation of a twister-like magnetic field structure
along the spin axis of the BH, but no net outflow was found, nor 
a magnetically dominated funnel. 

Our present simulations forming a BH-disk system (models with the H4 EOS)
are too short to assess if the post-collapse system would evolve 
in a similar way and possibly form an incipient jet at later times. 
As shown in Fig.~\ref{fig:vz_MHDratio} (right panels), a few ms after collapse matter is 
still largely infalling along the BH spin axis. Magnetic pressure is becoming 
comparable to the gas pressure at the edges of the disk, but it is still 
generally subdominant inside the baryon-poor funnel.

The main focus of this work, however, is on magnetized BNS mergers forming a long-lived NS.
The possibility that such a remnant could act as the central engine of a SGRB was put 
forward by the so-called magnetar model \cite{Zhang2001,Gao2006,Metzger2008}, 
which represents the most popular alternative to the standard BH-disk scenario. 
In this case, an accretion-powered jet is launched by a strongly magnetized 
NS surrounded by a massive accretion disk. While this is viable for BNS mergers,  
it clearly excludes NS-BH binaries as the possible progenitor. 
The magnetar model was recently revived, after the observation by the Swift satellite \cite{Gehrels2004-Swift} of
long-lasting ($\sim$minutes to hours) X-ray afterglows accompanying a significant 
fraction of all SGRB events \cite{Rowlinson2013,Lu2015}. 
This evidence poses a challenge to the BH-disk scenario, 
as the short accretion timescale onto the BH can hardly be reconciled with a sustained 
emission lasting $\gtrsim\!100-1000$~s.
Within the magnetar model, thanks to the EM spindown emission from the magnetized NS,
these afterglows might find instead a natural explanation.
Moreover, the observation of NSs with a mass of $\approx\!2 M_\odot$, by supporting 
the formation of a long-lived NS in a significant fraction of all BNS mergers, plays in favor 
of a magnetar central engine.

Nevertheless, this scenario has a potential difficulty in explaining the prompt SGRB emission.
Differently from the BH case, in which accretion along the BH spin axis rapidly evacuates  
a low density funnel, a long-lived merger remnant remains surrounded by a more isotropic 
baryon-loaded medium and the much higher rest-mass density along the spin axis
might be sufficient to choke a jet or to prevent its formation in the first place
\cite{Murguia-Berthier2014,Nagakura2014,Murguia-Berthier2017a}.

Our long-lived remnant models (with APR4 or MS1 EOS) reproduce the above situation and can 
thus provide useful hints into the viability of the magnetar model. 
As shown in Section~\ref{sec:dynamics} and Fig.~\ref{fig:rho_funnel}, towards the end of the simulations 
we find rest-mass densities along the orbital/spin axis of the order of $10^{10}$~g/cm$^3$
and slowly increasing (computed at $z\!\sim\!50$~km almost $50$~ms after merger). 
At the same time, the system is characterized by a quasi-stationary 
evolution showing no clear flow structure in the surrounding of the merger site, 
and in particular no net outflow along the axis
(cf.~Figs.~\ref{fig:apr4q10_mflux_xz},\,\ref{fig:ms1q10_mflux_xz} 
and left panels of Fig.~\ref{fig:vz_MHDratio}).
Moreover, we observe magnetic-to-fluid pressure ratios approaching unity inside a 
spherical region of radius $\sim\!100$~km, but no magnetically dominated funnel 
(Fig.~\ref{fig:vz_MHDratio}).
Finally, the magnetic field does not show a strong poloidal component along the 
axis (see Figs.~\ref{fig:field3d_apr4_ms1},\,\ref{fig:bhist_APR4_MS1}), 
which is necessary in order to launch a magnetically driven jet.
We conclude that {\it the systems studied in this work are unlikely to produce a jet on timescales 
of $\sim\!0.1$~s; either they do so on much longer timescales ($\gg\!0.1$~s) or 
they are simply unable to generate a collimated outflow}.

We stress, however, that our simulations cannot provide the final answer. 
First, we do not include neutrino radiation, which might provide support to the 
production of a jet. Second, we start with purely poloidal magnetic fields confined 
inside the NSs and we do not properly resolve all magnetic field amplification 
mechanisms, in particular the KH instability and MRI inside the remnant. 
We also note that while further increasing the strength of the initial magnetic 
fields ($\sim\!10^{15}$~G) would be difficult to motivate, simply changing the geometrical 
structure might still completely change the outcome. In \cite{Ruiz2016}, for instance, 
it is shown that initial 
(pre-merger) poloidal magnetic fields extending also outside the two NSs can help 
jet formation in the post-merger evolution.
Third, the emergence of an incipient jet probably requires simulations 
lasting ${\gtrsim}0.1$~s, i.e.~much longer than ours. 
All of the above elements will have to be reconsidered in future studies. 

As a final note on SGRB models, we recall that an alternative ``time-reversal" scenario
\cite{Ciolfi:2015:36,Ciolfi:2015:PoS} was proposed most recently to overcome the problems 
of the BH-disk and magnetar scenarios.
This model envisages the formation of a long-lived supramassive NS as the 
end product of a BNS merger, which eventually collapses to a BH on timescales of up to 
$\sim$minutes of even longer. 
During its lifetime, the strongly magnetized NS remnant injects energy into the 
surrounding environment via EM spindown. Then, it collapses to a BH and generates the 
necessary conditions to launch a jet. At that point, the merger site is surrounded by a 
photon-pair plasma nebula inflated by the EM spindown and by an external layer 
of nearly isotropic baryon-loaded ejecta (expelled in the early post-merger phase, but 
now diluted to much lower densities).
While the jet easily drills through this optically thick environment and escapes to 
finally produce the collimated gamma-ray emission, spindown energy 
remains trapped and diffuses outwards on much longer timescales. As a result, spindown 
energy given off by the NS {\it prior to collapse} powers an EM transient (in particular in the
X-rays) that can still be observed for a long time {\it after the prompt SGRB}.
This offers a possible way to simultaneously explain both the prompt emission and the 
long-lasting X-ray afterglows. 
Such a scenario covers timescales that extend far beyond the reach of present 
BNS merger simulations and thus it cannot be validated in this context. We do however 
note that the roughly isotropic matter outflows observed in our simulations would provide the 
required baryon-rich environment. On the other hand, the complicated field structures 
found in the remnants highlight that modeling the spindown radiation with a simple dipolar 
field can only serve as a toy model.


\section{Mass ejection}
\label{sec:ej}

We now discuss in more detail the ejection of matter during and after merger.
In order to compute the amount of unbound matter, we use the geodesic criterion 
$u_t\!<\!-1$ to estimate if a fluid element has the potential to escape to infinity. We then integrate 
the flux of unbound mass through spherical surfaces. The main source of error is the artificial 
atmosphere. Far away from the source, the ejecta are diluted enough such that the 
ejected matter with the lowest
density is lost to the artificial atmosphere, and the least unbound ejected matter 
becomes bound again because of the unphysical atmospheric drag (compare also
the discussion in \cite{Endrizzi:2016:164001}). Extracting at small radii on the other 
hand ignores matter that becomes unbound further out, i.e. the geodesic assumption is 
invalid in the more dynamic inner regions. As a best guess for the ejected mass, we use 
the maximum obtained from spherical surfaces placed at radii 
$148, 295, 443, 591, 738, 886$, and $1033 \kmeter$. We estimate those values to 
be accurate only within a factor 2, due to the errors described above. The results are 
reported in \Tref{tab:outcome}. We also note that those estimates do not include possible 
contributions from magnetically driven winds \cite{Siegel:2014:6}, since the geodesic 
criterion does not account for accelerations by magnetic fields.

According to our estimates, the APR4 models eject ${\sim}10^{-2}\usk M_\odot$, 
while the MS1 and H4 models only eject ${\sim}10^{-3}\usk M_\odot$.
The equal- and unequal-mass cases differ at most by a factor two (for the APR4 models).
This similarity should not come
as a surprise since our unequal-mass models have a mass ratio of $0.9$
and therefore tidal ejections are not as strong  as for the case of NS-BH 
binaries, where mass ratios as low as ${\sim}1/7$ are typically expected.

Non-magnetized versions of our MS1 and H4 equal-mass models have 
already been investigated in \cite{Hotokezaka:2013:24001} (we do not compare ejecta
masses for their APR4 model since our piecewise polytropic approximation of the 
APR4 EOS differs in the low density regime, which is more important for the ejecta
than for the general dynamics). As shown in \cite{Hotokezaka:2013:24001}, the thermal 
component of the EOS can have an impact as well.
Comparing to the models in \cite{Hotokezaka:2013:24001} using the same value 
$\Gamma_\text{th}=1.8$ as in our simulations, we find that our value is lower by a 
factor $1.9$ for the MS1 model, and higher by a factor $1.4$ for the H4 model.
The accuracy of these results is however not sufficient to attribute the differences to 
the presence of magnetic fields.  

\begin{figure}
  \centering
  \includegraphics[width=0.83\linewidth]{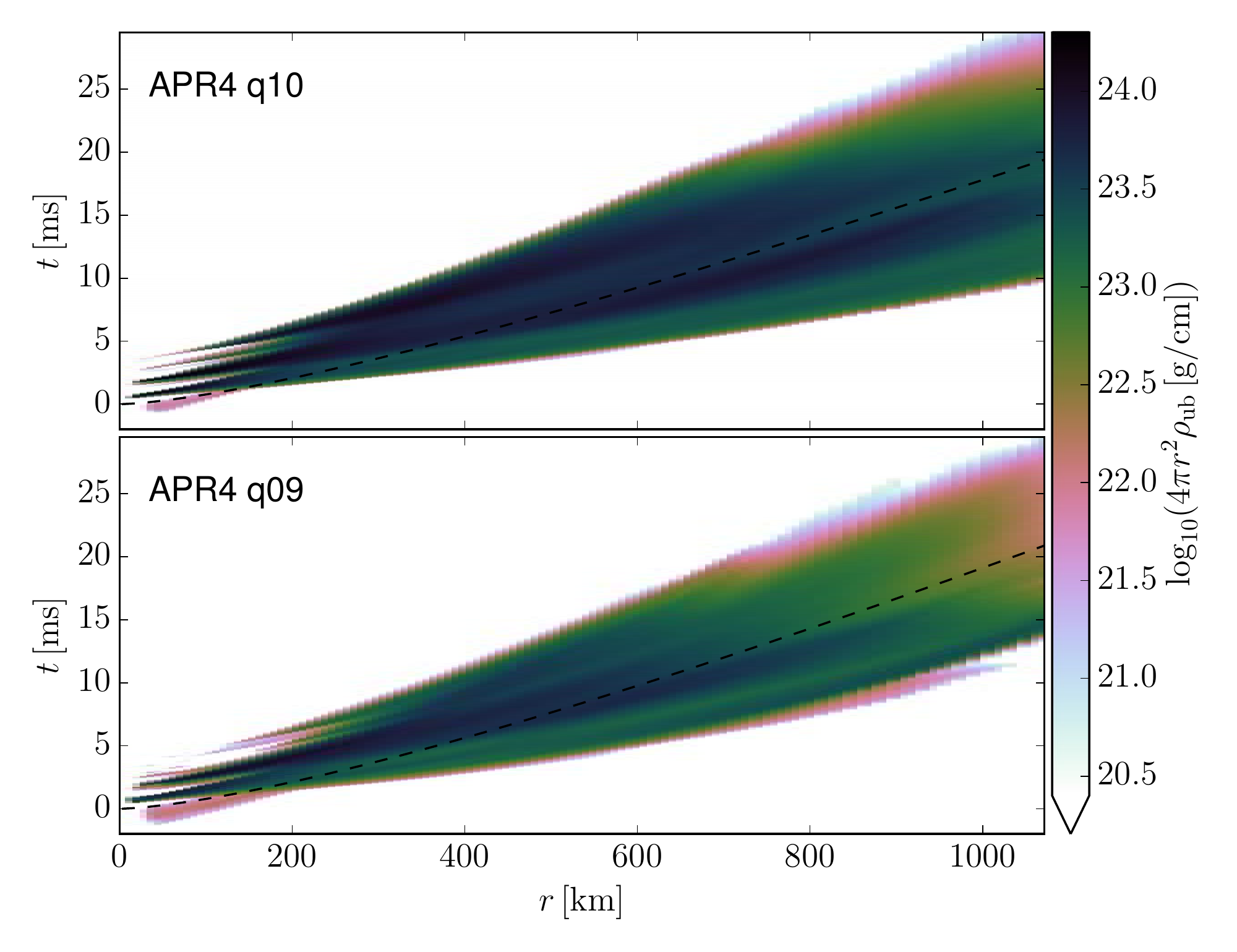}
  \includegraphics[width=0.83\linewidth]{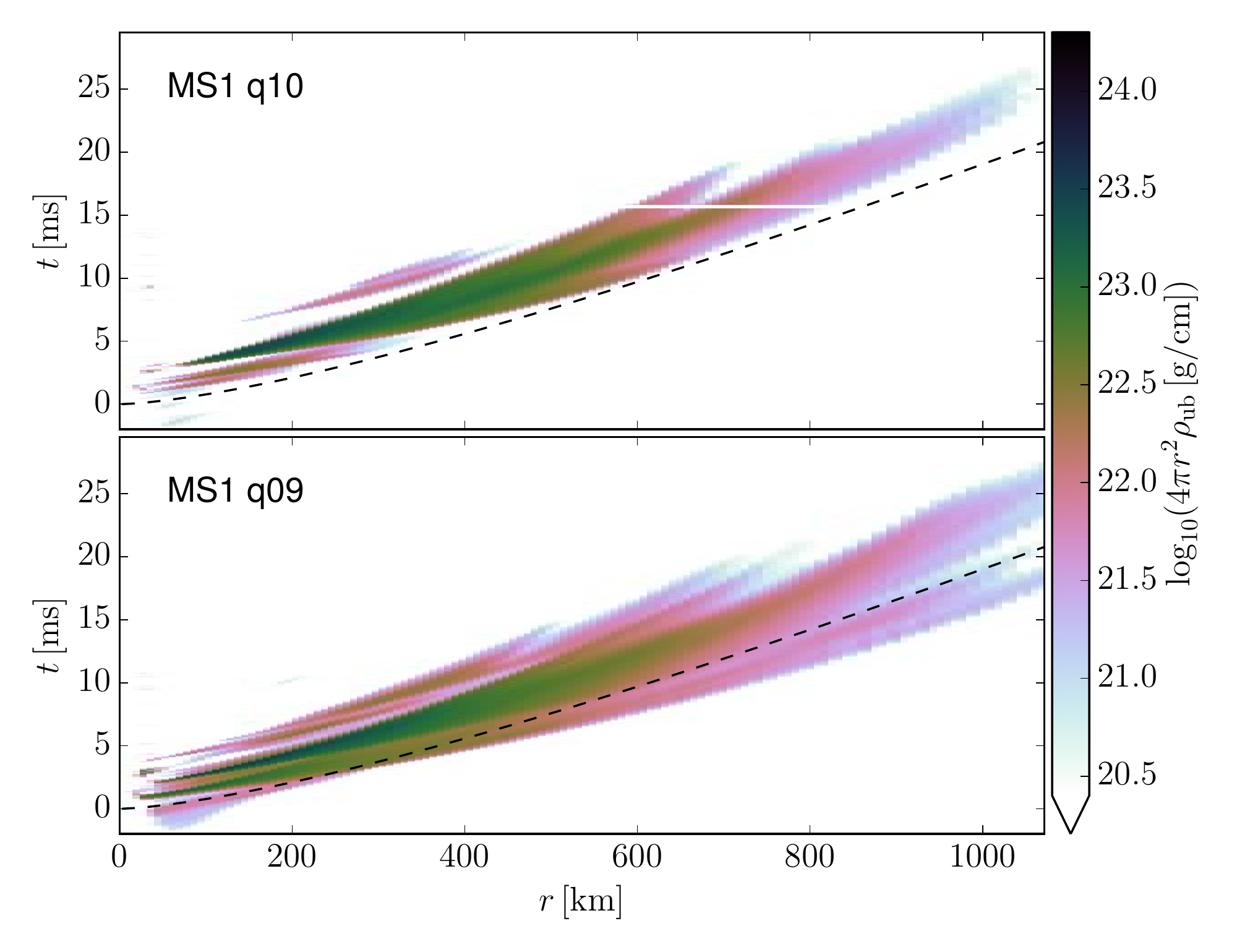}
  \includegraphics[width=0.83\linewidth]{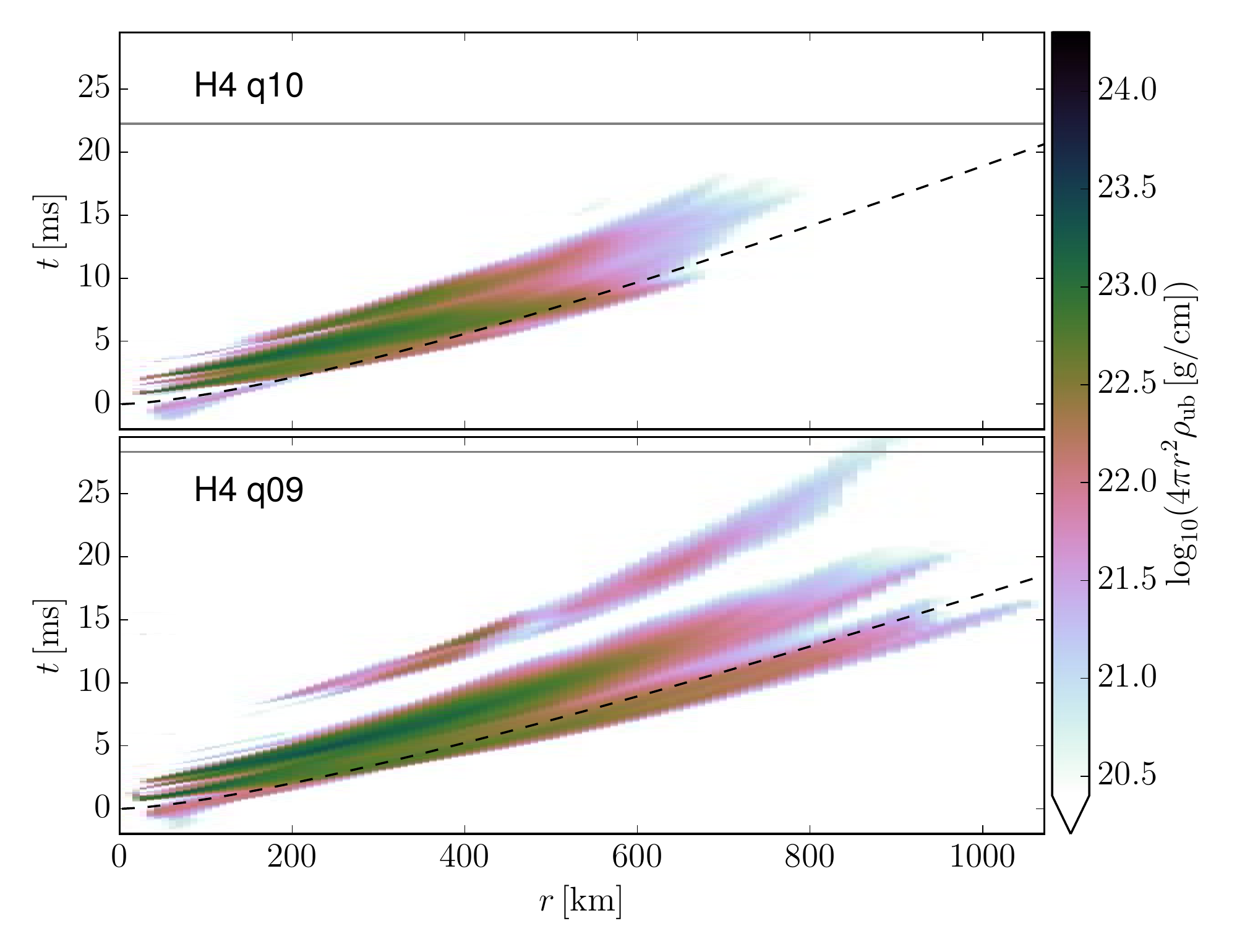}
  \caption{Radial distribution of unbound matter versus time (the white gap visible for the
  equal-mass MS1 model is an artifact caused by a corrupt data file). The color code
  corresponds to the increase of unbound mass inside spherical surfaces per 
  increase in radius. For comparison, we also show the 
  trajectory of a radially outgoing test mass with the escape velocity reported in \Tref{tab:outcome}, 
  estimated using Newtonian potential of a point mass corresponding to the ADM mass at the end of the simulation.
  Horizontal grey lines in the H4 panels mark the collapse to black hole.
    } 
  \label{fig:ejecta_rt}
\end{figure}

In order to judge more directly the impact of the magnetic field on this dynamic ejecta, 
we compare our APR4 equal-mass simulation to the corresponding unmagnetized case.
Those two models were evolved with same the code, grid setup, and artificial atmosphere, 
and the ejected mass was extracted with the same method. The only 
remaining error of the differences between the two cases is the discretization error. 
In this respect, we note that the 
difference between standard and high-resolution runs (see \Tref{tab:outcome}) is around
$3\%$. For the unmagnetized case, we find an ejecta mass of $0.0126 \usk M_\odot$,
i.e.~a difference around $4\%$ to the magnetized case (at the same resolution). 
In conclusion, within the numerical error we observe no impact of the magnetic 
field for this model.

To investigate the ejection mechanisms, we collect the unbound matter at regular time 
intervals during the simulation in 1D histograms binned by radial coordinate. 
From this, we produce spacetime diagrams of the ejection, shown in \Fref{fig:ejecta_rt}. 
For all models,
matter is ejected in several distinct waves and, with the exception of the equal-mass MS1
case, the first wave consists of material tidally ejected during merger. This can also be seen 
in the leftmost panels of Figures~\ref{fig:rho_xy_APR4} to~\ref{fig:rho_xy_H4}, showing 
the regions of unbound matter at merger. Not surprisingly, \Fref{fig:ejecta_rt} shows
that our unequal-mass models tidally eject more mass than the equal-mass ones. 

The second wave is more isotropic, as can be seen in \Fref{fig:rho_xz}, and is likely
the results of shock waves caused by the merger. 
Note that, although the breakout shock contributes significantly to the ejecta, there are further 
waves visible in \Fref{fig:ejecta_rt} (see also the discussion in \cite{Hotokezaka:2013:24001,Bauswein:2013:78}).
This sequence of non-tidal ejections also explains how the equal-mass APR4 model can eject more 
matter than the unequal-mass one. For the APR4 equal-mass case, the quasi-radial remnant 
oscillations are also stronger compared to the unequal-mass case, which provides a 
natural explanation for the higher non-tidal mass ejection.

Interestingly, the unequal-mass H4 model exhibits a wave emitted a few milliseconds after
the previous ones, which is not present for the equal-mass case.
We recall that those two models also showed differences in the evolution of the vortex 
structure (cf.~\Sref{sec:rot}). 
This last wave becomes unbound at a relatively large radius of $200\kmeter$.
Extrapolating back to the remnant, it seems plausible that the  
rearrangement of the remnant fluid flow starting at $t \approx 5\msec$ (see \Sref{sec:rot})  
launches a wave that unbinds material in the disk.

To estimate the escape velocity, we compute the volume integrals
\begin{align}
W_\infty &= \frac{1}{M_u} \int u_t W \rho_u \mathrm{d}V ,&
M_u &=\int W \rho_u \mathrm{d}V
\end{align}
where $\rho_u$ is the density of unbound matter in the fluid rest frame, $W$ the
Lorentz factor, and $\mathrm{d}V$ the proper volume element.
The integral is carried out over the computational domain outside a radius of $150\kmeter$
and $W_\infty$ is evaluated at the time where $M_u$ becomes maximal.
The average velocity of ejected matter at infinity then becomes 
$v_\text{esc} = \sqrt{1-W_\infty^{-2}}$. The results listed in \Tref{tab:outcome} are of
the order of $0.1\usk c$. As a cross check, we also computed for each model the trajectory 
of a radially outgoing test mass with the average escape velocity of the ejected matter.
The results shown in \Fref{fig:ejecta_rt} agree well with the ejecta, although the latter
naturally show a large spread.

\begin{figure}
  \centering
  \includegraphics[width=0.95\linewidth]{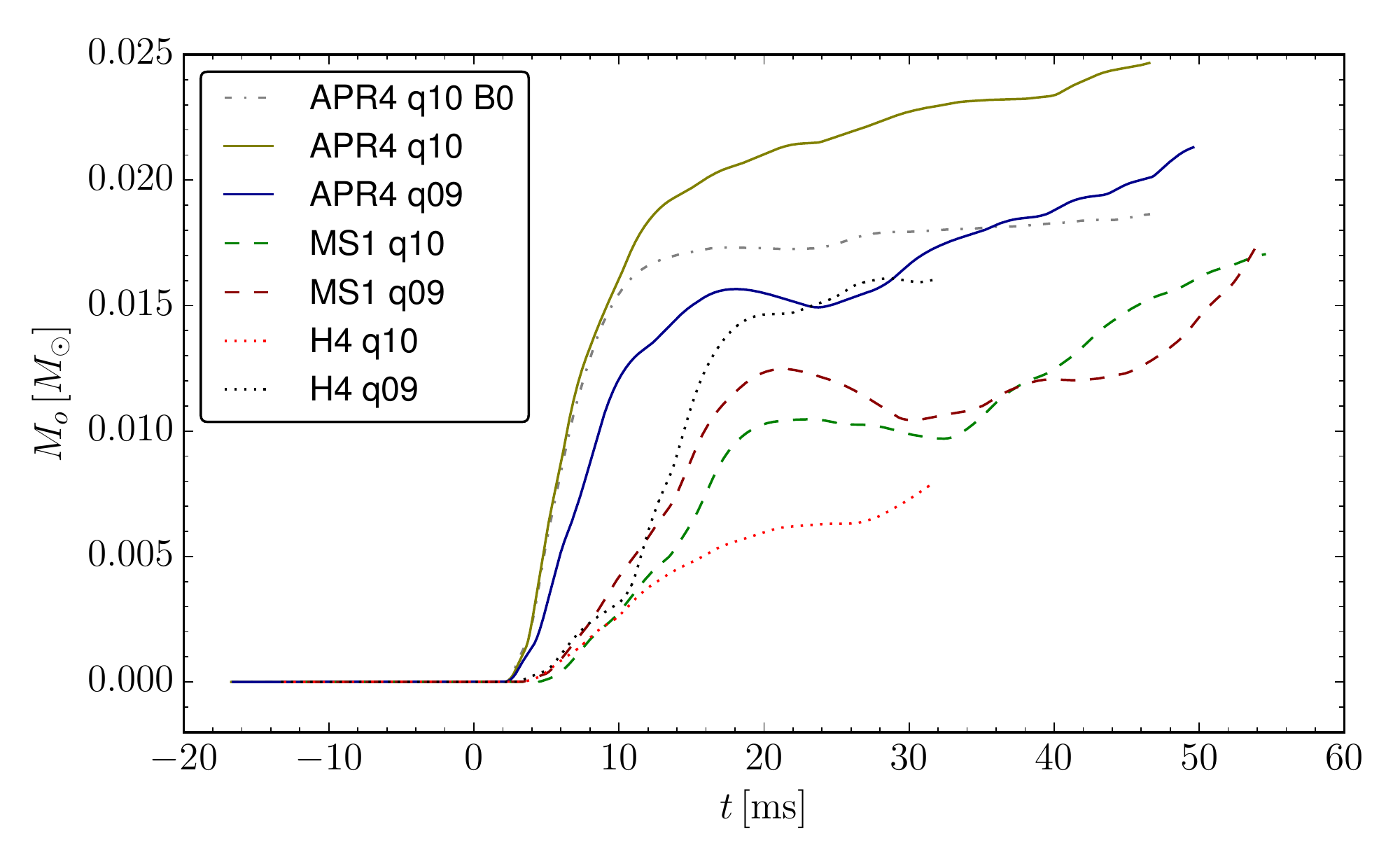}
  \caption{Cumulative outflow of matter through spherical surfaces with radius of $295\kmeter$
  for all models.
    } 
  \label{fig:outflows}
\end{figure}

As noted in Section~\ref{sec:dynamics}, in the post-merger phase there is also an 
outflow of matter that is bound according to the geodesic criterion
(cf.~Figures~\ref{fig:apr4q10_mflux_xz} to~\ref{fig:h4q10_mflux_xz}). 
In order to measure the corresponding mass flux, we compute the cumulative flux of all matter through a 
spherical surface with a radius of $295\kmeter$. \Fref{fig:outflows} shows the result for 
the different models. The flux is largest in the first ${\approx}15\msec$ after merger,
but also at later time we observe a net outflow. We also note that at $295\kmeter$ distance, 
there is no flux of unbound matter after $t\!\approx\!10\msec$ (see \Fref{fig:ejecta_rt}), 
and thus all the subsequent outflow accumulated is bound, at least according to the geodesic criterion.
For the long-lived NS cases, the system tends to approach a continuous outflow towards the end 
of the simulation, with rates around ${\sim}0.2$--$0.3 \usk M_\odot\per\second$.
 
\begin{table*}
\caption{Properties of kilonova/macronova transients associated with the dynamical ejecta of our BNS mergers, estimated from a simple analytical model given in \cite{Grossman2014} (see text).
$t_\mathrm{k-m}$, $L_\mathrm{k-m}$, and $T_\mathrm{k-m}$ are rough estimates for the peak time, bolometric luminosity and effective temperature of the signal.
The values for the APR4 model are taken from the high-resolution run.}
\begin{ruledtabular}
\begin{tabular}{lcccccc}
Model&
APR4 equal&
APR4 unequal&
MS1 equal&
MS1 unequal&
H4 equal&
H4 unequal
\\\hline 
$t_\mathrm{k-m}\,[\mathrm{days}]$&
$16$&
$4.2$&
$1.4$&
$1.5$&
$1.2$&
$1.4$
\\
$L_\mathrm{k-m}\,[10^{40}\,\mathrm{erg\,s}^{-1}]$&
$3.1$&
$2.2$&
$1.0$&
$1.1$&
$1.0$&
$1.3$
\\
$T_\mathrm{k-m}\,[10^{3}\,K]$&
$2.1$&
$2.3$&
$3.3$&
$3.3$&
$3.4$&
$3.1$
\end{tabular}
\end{ruledtabular}
\label{tab:kilonova}
\end{table*}

For magnetized models, it is natural to ask if the observed outflow is magnetically driven.
If this was the case, the geodesic criterion is invalid and also some of the formally bound 
material might escape the system, constituting a baryon-loaded wind. To answer this question, we compare the equal-mass
APR4 model to the corresponding non-magnetized (but otherwise identical) model.
The outflows shown in \Fref{fig:outflows} are very similar until around $10\msec$ after 
merger. After this time, however, the outflow for the magnetized model is significantly larger, with 
a flux about three times larger (${\approx}0.16 \usk M_\odot/\mathrm{s}$ compared to 
${\approx}0.06\usk M_\odot/\mathrm{s}$ of the non-magnetized case).
The cumulative outflow before this time is comparable to the ejected mass $M_\text{ej}$, 
i.e. the outflow is dominated by dynamic ejecta, while the subsequent outflow is formally bound.
We also recall that although the outflowing matter is not magnetically dominated, the magnetic
pressure at a radius $r=100\kmeter$ reaches around $0.1$ of the gas pressure (cf.~\Fref{fig:vz_MHDratio}), 
and therefore some influence on the dynamics of the outflows should be expected.
We conclude that, in the long-lived NS cases (APR4 and MS1),
{\it the main contribution 
to the matter outflows observed towards the end of our simulations ($t\!>\!20\msec$)
is magnetically driven}. 
We stress that we have no indication on whether these outflows 
correspond to matter that will remain bound and eventually fall back onto the central NS,
or escape to infinity as a baryon-loaded wind.

Neutrino emission and reabsorption, not considered in our present 
simulations, represent an additional mechanism to produce nearly isotropic baryon-loaded 
outflows \cite{Dessart:2009:1681}.
Therefore, properly accounting for neutrino emission would likely enhance the post-merger 
mass ejection reported here.

\bigskip

{\it Electromagnetic counterparts from dynamical ejecta}.
As pointed out in the Introduction, the ejecta of BNS mergers represent very promising sites 
for r-process nucleosynthesis and might provide an important contribution to the 
heavy element abundances observed in the local universe (e.g., \cite{Just2015a,Wu2016,Roberts2017a}). 
Moreover, the radioactive decay of these elements is expected to power a late-time 
EM transient, a so-called kilonova or macronova, which is among the most promising EM
counterparts to the GW signal from BNS mergers
\cite{Li:1998:L59,Kulkarni:2005:macronova-term,Rosswog2005,Metzger2010,Roberts2011,Barnes2013,Piran2013,Tanvir2013,Berger2013,Yang2015}.

Although a proper analysis is beyond the scope of this work, we can use 
a simple analytical model by Grossman et al. \cite{Grossman2014} to provide a rough, order-of-magnitude 
estimate of the peak time, peak bolometric luminosity and effective temperature of 
kilonova/macronova transients corresponding to the BNS mergers under investigation
(we refer to \cite{Grossman2014} for a discussion on the limitations of the model):
\begin{eqnarray}
t_\mathrm{k-m}&=&4.9\,\bigg{(}\frac{M_\mathrm{ej}}{10^{-2} M_\odot}\bigg{)}^{\!1/2}\bigg{(}\frac{v_\mathrm{esc}}{0.1\,c}\bigg{)}^{\!-1/2}\,\mathrm{days}\, ,
\nonumber \\
L_\mathrm{k-m}&=&2.5\,\times 10^{40}\bigg{(}\frac{M_\mathrm{ej}}{10^{-2} M_\odot}\bigg{)}^{\!1-\alpha/2}\bigg{(}\frac{v_\mathrm{esc}}{0.1\,c}\bigg{)}^{\!\alpha/2}\,\mathrm{erg\,s}^{-1}\, ,
\nonumber \\
T_\mathrm{k-m}&=&2200\,\bigg{(}\frac{M_\mathrm{ej}}{10^{-2} M_\odot}\bigg{)}^{\!-\alpha/8}\bigg{(}\frac{v_\mathrm{esc}}{0.1\,c}\bigg{)}^{\!(\alpha-2)/8}\,K\, .
\nonumber 
\end{eqnarray}
The above formulas are obtained from \cite{Grossman2014} by fixing the ejecta opacity 
to the fiducial value $\kappa=10\,\mathrm{cm}^2\,g^{-1}$ \cite{Barnes2013}.
Moreover, we set $\alpha=1.3$ as in \cite{Grossman2014}.
Note that here we are only considering the contribution from the dynamical ejecta that 
are formally unbound in our simulations. 
Further mass outflows (including magnetically/neutrino driven winds) can also 
contribute to the kilonova/macronova emission, 
although with a higher effective temperature and shorter 
timescale due to the lower opacity \cite{Metzger2014c,Grossman2014}.

Results are given in Table~\ref{tab:kilonova}. We find the MS1 and H4 models, 
both with equal and unequal mass, to have similar estimates for the kilonova/macronova 
parameters: peak time of $\sim\!1$~day, peak luminosity of $\sim\!10^{40}$~$\mathrm{erg\,s}^{-1}$, 
and effective temperature around 3000~$K$. The APR4 models eject instead around 
one order of magnitude more mass (see \Tref{tab:outcome} and \Fref{fig:ejecta_rt}),
which results in much longer timescales, higher luminosity, and slightly lower effective temperature.
In this case, equal- and unequal-mass models also show appreciable differences, in particular in the 
peak time (16 days and 4 days for the equal- and unequal-mass models, respectively). 

As a final note, we recall that the interaction of the ejecta with the interstellar medium can also produce an 
EM transient via non-thermal synchrotron emission, which 
typically falls in the radio band and emerges on much longer timescales, 
up to $\sim$\,years \cite{Nakar2011}.


\section{GW emission}
\label{sec:gw}

\begin{figure*}[!ht]
  \centering
  \includegraphics[width=0.99\linewidth]{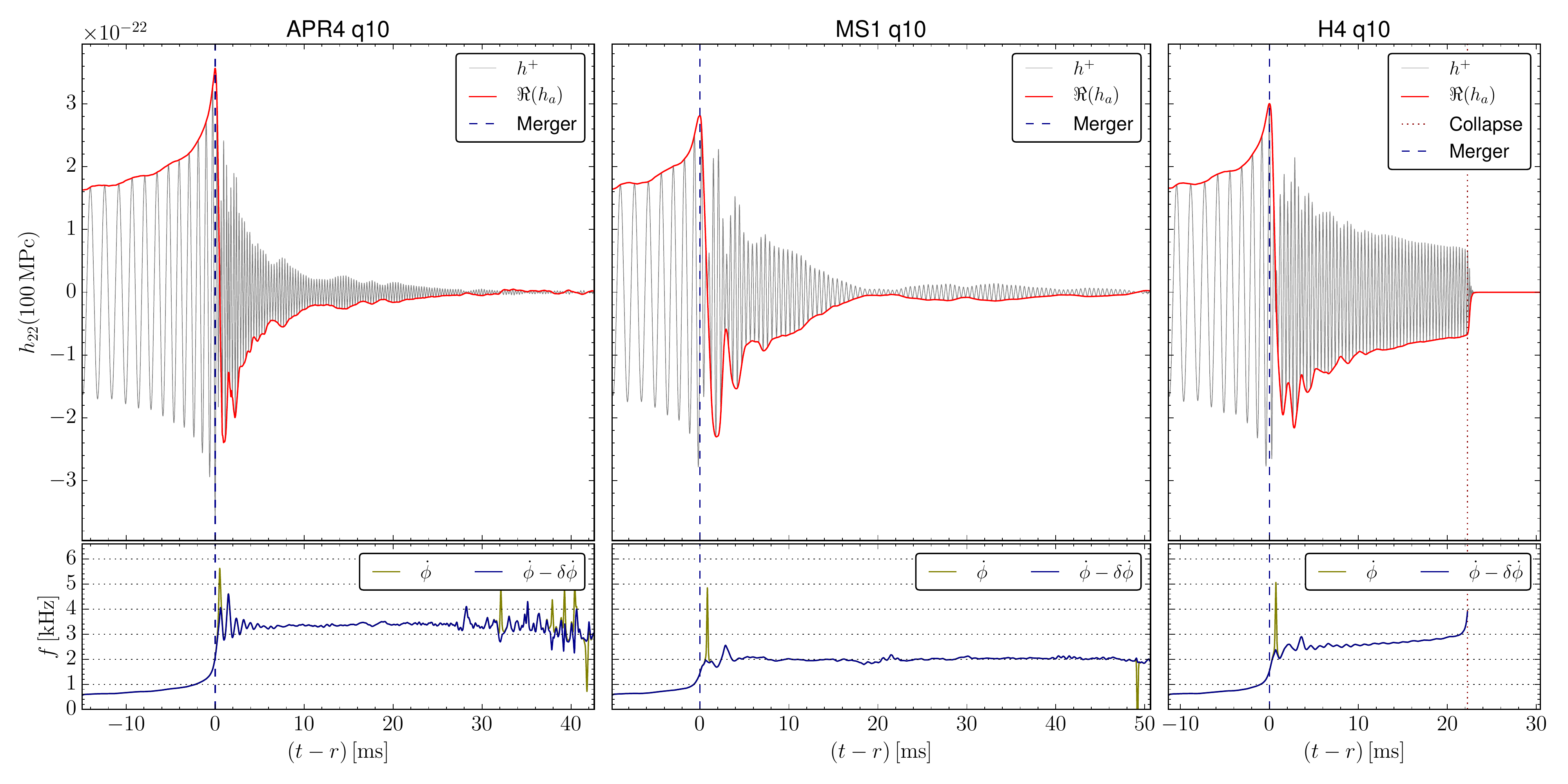}
  \caption{Gravitational wave signal for the equal-mass APR4, MS1, and H4 models 
  (from left to right). Top panels: the thin gray 
  lines show the strain amplitude $h^+_{22}$ at $100\usk\mega\parsec$, the red line
  the real part of the complex amplitude $h_a$ (see text). 
  Bottom panels: the phase velocity of the complex strain 
  $h_{22}$ before and after correcting for phase jump contributions (in yellow
  and blue, respectively). Note the phase velocity after $30\msec$ is
  not meaningful due to the low amplitude and lack of a clearly dominant mode.
  The vertical lines mark the times of merger (dashed)
  and of black hole formation (dotted).} 
  \label{fig:gw_strain_q10}
\end{figure*}

\begin{figure*}[!ht]
  \centering
   \includegraphics[width=0.99\linewidth]{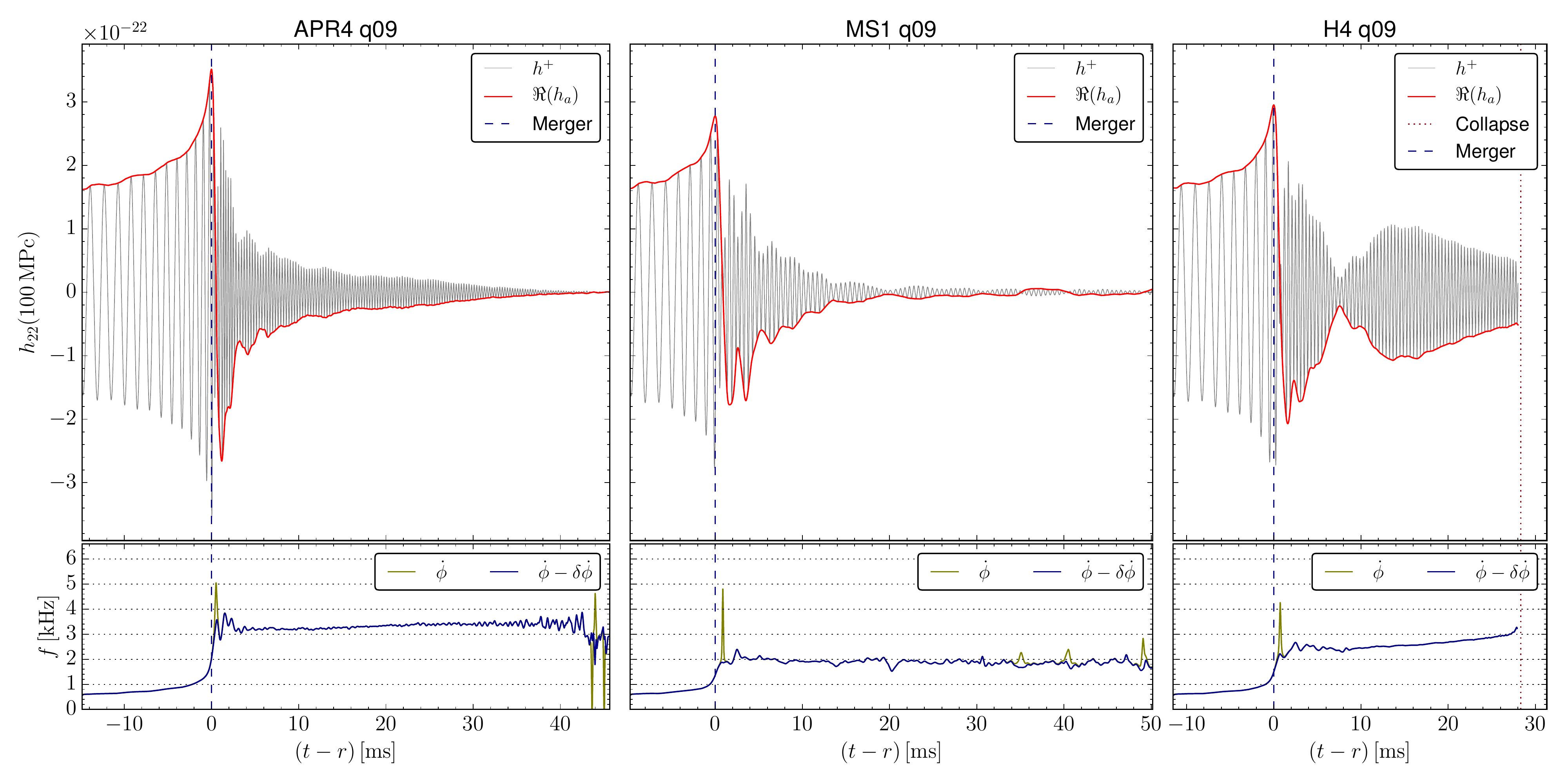}
  \caption{Like \Fref{fig:gw_strain_q10}, but showing the unequal-mass models.
  Note that the H4 unequal-mass model formed a black hole (vertical dotted line), but the 
  simulation was not carried on for long enough to extract the gravitational 
  wave signal of the collapse.} 
  \label{fig:gw_strain_q09}
\end{figure*}

In this Section, we conclude our analysis by discussing the GW emission of our BNS mergers.
For all our simulations, we extract the GW strain from the Weyl scalar $\Psi_4$
at a fixed radius of $1181 \kmeter$, without extrapolating to infinity.
The numerical accuracy is discussed in the Appendix.
The GW strains given in the following are the coefficients of the decomposition 
into spin-weighted spherical harmonics $_{-2}Y_{lm}$, and the strain at a particular 
viewing angle can be obtained by multiplication with $\left|_{-2}Y_{lm}(\theta, \phi)\right|$. 
For the time integration, we developed a new method which is described in 
\cite{Kastaun:2016b:arxiv}; the advantage
is that the improved removal of offsets results in centered waveforms also for 
low-amplitude parts, i.e. minima and tails.

\begin{figure}[!ht]
  \centering
   \includegraphics[width=0.99\linewidth]{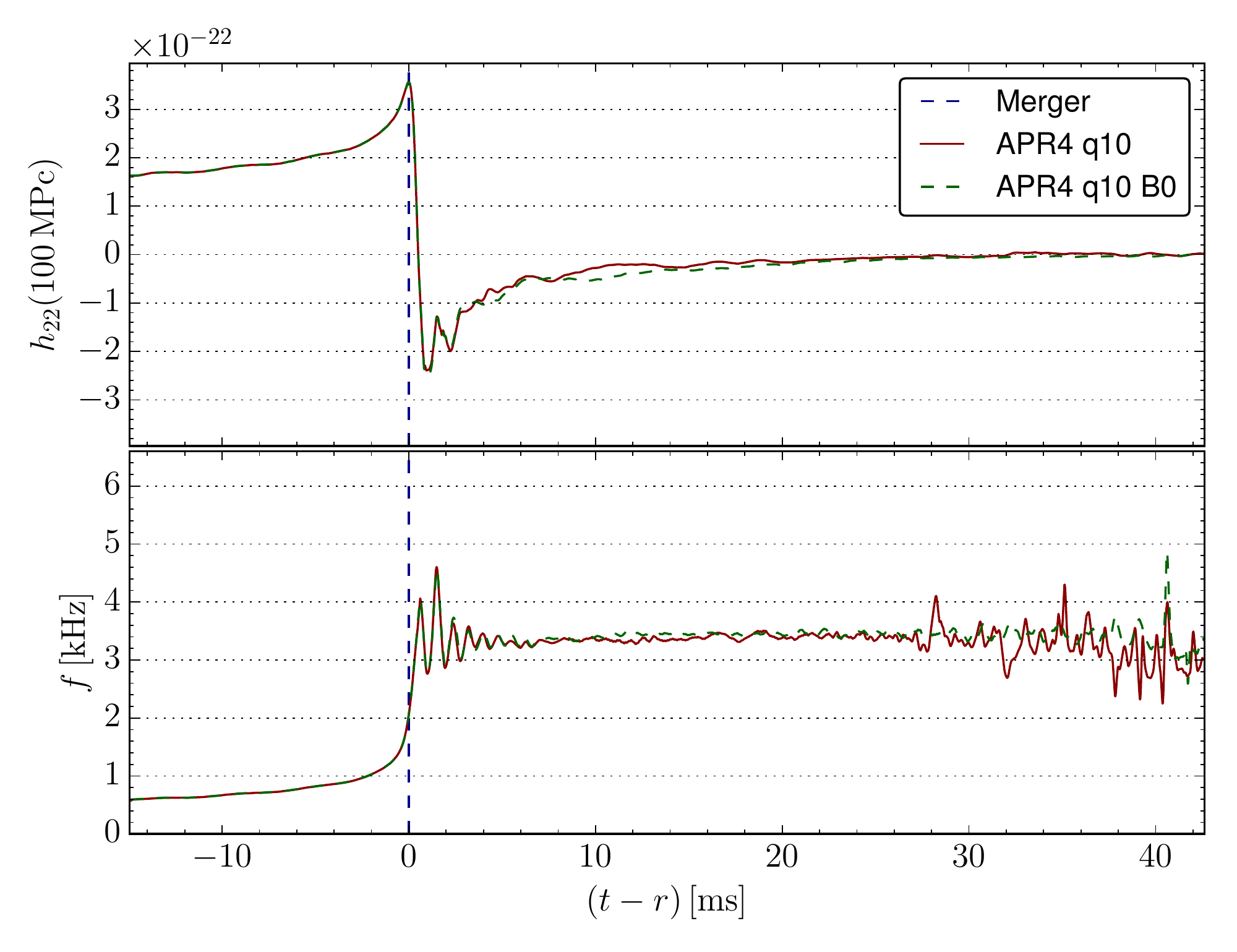}
  \caption{Comparison of gravitational wave strain $\Re(h_a)$ (top panel) 
  and jump-corrected phase velocity (bottom panel), between the magnetized 
  APR4 equal-mass model and the corresponding  
  non-magnetized model.} 
  \label{fig:gw_strain_bcomp}
\end{figure}

More importantly, we also employ a scheme to detect phase jumps caused by 
over-modulation. This term denotes signals in 
the form $A(t) e^{i \phi(t)}$, where $A \in \mathbb{R}$ is slowly changing 
compared to $\phi$, but can have zero crossings. A sign change of $A$ then corresponds 
to a phase jump by $\pi$ of the signal.
More generally, if $A$ is complex valued a rapid phase change 
occurs when $A$ passes close to the origin in the complex plane.
Our scheme decomposes the complex-valued strain amplitude $h = h^+ - i h^\times$
as $h(t) = h_a(t) e^{i \phi(t)}$, such that $h_a$ has a significant imaginary part
only near phase jumps, and a real part that can cross zero.
This is expressed by a phase correction $\delta\phi$, with 
$h_a(t) = |h(t)| e^{-i\delta\phi(t)}$. 
For further details of the method, we refer to \cite{Kastaun:2016b:arxiv}.

The GW strain and the phase velocity for all magnetized models are shown in 
Figures~\ref{fig:gw_strain_q10} and~\ref{fig:gw_strain_q09}. 
In addition, we visualize the phase jumps using the real part of $h_a$ and the 
jump-corrected phase velocity. All models show the characteristic 
amplitude minimum seen at merger in many BNS simulations. 
Using our heuristic phase jump detection,
we find that those minima are caused by over-modulation. This is clearly visible
in the phase velocities, which exhibit a sharp peak (coincident with the time 
of the amplitude minimum) before subtracting the correction $\delta\phi$.
This observation is relevant for GW astronomy, where the data analysis 
is very sensitive to the phasing. 
For all the cases at hand, the phase around merger 
can be well described by two relatively smooth parts separated by a rapid jump
by $\pi$. The phase velocity at merger, i.e.~at the time of maximum strain amplitude,
is given in \Tref{tab:outcome}.
The jump-corrected phase velocity can still show a modulation lasting a few ms 
after merger (most evident in the APR4 equal-mass case). This is most likely 
caused by quasi-radial oscillations, which we also observe in the maximum density.

After merger, the instantaneous GW frequency increases slightly and slowly for the 
APR4 models, and remains almost constant for the MS1 models. For the H4 models, 
it increases significantly until the system starts collapsing into a BH. 
For the equal-mass model, the frequency quickly increases to $4\usk\kilo\hertz$
at the time of the collapse. For the unequal-mass H4 model, 
the simulation was ended before the signal of the collapse reached the extraction 
radius. 
For the cases at hand, we find that the frequency drift becomes larger the closer
the remnant is to the collapse threshold.

\begin{figure}[!ht]
  \centering
  \includegraphics[width=0.99\linewidth]{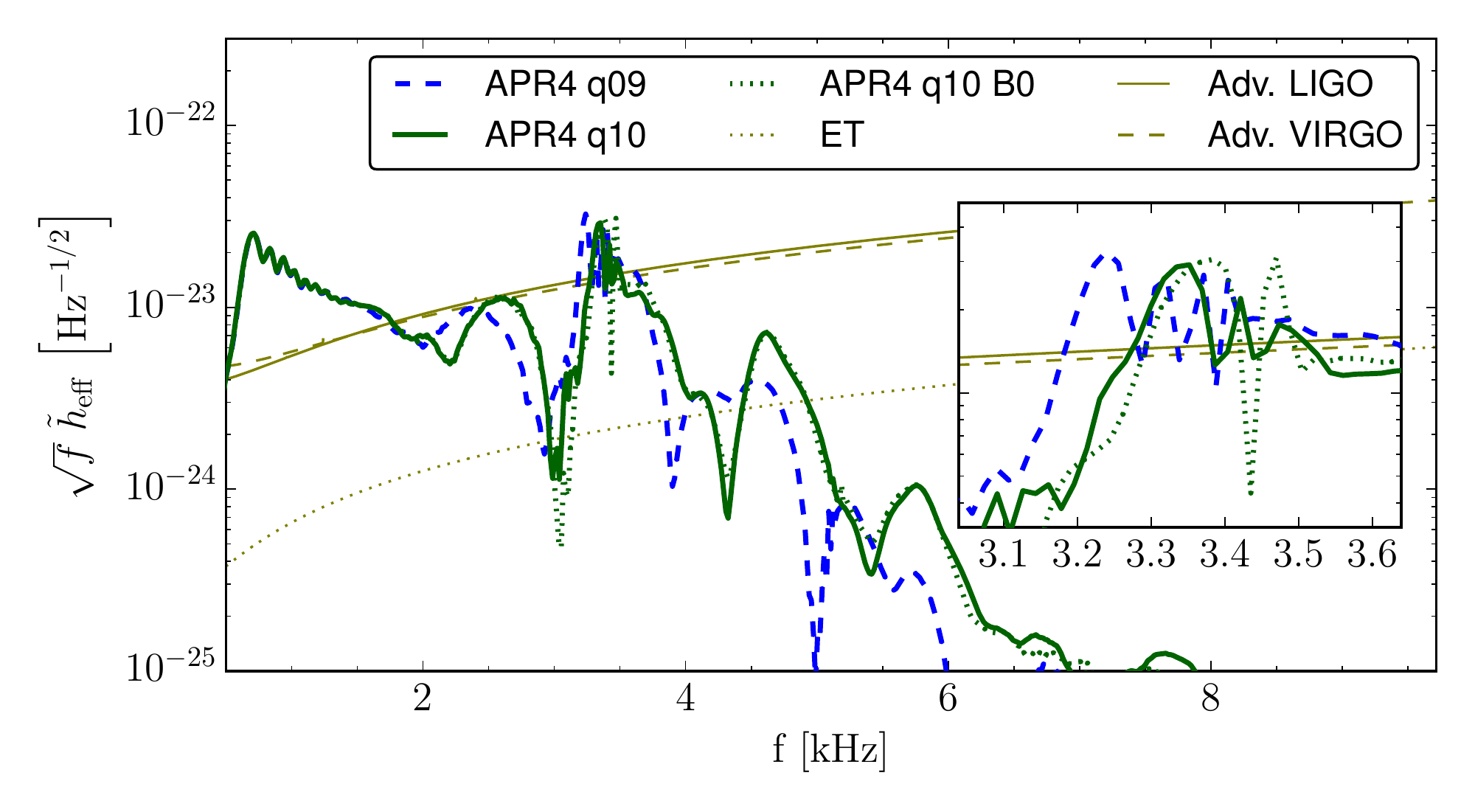}
  \caption{Power spectrum of the gravitational wave strain at $100\usk\mega\parsec$ for 
   the APR4 models, compared to the sensitivity curves of current and planned
   gravitational wave detectors.} 
  \label{fig:gw_spec_apr4}
\end{figure}

\begin{figure}[!ht]
  \centering
  \includegraphics[width=0.99\linewidth]{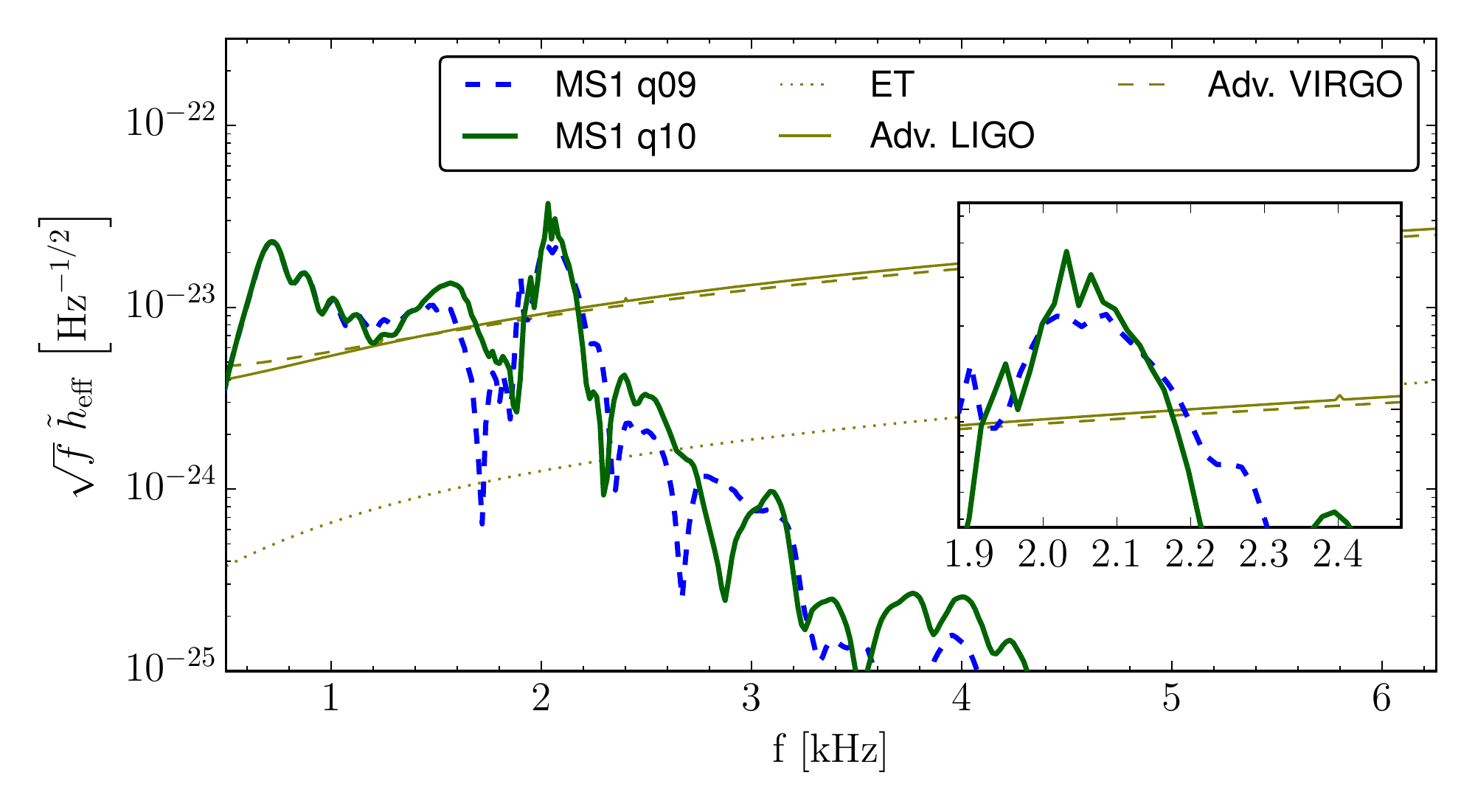}
  \caption{Like \Fref{fig:gw_spec_apr4}, but for the MS1 models.} 
  \label{fig:gw_spec_ms1}
\end{figure}

\begin{figure}[!ht]
  \centering
  \includegraphics[width=0.99\linewidth]{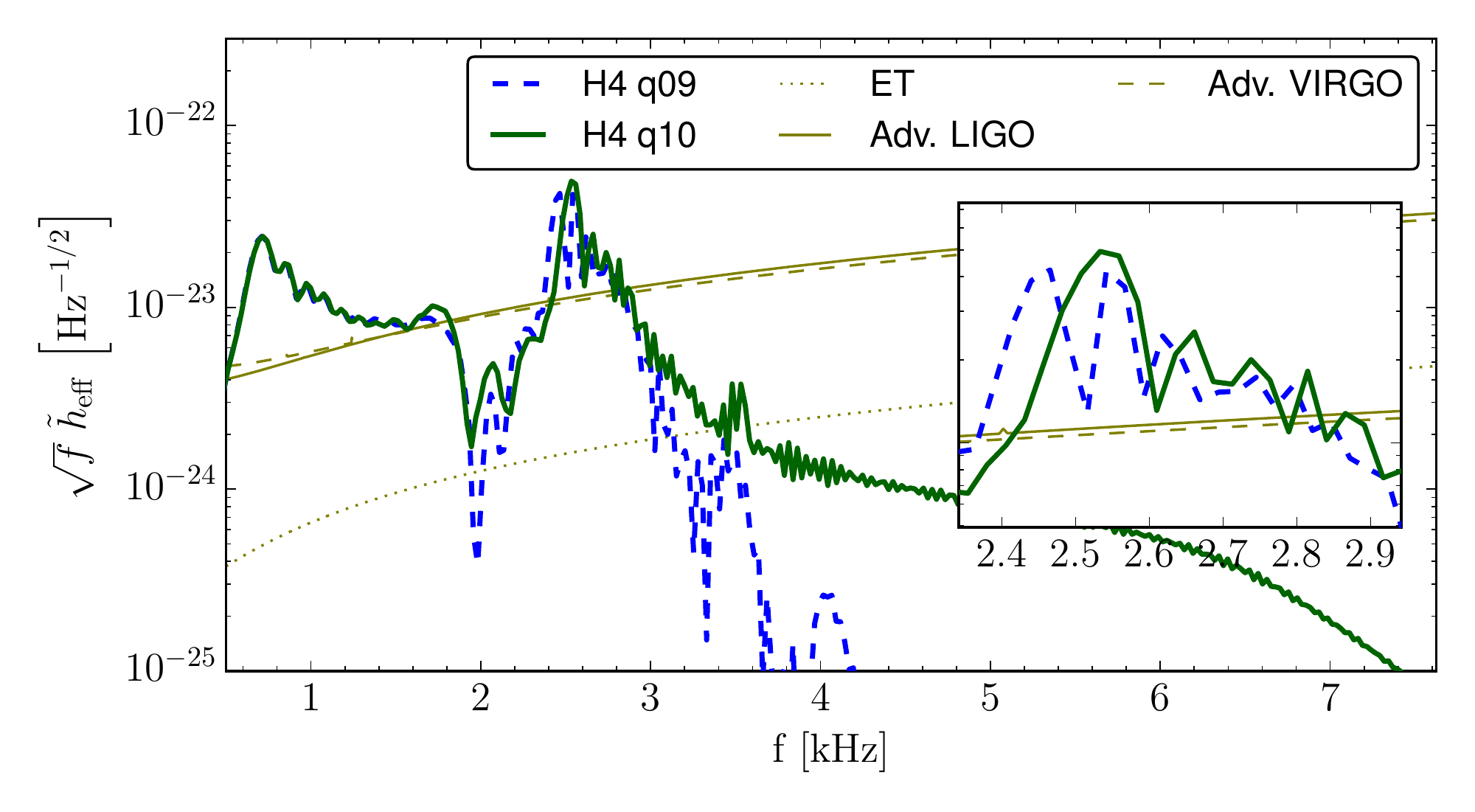}
  \caption{Like \Fref{fig:gw_spec_apr4}, but for the H4 models. Note 
  the differences in the high-frequency part are simply due to the fact that  
  the unequal-mass case was not evolved long enough to obtain the 
  part of the signal corresponding to the collapse to a black hole.} 
  \label{fig:gw_spec_h4}
\end{figure}

In terms of post-merger waveforms, long-lived remnants (APR4, MS1 EOS) are 
characterized by comparable amplitudes that decay significantly within $10$--$20\msec$.
The HMNS cases (H4 EOS) have instead a stronger and more persistent emission 
until the sudden drop of amplitude associated with the collapse. 
The largest difference between equal and unequal-mass cases is also found for the 
H4 EOS. The amplitude for mass ratio $q=0.9$ shows a pronounced second 
minimum, while for the equal-mass model it decreases monotonically.
One possible explanation for this general type of behavior would be the excitation of
an unstable oscillation mode while the original mode excited during merger is damped.
This seems unlikely since the phase velocity remains smooth, which would not
be the case when two different modes with comparable amplitude are present at 
the same time. Another possibility is that the original mode becomes unstable 
due to an increase of compactness and frequency. This is also unconvincing
since the mode frequencies span the same range for both mass ratios. In case 
of a CFS unstable mode, the inertial-frame frequency should also be small
near the critical rotation rate. 

We favor an explanation recently proposed
in \cite{Kastaun:2016b:arxiv}, namely that the density deformation is partly 
due to vortices in the fluid flow, and that these can undergo both smooth and
sudden rearrangements. This could also explain smaller irregularities 
of the strain amplitude. The hypothesis is not proven, in particular it is 
possible that vortex rearrangements and frequency changes have a common 
cause instead. However, as discussed in \Sref{sec:rot}, we do see for the H4 
unequal-mass case a clear rearrangement of the remnant structure in the 
frame corotating with the $m=2$ density deformation (cf.~\Fref{fig:h4q09_traj_xy}). 
Also, we found that the contributions of the outer layers and 
the core to the quadrupole moment have opposite sign. This might lead to 
cancellation effects amplifying the impact of rearrangements on the GW 
amplitude.

The effect of the magnetic field on GW strain and phase velocity is shown
in \Fref{fig:gw_strain_bcomp} for the equal-mass APR4 model. We find very 
little difference in this case. Note that the impact for a remnant 
closer to collapse could be larger since near the threshold for BH formation the 
system tends to be very sensitive to small changes. In particular, 
the lifetime of the remnant could be altered significantly. 

The Fourier spectra of the GW signals are shown in Figures~\ref{fig:gw_spec_apr4} 
to~\ref{fig:gw_spec_h4}, each comparing the equal- and unequal-mass models for 
one EOS. The main peak caused by the post-merger phase shows only minor changes 
for different mass ratios, compared to the width of the peak. The impact of
the EOS exceeds by far that of the mass ratio, at least in the range $q\!=\!0.9$ to $1$. 
We note that a small influence of the mass ratio makes it easier to constrain
the EOS from the post-merger frequency. 
Correlations between EOS, initial NS properties,
and post-merger frequencies have been studied by different groups, e.g. 
\cite{Bauswein:2012:11101,Bernuzzi:2015:1101B,Rezzolla:2016:93l4051R}, for a 
large number of models.

In all cases, the post-merger peak as well as the inspiral contribution are above the 
(design) sensitivity curves of the advanced LIGO and Virgo detectors. 
Nevertheless, the corresponding 
signal-to-noise ratio (SNR) is likely insufficient for a confident detection of the 
post-merger signal at $100\usk\mega\parsec$ distance.
Of the three EOS, the APR4 EOS leads to the post-merger signal with the smallest SNR. 
Although the H4 models emit the strongest post-merger signals (see discussion above), 
their frequency is also higher, such that the MS1 and H4 cases result in comparable SNRs.

The dominant frequency of the post-merger phase for each model is given in \Tref{tab:outcome}.
We report both the location $f_\mathrm{pm}$ of the maximum in the Fourier spectrum as well as
a measure defined in \cite{Hotokezaka:2013:44026} using the instantaneous frequency
$f$ to compute
\begin{align}
f_{10} &= \left( \int|h(t)| \,\mathrm{dt} \right)^{-1} \int f(t)|h(t)| \,\mathrm{dt},
\end{align}
where the time integrals are carried out over the first $10\msec$ after merger.
Interestingly, the GW frequency in the post-merger phase is approximately
twice the maximum rotation rate inside the remnant (compare $2\nu_\mathrm{max}$ and 
$f_\mathrm{pm}$ in \Tref{tab:outcome}, as well as \Fref{fig:rot_max}). 
As was already observed in 
\cite{Kastaun:2015:064027, Endrizzi:2016:164001, Kastaun:2016, Kastaun:2016b:arxiv, 
Hanauske2016arXiv},
the maximum rotation rate is apparently limited by the angular velocity of the 
$m=2$ density deformation, which is in turn half of the GW frequency.
The frequency of the main post-merger peak increases with the bulk compactness
of the remnant (as does the rotation rate, see \Sref{sec:rot}), which depends 
on the EOS. 

When considering the characteristic low- and high-frequency side peaks appearing around the 
main post-merger peak, we find more significant differences between the equal and unequal-mass 
cases. We caution however that those peaks are not necessarily related directly to physical oscillations. 
As was already shown in \cite{Kastaun:2016b:arxiv}, their location can change drastically when 
removing the aforementioned phase jumps. This can be explained in terms of cancellations 
between the contributions of different parts of the signal to the Fourier spectrum. 

The impact of the magnetic field on the spectrum is rather small, as shown in
Fig.~\ref{fig:gw_spec_apr4}. 
We observe a slight shift of the main peak, which is however less than the peak width. 
The sub-structure of the peak also changes slightly, such that
a sub-peak at $3.47\usk\kilo\hertz$ becomes the new global maximum for the non-magnetic case.
The average frequency $f_{10}$ changes less than $0.5\%$.
Also the amplitude of the peak and the corresponding SNR is essentially unaffected by the magnetic 
field. Overall, we conclude that {\it magnetic fields up to the strength considered here are 
unlikely to cause any detectable changes in the GW signal for BNS mergers forming a long-lived NS}.


\section{Summary and Conclusions}
\label{sec:conclusion}

In this paper, we investigated the merger of BNS systems by means of GRMHD 
simulations, with special attention devoted to mergers producing a long-lived NS 
remnant (i.e.~a supramassive or stable NS). We considered equal and unequal 
mass binaries with mass ratios $q=1$ and $0.9$, keeping a fixed total 
gravitational mass at infinity of $2.7\usk M_\odot$. We considered three 
different EOS known in the literature: APR4, MS1, and H4. For the given 
total mass, these EOS lead to the formation of supramassive, stable, and 
hypermassive NS remnants, respectively. Only the latter models (H4 EOS) 
collapse to a BH by the end of our simulations, which cover the evolution 
up to ${\sim}30$--$50$~ms after merger. 

\bigskip
{\it Remnant structure, rotation profile, accretion disk}. We studied in detail 
the structure and the fluid flow of the merger remnants. In a frame corotating 
with the dominant $m=2$ density deformation, the remnant structure 
appears much more complex than simple differential rotation. In particular, 
we found long-standing vortices correlated with density perturbations, 
which slowly evolve towards axisymmetry. In the H4 unequal-mass case, 
we also found a sudden rearrangement of the internal flow starting $\sim5$~ms 
after merger, which seems to have an impact on the HMNS lifetime and to 
leave a distinctive signature on GW signal and mass ejection. 

For the long-lived models, a quasi-stationary state is reached around 
$20\msec$. For all models, the rotation profiles on the equatorial plane 
around this time shows a generic structure with a slowly rotating core, a 
maximum rotation rate at a radius of ${\sim}15$--$20\kmeter$, and an 
approximately Keplerian rotation profile in the outer layers. This confirms 
previous indications suggesting that the collapse is not prevented by a 
rapidly rotating core, but rather by the centrifugal support of the outer 
layers of the remnant. The EOS is found to have a much stronger impact 
than the mass ratio on the maximum angular velocity, which is 
approximately given by the angular velocity of the $m=2$ density 
perturbation. Moreover, we found that the slowly rotating core is well 
approximated by the core of a TOV (i.e.~non-rotating NS) solution and 
that BNS merger remnants seem to resist the collapse as long as a 
TOV core equivalent is admitted. Our H4 models indeed collapse to 
BH as soon as this condition is no longer satisfied.

For our collapsing (H4) models, we found BHs with spin parameter of 
${\sim}0.6$ surrounded by accretion disks of $0.1-0.2\usk M_\odot$. 
In the long-lived NS cases (APR4 and MS1), we found a significant 
amount of mass outside the remnant at radii $r>20\kmeter$: $\sim0.2$ 
and $\sim0.4\usk M_\odot$ for the APR4 and MS1 models, respectively 
(roughly half of which outside a radius of $60$~km). We note that further 
away from the remnant the matter is distributed more isotropically 
(i.e.~also along the orbital axis) and its internal flow is rather unordered 
and does not correspond to simple accretion, at least on the timescales 
covered by our simulations (${\sim}50$~ms after merger).
We also note that a small fraction of this mass will be ejected from the 
system. As a general trend, unequal-mass systems are found to produce 
more massive disks (by ${\sim}25$\% in the long-lived NS cases, and 
${\sim}60$\% in the collapsing cases).

\bigskip
{\it Magnetic fields}. The evolution of magnetic fields is characterized by 
different stages of amplification. We started from initial poloidal fields of 
${\sim}10^{15}$~G confined inside the two NSs and we observed a first 
stage of amplification taking place already before merger. Interestingly, 
all models started with the same total magnetic energy and gravitational 
mass at infinity, and all ended up with roughly the same magnetic energy 
at the time of merger, which is about one order of magnitude higher than 
the initial one. Nevertheless, it is still unclear whether this amplification 
corresponds to a well-resolved physical mechanism, although our 
analysis ruled out a number of physical and numerical causes. A possible 
explanation might be that the time-changing tidal deformations during 
inspiral induce fluid flows inside the two NSs that might amplify the magnetic 
field. This effect will be further investigated in future studies. 

After merger, magnetic fields are strongly amplified for $5$--$10\msec$, 
most likely by the KH instability. Further amplification continues at later 
times, although at a lower rates. In this last phase, the MRI outside the 
NS remnant is likely playing a major role in the amplification. 
From our resolution study, it is clear that the KH phase is not well resolved. 
However, the magnetic energy achieved in last stage (up to ${\sim}50\msec$ 
after merger for the long-lived NS models) shows a much better convergence. 

The overall dependence on EOS and mass ratio is non-trivial and no 
general trend is observed. At all stages, the magnetic field amplification 
is mostly in the toroidal component and takes place mostly on the 
equatorial plane. We studied the geometrical distribution of magnetic 
fields in 3D for the long-lived NS cases, and found that no ordered 
configuration has emerged around the orbital axis by the end of our 
simulations, although we note a slight tendency to helical structures.

\bigskip
{\it Short gamma-ray bursts}. We discussed how our results compare 
with different scenarios linking BNS mergers to the central engine of 
SGRBs. In particular, we considered the leading BH-disk scenario and 
the alternative magnetar scenario. Both models envisage the formation 
of an accretion-powered jet launched by the post-merger system, i.e.~a 
BH surrounded by a massive accretion disk in the former case and a 
strongly magnetized long-lived NS also surrounded by an accretion disk 
in the latter case.
While not much can be added on the standard BH-disk scenario from 
our collapsing (H4) models, since the simulations were interrupted only 
a few ms after BH formation, our long-lived NS (APR4 and MS1) models 
provided useful indications on the viability of the magnetar scenario. We 
note that so far this case has been poorly investigated in numerical 
relativity, with only very few studies reporting on GRMHD simulations 
of BNS mergers with long-lived NS remnants. We found that $\sim50$~ms 
after merger, the long-lived NS is still surrounded by a dense and nearly 
isotropic environment. In particular, baryon pollution along the orbital axis 
is substantial (densities of $\sim10^{10}$~g/cm$^3$) and could easily 
prevent the formation of an incipient jet. In addition, there is no well 
defined accretion flow nor an ordered magnetic field structure that could 
favor the launch of a collimated outflow. We thus concluded that the 
long-lived NS systems considered are not able to produce a jet, at least 
on timescales of $\sim0.1$~s. As we discussed, however, such a 
conclusion could be affected by our present limitations. 

\bigskip
{\it Matter ejection}. We carried out a detailed analysis of the matter 
ejected during and after merger. We estimated the outflow of matter 
that is unbound according to the geodesic criterion and we found 
dynamical ejecta composed by (i) initial tidal tails launched right before 
merger that are more massive for the unequal-mass models, (ii) a strong 
ejecta wave, most likely due to the breakout shock generated when the 
two NS cores collide, and (iii) additional ejecta waves launched by the 
first oscillations of the remnant NS. In total, these ejecta amount to 
$\sim10^{-2}\usk M_\odot$ for the APR4 models and 
$\sim10^{-3}\usk M_\odot$ for the others~\footnote{For a collection of 
quantitative results on ejecta from BNS mergers, see the recent work of 
Dietrich \& Ujevic \cite{Dietrich2017} and refs. therein}. Within the errors, 
magnetic fields have negligible effect on these results. 
Using a simple analytical 
model by Grossman et al. (2014), we also obtained order-of-magnitude 
estimates for the corresponding kilonova/macronova signals. We found 
electromagnetic transients peaking around $1$--$10$ days after merger, 
with peak luminosities of ${\sim}10^{40}$~erg/s and effective 
temperatures of ${\sim}2000$--$3000$~K.

In addition to the formally unbound ejecta, we observed further matter 
outflows. These become dominant $15$--$20\msec$ after merger and, 
although slower, they can contribute significantly to the total flux 
accumulated by the end of the simulations across a spherical surface 
of radius ${\approx}300\kmeter$. In particular, for the long-lived NS 
cases ${\sim}50\msec$ after merger, the cumulative flux of formally 
bound matter can be comparable to the unbound ejecta (APR4) or 
even dominant (MS1). 
Moreover, by comparing results obtained with and without magnetic 
fields, we found that the main contribution to these outflows is magnetically 
driven. This indicates that the geodesic criterion does not apply and 
leaves the possibility that a relevant fraction of this matter could also 
become unbound at later times. 
Finally, our simulations suggest that the ongoing matter ejection will 
persist for much longer. 

\bigskip
{\it Gravitational wave emission}. 
For all our models, we analyzed the GW signal, with particular attention 
to the post-merger waveform and spectrum.
Systems forming a long-lived NS (APR4 and MS1 models) have post-merger 
waveforms of similar amplitudes which rapidly decay within ${\sim}20\msec$. 
The collapsing (H4) models show a stronger post-merger GW emission 
that is however shut off as soon as the HMNS collapses to a BH. 
We note that all models exhibit a phase jump during merger, which might 
be relevant for GW analysis.
In agreement with well established results in the literature, we found 
post-merger spectra characterized by a main peak at a frequency of 
$2$--$3$~kHz. 
While the mass ratio has minor influence on this frequency, differences 
are significant for different EOS. In particular, more compact remnants 
have a higher peak frequency. We recall that all our BNS systems have 
the same total mass at infinite separation. 
By comparing the spectra of the magnetized and non-magnetized APR4 
equal-mass models, we concluded that for BNS merger forming a 
long-lived NS, magnetic fields up to $\sim10^{16}$~G are unlikely to 
alter the GW spectrum in a detectable way. 

Although for all our models the main post-merger peak lies above the 
sensitivity curves of advanced LIGO and Virgo, the SNRs are most 
probably not sufficient for a confident detection of the post-merger 
part of the GW signal at a distance of $\sim100$~Mpc or more. 

\bigskip
{\it Outlook}. 
With the present work, we initiated a systematic investigation of BNS 
mergers leading to the formation of a long-lived NS. As suggested by 
recent observations, this case might represent a significant fraction of 
all BNS mergers. Nevertheless, it remains poorly studied in numerical 
relativity and thus more effort in this direction is urgently needed. 

The results presented here are affected by various limitations that 
should be overcome step by step in the future. 
In particular, a higher resolution is needed to better resolve the KH 
instability and possibly the MRI also inside the remnant. Moreover, an 
improved description of the microphysics including composition and 
neutrino radiation is likely to affect the structure of the NS remnant
and surrounding disk/environment, and the matter outflows. Both
improvements are also required to make conclusive statements 
about jet formation.

See Supplemental Material at http://link.aps.org/ \!supplemental/10.1103/PhysRevD.95.063016 for initial data (\texttt{LORENE} input/ouput files) and data files of the obtained gravitational waveforms.


\section*{Acknowledgments}

This work is supported by the MIUR FIR Grant No. RBFR13QJYF. 
We acknowledge PRACE for awarding us access to SuperMUC based in 
Germany at LRZ (Grant GRSimStar). Numerical simulations were also run 
on the cluster Fermi at CINECA (Bologna, Italy) via the ISCRA Grant IsB11\_MagBNS.
Support for this work was provided by the National Aeronautics and Space 
Administration through Einstein Postdoctoral Fellowship Award Number PF6-170159 
issued by the Chandra X-ray Observatory Center, which is operated by the 
Smithsonian Astrophysical Observatory for and on behalf of the National 
Aeronautics Space Administration under Contract No.~NAS8-03060.
R.P. acknowledges partial support from the NSF Grant No.~AST-1616157.

\section*{Appendix: Resolution study}

\begin{figure}
  \centering
  \includegraphics[width=0.49\textwidth]{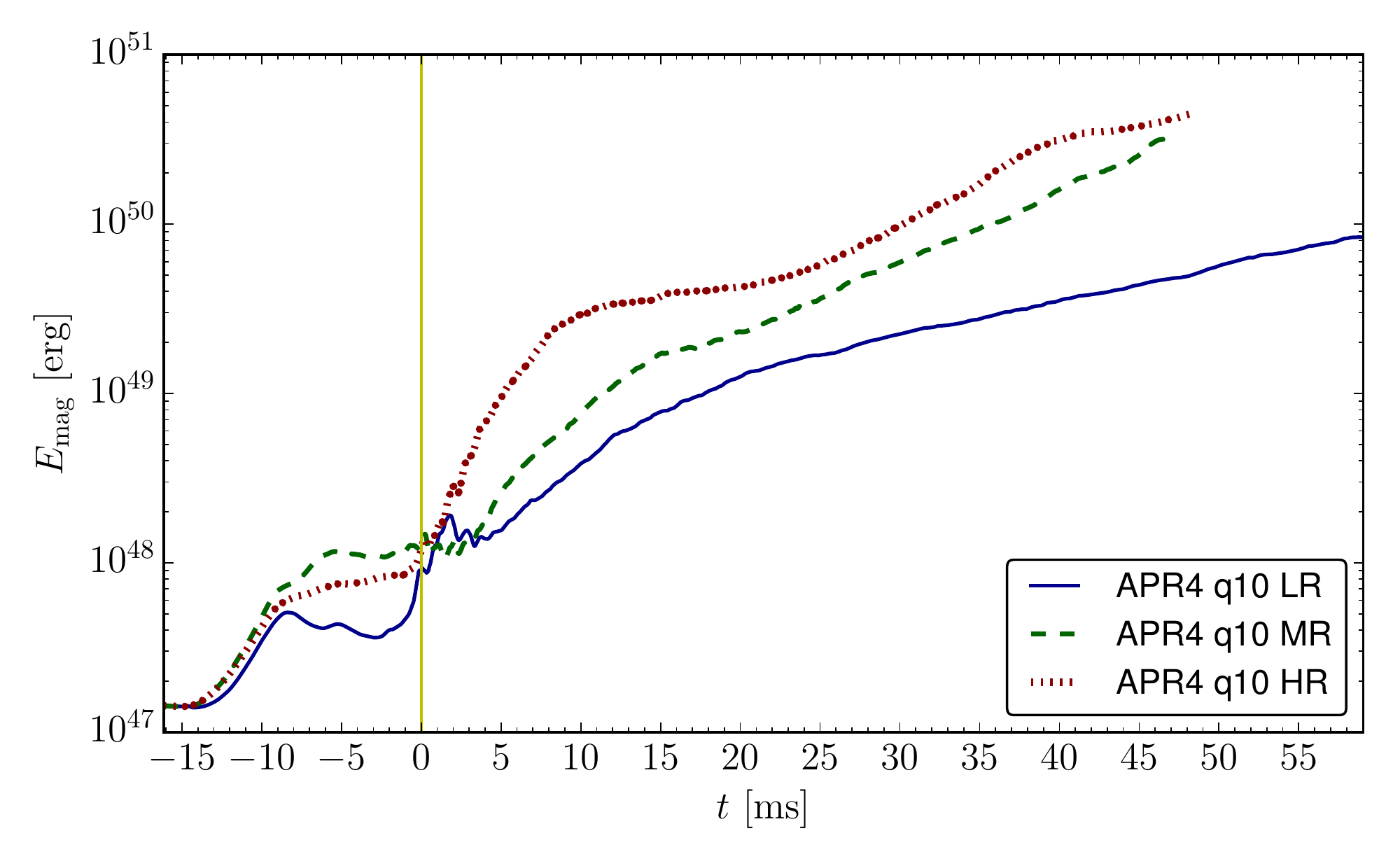}
  \caption{Evolution of magnetic energy for the equal-mass APR4 model at different resolutions: low resolution $dx=277\usk\meter$ (LR), fiducial/medium resolution $dx=222\usk\meter$ (MR), and high resolution $dx=177\usk\meter$ (HR). The vertical line marks the time of merger. 
  } 
  \label{fig:mag_1D_res}
\end{figure}

In order to estimate the numerical errors, we evolved the equal-mass APR4 model
with two additional resolutions, one higher and one lower by a factor $1.25$ than the fiducial (medium) 
resolution. The corresponding spacing of the finest grids is $277, 222$, and $177\usk\meter$.
First, we compute the error of the maximum rest-mass density during the evolution. 
We define the difference between two resolutions as
\begin{align}
  \delta_\rho \equiv \frac{\int \left(\rho_1\left(t\right) - \rho_2\left(t\right) \right)^2 \, \mathrm{d}t}{
                            \int \frac{1}{4}\left(\rho_1\left(t\right) + \rho_2\left(t\right) \right)^2 \, \mathrm{d}t}
  ,
\end{align}
where the integrals are carried out over the full duration of the simulations, and
the time coordinates are aligned at the time of the merger for each run.
We obtain relative differences of $\delta_\rho \approx 3.5\%$ between low and medium resolution and 
${\approx}1.0\%$ between medium and high resolution. This would correspond to a convergence order 
of $5.7$. Similarly, the minimum of the lapse would converge with order $4.6$.
Nevertheless, both convergence orders are clearly misleading, since the hydrodynamic evolution 
scheme is second order accurate at best, and in practice between first and second order due 
to the presence of shock waves. 
In the following, we provide error estimates under the assumption that only 
the lowest resolution is too low and that results show first order convergence starting from 
medium resolution.

We now estimate the error on the GW frequencies for the APR4 case. The average post-merger frequency $f_{10}$
differs by $0.31\%$ between low and medium resolution, and by $0.24\%$ between medium and high resolution.
From the latter results and the above assumption of linear convergence, we estimate the error 
of $f_{10}$ to be below $2\%$. 
We also note that the frequency range relevant for our results (up to $4\kHz$) corresponds 
to wavelengths resolved by at least 10 grid points at the extraction radius,
which is sufficient to prevent signal loss.

Our MS1 and H4 equal-mass models have also been studied in \cite{Hotokezaka:2013:44026},
using the same piecewise polytropic approximation of the EOS and the same thermal part, 
but without magnetic fields. 
Assuming that the impact of the magnetic field is as small as for the APR4
case, we expect to obtain similar frequencies. 
For those models, the post-merger frequency $f_{10}$ indeed agrees
within $1.3\%$ and $0.4\%$, respectively.
Our unmagnetized APR4 equal-mass model is almost the same as another model 
studied in \cite{Hotokezaka:2013:44026}, apart from a slightly different piecewise 
polytropic approximation (see \cite{Endrizzi:2016:164001}) of the APR4 EOS 
used in our work. For this model, $f_{10}$ agrees within $1.5\%$.
We conclude that within the numerical error and neglecting the influence of magnetic fields, 
our results agree well with \cite{Hotokezaka:2013:44026}.

Next, we consider the finite difference error in our estimates of the (unbound) ejected mass. 
Again, a direct measure of the convergence order yields an unrealistically large value (${\approx}10$).
The difference between medium and high resolution is $3.5\%$, and under the 
assumption of first order convergence, we obtain a total error of $17\%$. 
Note, however, that this does not include the effects of the artificial atmosphere and the 
assumptions used in the extraction. In total, we roughly estimate the mass of the unbound ejecta 
to be accurate within a factor of 2.

Finally, we consider the impact of resolution on magnetic field evolution and amplification. 
Figure~\ref{fig:mag_1D_res} shows the evolution of the total magnetic energy for 
low, medium and high resolution. The initial amplification during the inspiral seems to 
converge until the saturation phase, where the absolute differences become suddenly larger and 
convergence is gradually lost. This could mean that the saturation is due to the finite resolution
or that a physical effect causing the saturation is more difficult to resolve.
We note that a resolution study does not allow us to exclude the remote possibility that 
interaction with the artificial atmosphere is responsible for the amplification.
As expected, in the post-merger phase we are not in 
a regime of convergence. This is likely due to the unresolved small-scales at which the key amplification 
mechanisms act (in particular the KH instability). Nevertheless, 
for $t>30\msec$ we find a much better agreement between the medium and high resolutions compared 
to the low and medium resolutions. A possible explanation is that magnetic energy in this late phase is 
dominated by the contributions of MRI and winding outside the remnant, which are much better resolved
(as shown in \Fref{fig:MRI}, the resolution should be sufficient to resolve the fastest growing MRI modes).

\bibliographystyle{apsrev4-1-noeprint}

\end{document}